\documentclass[a4paper,11pt]{article}
\pdfoutput=1 

\usepackage{jcappub} 

\usepackage[T1]{fontenc} 

\title{Galaxy rotation curves in modified gravity models}

\author[a,b]{\'Alefe de Almeida}
\author[b]{Luca Amendola}
\author[b]{Viviana Niro}

\affiliation[a]{N\'ucleo Cosmo-ufes \& Departamento de F\'isica, CCE, Universidade Federal do Esp\'irito Santo,
	Av. Fernando Ferrari 514, Vit\'oria, ES, 29075-910 Brazil}
\affiliation[b]{Institute for Theoretical Physics, University of Heidelberg, Philosophenweg
	16, D-69120 Heidelberg, Germany}

\emailAdd{alefe@cosmo-ufes.org}
\emailAdd{l.amendola@thphys.uni-heidelberg.de}
\emailAdd{niro@thphys.uni-heidelberg.de}

\abstract{In this work, we investigate the possibility that the galaxy rotation curves 
	can be explained in the framework of modified gravity models that introduce a Yukawa term in the gravitational potential.  We include dark matter and 
	assume that the fifth-force couples differently  to dark matter and to baryons. 
	We aim at constraining the modified gravity parameters $\beta$ and $\lambda$, that is, 
	the strength   and the range of the Yukawa fifth force, respectively,
	using a  set of 40 galaxy rotation curves data from the SPARC catalogue. 
	We include baryonic gas, disk and bulge components, along with a NFW halo of dark matter. Each galaxy rotation curve is modeled with  three free parameters, beside the two global Yukawa parameter. We find that
	the inclusion of the Yukawa term improves the  $\chi^2$ from $680.75$ to $536.23$ for $655$ degrees of freedom. As global best-fit we obtain $\beta = 0.34\pm0.04$ and $\lambda = 5.61\pm0.91$kpc and a dark matter content on average 20\% smaller than without the Yukawa term. The Bayesian evidence in favor of a NFW profile plus Yukawa term is higher than 8$\sigma$ with respect to the standard gravity parametrization.}

\arxivnumber{1805.11067}

\begin{document}
\maketitle
\flushbottom

\section{Introduction }

The observations of dynamics of the galaxies, in particular their
rotation curves, constitute one of the main evidences for dark matter.
However, also alternatives to Newtonian gravity have been employed
to address the dark matter problem (e.g. ref.\cite{1963MNRAS.127...21F,1983ApJ...270..371M,1998ApJ...508..132D,
	2002ARA&A..40..263S,milgrom2006mond,2006JCAP...03..004M,rodrigues2010galaxy,
	2006ApJ...636..721B,2007MNRAS.375.1423C,2007MNRAS.381.1103F,2008PhRvD..77l3534B,
	2012PhRvD..85l4020M}),
indicating that the underlying gravity at galactic scales may be not
Newtonian and a different gravity should be considered instead of a new
material component. In these models, since galaxies can be approximated
within the weak-field regime of relativity, the problem can be expressed
as a modified Newtonian potential. In particular, most work has been
performed adopting a potential with a Yukawa-like correction
\begin{equation}
\Phi(\textbf{x})=-G\int\frac{\rho(\textbf{x}')}{|\textbf{x}-\textbf{x}'|}\left(1+\beta e^{-|\textbf{x}-\textbf{x}'|/\lambda}\right)d^{3}\textbf{x}'\;.\label{yukawa1}
\end{equation}
where $\rho(x)$ is the matter density distribution.
The parameter $\beta$ measures
the strength of the `fifth-force' interaction while the second parameter, $\lambda$,
gives its range. A Yukawa-like correction to the Newtonian potential is predicted by 
several  modified gravity theories, see 
e.g.\cite{1984A&A...136L..21S,1979PhLB...88..265S,2001PhRvD..63d3503D,2006JCAP...03..004M,2007JPhCS..91a2007C,2008PhRvD..77b4041C,2010CQGra..27w5020C,2011MNRAS.414.1301C,2011PhRvD..84l4023S,2013MNRAS.436.1439M,2013PhRvD..87f4002S}). 
The extra interaction  could
be mediated by scalars \cite{2007JPhCS..91a2007C,2008PhRvD..77b4041C,2011MNRAS.414.1301C,2011PhRvD..84l4023S,2013PhRvD..87f4002S},
massive vectors\cite{1984A&A...136L..21S,1979PhLB...88..265S,2006JCAP...03..004M,2013MNRAS.436.1439M} or
even with two rank-2 tensor field\cite{2001PhRvD..63d3503D,2010CQGra..27w5020C}.

Let us review some of the previous research in this field, in order to
highlight the differences to our work. 
Ref.\cite{2011PhRvD..84l4023S} investigated the sum of a repulsive and an attractive Yukawa interaction obtained as Newtonian limit of a higher-order gravity model. 
Ref.\cite{2011PhRvD..84l4023S}  did not fit the Yukawa parameters to the data but fixed them to   $\lambda_1=100\;\text{kpc},\lambda_2=10^{-2}\;\text{kpc}$, and  $\beta_1=1/3, \beta_2=-4/3$, showing that these values provide a good fit to the Milky Way and to NGC 3198. They also find that modified gravity interacting  with baryons only is not sufficient to reproduce the observed behaviour of galaxy rotation curves and some amount of dark matter is still  needed.
In contrast, in ref.\cite{1984A&A...136L..21S} (see also  ref.\cite{2013PhRvD..87f4002S})
it was shown that repulsive Yukawa corrections could produce constant
profiles for rotation curves at large radii without  
dark matter, if $-0.95\leq \beta \leq-0.92$ and $\lambda \simeq 25-50$~kpc.

Following ref.\cite{1984A&A...136L..21S}, most other works did not include dark matter. While refs.\cite{2001PhRvD..63d3503D,2011PhRvD..84l4023S}  applied the prediction from modified gravity to just one or two 
galaxy, ref.\cite{2013MNRAS.436.1439M} performed a full comparison against observed rotation curves datasets of nine galaxies  belonging to THe HI Nearby Galaxy Survey 
catalogue (THINGS). Different values for $\beta$ and $\lambda$ were used for the different galaxies and then the averaged values, 
$\beta=-0.899 \pm 0.003$  and $\lambda=23.81 \pm 2.27$~kpc, were finally employed to fit the galaxies of the Ursa Major 
and THINGS catalogue. 
This analysis was carried out without  a dark matter component. The THINGS catalogue was also implemented in 
ref.\cite{2014PhRvD..89j4011R}, obtaining similar results, $\beta=-0.916 \pm 0.041$ and $\lambda=16.95 \pm 8.04$~kpc, again without  dark matter.

Positive values of $\beta$ (i.e. an attractive fifth force) were instead obtained in a few works. In 
ref.\cite{2011PhRvD..83h4038M}, Low Surface Brightness (LSB) galaxies were used.  
Considering only positive $\beta$ in the fit, values from 1.83 to 11.67 were found, while the $\lambda$ values ranged from 
0.349~kpc to  75.810~kpc. 
In ref.\cite{2011MNRAS.414.1301C}, instead, the value of $\beta$ was kept fixed, either to 1/3 as predicted in $f(R)$ gravity models, or to $1$ as a comparison test, and the Yukawa potential was employed to analyse simulated datasets. 
The main aim of ref.\cite{2011MNRAS.414.1301C} was to investigate the bias induced by the assumption of a wrong gravitational theory: they conclude that the disc mass is underestimated and there is a high bias 
on the halo scale length and the halo virial mass. A strength $\beta=1/3$  has been found to be compatible also with the dynamics of clusters  (\cite{Capozziello:2008ny}), again without dark matter. 

Our work however  is closer in spirit to ref.\cite{2003PhRvL..91n1301P}, one of the few papers including dark matter. There, it was shown that a repulsive Yukawa interaction between baryons and dark matter gives a good fit  for the following galaxies: 
UGC 4325, $\beta$=$-1.0\pm 0.25$ and $\lambda$=$1.7 \pm 0.6$~kpc; NGC 3109, $\beta$=$-1.1\pm 0.16$ and $\lambda$=$1.2 \pm 0.17$~kpc; 
LSBC F571-8, $\beta$=$-0.9\pm 0.18$ and $\lambda$=$1.1 \pm 0.1$~kpc; NGC 4605, $\beta$=$-1.1\pm 0.3$ and $\lambda$=$0.2 \pm 0.02$~kpc. In this paper, however, no baryon component is included and the dark matter profile is taken as a simple power law.

As in previous work, we also wish to constrain  $\beta$ and $\lambda$
with galaxy rotation curves data. However, the present paper differentiates itself from most earlier research because  \emph{a}) we do not exclude
dark matter, \emph{b}) we assume that the fifth-force couples differently
to dark matter and  to baryons, without restriction on the coupling sign, \emph{c}) we include baryonic
gas, disk and bulge components, according to observations, separately
for each galaxy, along with a NFW dark matter halo profile,  \emph{d}) we do not fit $\beta,\lambda$ individually
to each galaxy, but rather look for a global fit, and finally, \emph{e})  we adopt a much larger datasets than earlier work, namely 40 galaxies from the Spitzer Photometry $\&$ Accurate
Rotation Curves (SPARC)\cite{2016AJ....152..157L}.  In the next section we discuss some aspects of  these points.

We anticipate our main result. While most previous analyses both without dark matter \cite{1984A&A...136L..21S, 2013MNRAS.436.1439M,2014PhRvD..89j4011R} and with dark matter \cite{2003PhRvL..91n1301P}, find a best fit for $\beta\approx -1$, i.e. repulsive fifth force,  we find $\beta=0.34\pm0.04$, corresponding to an attractive fifth force. Moreover we find that the dark matter mass is on average 20\% lower than without the Yukawa correction.
It is clear however that this result cannot be straightforwardly interpreted as a rejection of standard gravity, since 
it is based on a set of parametrizations of the gas, bulge, disk and dark matter profiles that, although realistic, is not yet general enough to include all possibilities.

\section{A species-dependent coupling}

If the fifth force is felt differently by baryons (subscript $b$) and dark matter ($dm$), one needs to introduce two coupling constants, say $\alpha_b$ and $\alpha_{dm}$. To fix the ideas, let us assume the fifth force is carried by a scalar field with canonical kinetic term and conformal coupling. Then the particles will obey geodesic equations of the form
\begin{eqnarray}
T^{\mu}_{(b)\nu;\mu}&=&-\alpha _b T_{(b)} \phi_{;\nu}\\
T^{\mu}_{(dm)\nu;\mu}&=&-\alpha_{dm} T_{(dm)} \phi_{;\nu}
\end{eqnarray} 
where $T_{(x)\nu}^{\mu}$ is the energy-momentum tensor of component $x$ and $T_{(x)}$ its trace.
The scalar field obeys instead a Klein-Gordon equation which can be written as
\begin{equation}
T^{\mu}_{(\phi )\nu;\mu}=(\alpha_b T_{(b)}+\alpha_{dm} T_{(dm)}) \phi_{;\nu}
\end{equation} 
The total energy-momentum tensor is clearly conserved.

In the so-called linear quasi-static approximation, i.e. when we can disregard the propagation of $\phi$ waves,  the total potential  between two particles of species $x,y$ acquires a Yukawa term as in Eq. \ref{yukawa1}, with strength \cite{1992CQGra...9.2093D,Amendola:2003wa,2003PhRvL..91n1301P}
\begin{equation}
\beta=\alpha_x \alpha_y
\end{equation}
and universal range $\lambda=m^{-1}$, where $m$ is the  scalar field mass. In a galaxy, the baryonic component follows rotation curves that, in equilibrium, are determined by the sum of the potentials produced by the baryons themselves and by the dark matter component. As a consequence, baryons feel a fifth force which is the sum of the baryon-baryon force and the baryon-dark matter one. The first is proportional to $\alpha_b^2$, while the second one to $\alpha_b\alpha_{dm}$. Local gravity experiment, however, show that $|\alpha_b|$ has to be very small, typically less than $10^{-2}$ \cite{2001LRR.....4....4W,2016ChPhC..40j0001P}.   We can therefore neglect the baryon-baryon fifth force, i.e. assume that baryons exert just the standard gravitational force on the other baryons. Notice that if we cannot invoke a screening mechanism to screen the fifth force in, say, the solar system, and therefore evade the local constraints, because then the same mechanism would presumably also screen stars, which then would follow standard rotation curves, rather than those modified by the fifth force. All this means that the rotation curves only depend on the Yukawa strength $\beta=\alpha_b\alpha_{dm}$. This, along with $\lambda$, is the parameter we wish to determine. Since it turns out that $\beta$ is of order unity, we conclude that $\alpha_{dm}$ must be very large, ${\cal O}(100)$. We will not discuss whether this large value is compatible with other constraints, e.g. from cosmology. However we notice that since we find $\lambda\approx 6$ kpc, any observation involving scales much larger than this will see a Yukawa force suppressed as $\exp(-r/\lambda)$, and therefore negligible.
Finally, we notice that since we are assuming no screening, the values of $\beta,\lambda$ do not change from galaxy to galaxy. That is why we do not try to fit the values individually to each galaxy, but rather seek a global fit.  For technical
reasons, to be discussed later, however, we decided to select 40 galaxies and group them in four datasets
of ten each, and find the best Monte Carlo fit for each group. Even in this way, the complexity is quite high, since we deal
with a number of simultaneously-varying parameters for each group ranging from 23 up to 25.

In many previous works, as we have seen in the previous section, dark matter was not considered. Therefore, following the interpretation given above, what has been measured was the baryon-baryon strength $\beta=\alpha_b^2$. Since the fits to rotation curves  have  mostly provided a negative value, one has to modify the picture above by introducing either a non-canonical kinetic term (actually, a field with imaginary sound speed, which then suffers of a gradient instability), or a vector boson rather than a scalar one. In any case, a value $|\alpha_b|$ of order unity is in contrast with local gravity constraints. 
The only way to make these models consistent with local gravity constraints is to assume then
that experiments on Earth are screened while stars are not. 
For example, in ref.\cite{2017PhRvD..95f4050B} is considered that the fifth force, at galactic scales, is produced by 
the symmetron scalar field \cite{2010PhRvL.104w1301H}. 
Four galaxies from the SPARC data set were used for the fit and no dark matter was considered. 
The deviations from standard gravity at galactic scales are screened in high density environments, e.g. solar system \cite{2018arXiv180505226O}. 
Furthermore, in ref.\cite{2018arXiv180505226O} it is shown that the symmetron field can explain the vertical motion of stars in the Milky Way, without dark matter.

\section{The Yukawa correction }

\label{section2}

The Yukawa-like corrections to the Newtonian potential has the general
form 
\begin{equation}
\Phi(\textbf{x})=-G\int\frac{\rho(\textbf{x}')}{|\textbf{x}-\textbf{x}'|}\left(1+\beta e^{-|\textbf{x}-\textbf{x}'|/\lambda}\right)d^{3}\textbf{x}'\;.\label{yukawa}
\end{equation}

Clearly, we recover Newtonian gravity when $\beta=0$, or at scales
much larger than $\lambda$. In the case of scales much smaller
than $\lambda$, gravity could be stronger or weaker than Newtonian,
depending on the sign of $\beta$, that we leave free in our analysis.

We will assume a spherical distribution for dark matter derived from
$N$-body simulations of cold dark matter (CDM), the Navarro-Frenk-White
profile (hereafter NFW)\cite{1996ApJ...462..563N}
\begin{equation}
\rho_{\text{NFW}}(r)=\frac{\rho_{s}}{\frac{r}{r_{s}}(1+\frac{r}{r_{s}})^{2}}\;,\label{nfw}
\end{equation}
where $\rho_{s}$ is the characteristic density and $r_{s}$ is the
scale radius. In principle, these parameters are independent but several
$N$-body simulations(e.g.\cite{2008MNRAS.391.1940M,Maccio:2008pcd})
claims that there is a relation between them. This relation is usually
parametrized by the concentration parameter $c\equiv r_{200}/r_{s}$
and $M_{200}\equiv(4\pi/3)200\rho_{\text{crit}}r_{200}^{3}$, where
$\rho_{\text{crit}}$ is the critical density. Hence, the NFW profile
can be written in terms of a single parameter, namely $M_{200}$.
Thus, we can relate $(\rho_{s},r_{s})\rightarrow(c,M_{200})$ via\cite{mo2010galaxy}
\begin{align}
\rho_{s} & =\frac{200}{3}\frac{c^{3}\rho_{\text{crit}}}{\ln(1+c)-\frac{c}{1+c}}\;,\label{rhos}\\
r_{s} & =\frac{1}{c}\left(\frac{3M_{200}}{4\pi200\rho_{\text{crit}}}\right)^{1/3}\;\label{rs}
\end{align}
We will then assume, for galaxy-sized halos, the following $c-M_{200}$ relation\cite{2008MNRAS.391.1940M} 
\begin{equation}
c(M_{200})=10^{0.905}\left(\frac{M_{200}}{10^{12}h^{-1}\text{M}_{\odot}}\right)^{-0.101}\;.
\end{equation}
Hence, the equation (\ref{rs}) becomes 
\begin{equation}
r_{s}\approx28.8\left(\frac{M_{200}}{10^{12}h^{-1}\text{M}_{\odot}}\right)^{0.43}\text{kpc}\;,
\end{equation}
where we used $\rho_{\text{crit}}=143.84\;\text{M}_{\odot}/\text{kpc}^{3}$
and $h=0.671$\cite{2016A&A...594A..13P}.

The gravitational potential can
be computed inserting the equation (\ref{nfw}) in the equation (\ref{yukawa}).
The potential $\Phi$ can be written as a sum of the usual Newtonian
potential for a NFW profile, $\Phi_{\mathrm{NFW}}$, plus a modified
gravity part $\Phi_\text{mg}$ ($\propto\beta$) that can be integrated
analytically for NFW \cite{Pizzuti:2017diz}
\begin{align}
\Phi_{\text{mg}}(r) & =\frac{2\pi G\beta\rho_{s}r_{s}^{3}}{r}\Bigg\{\exp\left(-\frac{r_{s}+r}{\lambda}\right)\left[\text{Ei}\left(\frac{r_{s}}{\lambda}\right)-\text{Ei}\left(\frac{r_{s}+r}{\lambda}\right)\right]+\nonumber \\[1ex]
& -\exp\left(\frac{r_{s}+r}{\lambda}\right)\text{Ei}\left(-\frac{r_{s}+r}{\lambda}\right)+\exp\left(\frac{r_{s}-r}{\lambda}\right)\text{Ei}\left(-\frac{r_{s}}{\lambda}\right)\Bigg\}\;,
\end{align}
where  $\text{Ei}(x)$ is defined as 
\begin{equation}
\text{Ei}(x)=-\int_{-x}^{\infty}\frac{e^{-t}}{t}dt\;,
\end{equation}
In the following, we refer to the parametrization discussed in this section  (specifically, the gas, disk, and bulge baryonic components, plus the NFW profile for the 
dark matter with mass-concentration relation, plus the Yukawa term coupled to dark matter only) simply as the Yukawa model, and we will compare it to the ``standard model'', 
i.e. the same parametrization but without the Yukawa term.

\section{The data sample}\label{section3}

As already mentioned, the observational data for the rotation curves considered in this
work are taken from the catalogue Spitzer Photometry \& Accurate Rotation Curves (SPARC)\cite{2016AJ....152..157L}, which contains 175 disk galaxies. This catalogue includes observations at near-infrared (3.6$\mu$m), which can trace the stellar distribution, and high-quality rotation curves from HI/H$\alpha$ measurements. The HI regions are measured by the 21-cm line hyperfine transition of the neutral atomic hydrogen while the HII regions are visible due the emission of the H$\alpha$ line of the ionized gas. 

To evaluate the kinematics of disk galaxies, the components which
are relevant to the analysis are: gas, disk, bulge and dark matter
halo. The main physical quantity is the total circular velocity at
the galactic plane, $V_{c}$, which is related to  the total gravitational
potential, $\Psi$, via 
\begin{equation}
V_{c}^{2}=r\frac{d\Psi}{dr}\;.\label{circularvel}
\end{equation}
Here, as already discussed, we will assume that the potential which
produces the acceleration for the baryonic components is the Newtonian
one, while  for dark matter is given by equation (\ref{yukawa}).
Thus, the total gravitational potential reads 
\begin{equation}
\Psi=\Phi_{\text{gas}}+\Phi_{\text{disk}}+\Phi_{\text{bulge}}+\Phi_{\text{NFW}}+\Phi_{\text{mg}}\;.
\end{equation}
The linearity of equation (\ref{circularvel}) allows us to split
the total circular velocity in terms of each component. We have therefore
the following formula 
\begin{equation}
V_{c}^{2}(r)=V_{\text{gas}}^{2}(r)+\Upsilon_{*\text{D}}V_{\text{disk}}^{2}(r)+\Upsilon_{*\text{B}}V_{\text{bulge}}^{2}(r)+V_{\text{NFW}}^{2}(r)+V_{\text{mg}}^{2}(r)\;,\label{circular}
\end{equation}
where $\Upsilon_\text{*D}$ ($\Upsilon_\text{*B}$) is the stellar mass-to-light ratio for the disk (bulge) and 
it is equivalent to the mass $M_\text{D}$ of the 
disk ($M_\text{B}$ of the bulge) divided by the luminosity $L_\text{D}$ of the disk ($L_\text{B}$ of the bulge).

The rotation velocities $V_{\text{gas}},V_{\text{disk}}$ and $V_{\text{bulge}}$
are already available on SPARC for each galaxy on \href{http://astroweb.cwru.edu/SPARC}{astroweb.cwru.edu/SPARC}.
As an example we show in the table \ref{tab:sparcex} the values of
each component at the observed radius for a single galaxy. The details about how each baryonic component was derived can be found in ref.\cite{2016AJ....152..157L} but we will discuss briefly here. 
Using the redshift or blueshift of the HI emission line for the outer region of the galaxy and also the H$\alpha$ measurements for the inner region it is possible to obtain a high-quality rotation curve data. The gaseous component, $V_\text{gas}$, is obtained from the superficial density distribution, $\Sigma_\text{gas}$, inferred also by HI observations. Then, solving the Poisson equation for $\Phi_\text{gas}$ and using  equation (\ref{circularvel}) we have $V_\text{gas}$. The computation of $V_\text{disk}$ and $V_\text{bulge}$ is analogous:  the surface brightness, $I(r)$, of each stellar component (disk and bulge) is directly measured at the near-infrared (3.6$\mu$m) band. At this band, the surface density of each stellar component, namely $\Sigma_\text{disk}$ and $\Sigma_\text{bulge}$, is proportional to the surface brightness, i.e. $\Sigma(r)=\Upsilon_*I(r)$, where $\Upsilon_*$ is the mass-to-light ratio of the respective component. Hence, from $\Sigma$ one obtains $V_\text{disk}$ and $V_\text{bulge}$ in the same way as for the gaseous component. For practical reasons, in the derivation of $V_\text{disk}$ and $V_\text{bulge}$ it is assumed $\Upsilon_\text{*D}=\Upsilon_\text{*B}=1$, hence with this normalization of the mass-to-light ratios the problem can be rescaled trivially for any $\Upsilon_\text{*D},\Upsilon_\text{*B}
$, which are then additional free parameters.

Since we are considering that the baryons are not coupled to the fifth force, the procedure  by ref.\cite{2016AJ....152..157L} can be directly adapted to our purpose. Finally, in order to obtain the velocities of each component at every radius $r$ we perform a cubic spline interpolation.

\begin{table}[tbp]
	\centering 
	\begin{tabular}{|l|c|c|c|c|}
			\hline 
			Radius  & $V_{\text{obs}}$  & $V_{\text{gas}}$  & $V_{\text{disk}}$ & $\Sigma_{\text{disk}}$ \\
			(kpc)  & $(\text{km s}^{-1})$  & $(\text{km s}^{-1})$  & $(\text{km s}^{-1})$  & $(L_{\odot}\text{pc}^{-2})$\\
			\hline 
			0.42  & $14.2\pm1.9$  & 4.9  & 4.8  & 11.0  \\\hline
			1.26  & $28.6\pm1.8$  & 13.1  & 10.8  & 5.8  \\\hline
			2.11  & $41.0\pm1.7$  & 19.6  & 13.6  & 2.7  \\\hline
			2.96  & $49.0\pm1.9$  & 22.4  & 13.3  & 1.0  \\\hline
			3.79  & $54.8\pm2.0$  & 22.8  & 12.6  & 0.7  \\\hline
			4.65  & $56.4\pm3.1$  & 21.4  & 12.3  & 0.4  \\\hline
			5.48  & $57.8\pm2.8$  & 18.7  & 12.0  & 0.2  \\\hline
			6.33  & $56.5\pm0.6$  & 16.7  & 10.6  & 0.0  \\
			\hline 
	\end{tabular}%
	\caption{\label{tab:sparcex} Table for the galaxy UGCA442 emphasizing each baryonic component (there is no bulge in this galaxy ).  $\Sigma_{\text{disk}}$ is the surface density for the disk.}
\end{table}

Finally, the components $V_{\text{NFW}}$ and $V_{\text{mg}}$ are
given by 
\begin{equation}
V_{\text{NFW}}^{2}(r)=\frac{4\pi Gr_{s}^{3}\rho_{s}}{r}\left[-\frac{r}{r+r_{s}}+\ln\left(1+\frac{r}{r_{s}}\right)\right]
\end{equation}
and \footnote{Unfortunately the  version of this equation published on JCAP contained some typos.}
\begin{align}
V_{\text{mg}}^{2}(r) & =-\frac{2\pi G\beta\rho_{s}r_{s}^{3}}{r}\Bigg\{\frac{2r}{r_{s}+r}+\exp\left(\frac{r_{s}+r}{\lambda}\right)\left(\frac{r}{\lambda}-1\right)\text{Ei}\left(-\frac{r_{s}+r}{\lambda}\right)+\nonumber \\[1ex]
& +\exp\left(-\frac{r_{s}+r}{\lambda}\right)\left(1+\frac{r}{\lambda}\right)\left[\exp\left(\frac{2r_{s}}{\lambda}\right)\text{Ei}\left(-\frac{r_{s}}{\lambda}\right)+\text{Ei}\left(\frac{r_{s}}{\lambda}\right)-\text{Ei}\left(\frac{r+r_{s}}{\lambda}\right)\right]\Bigg\}\;.
\end{align}

\section{Fitting the rotation curves }

\label{section4} We present now the procedure to find the best-fit
values for the set of free parameters, namely $\{\Upsilon_{*\text{D}},\Upsilon_{*\text{B}},M_{200}\}$
for each galaxy plus $\{\beta,\lambda\}$, against the observational
data. It is assumed here that the errors of the observed rotation
curve data follow a Gaussian distribution, so that we can build the
likelihood for each galaxy as follows 
\begin{equation}
\mathcal{L}_{j}(p_{j},\beta,\lambda)=(2\pi)^{-N/2}\Bigg\{\prod_{i=1}^{N}\sigma_{i}^{-1}\Bigg\}\exp\Bigg\{-\frac{1}{2}\sum_{i=1}^{N}\Bigg(\frac{V_{\text{obs},j}(r_{i})-V_{c}(r_{i},p_{j},\beta,\lambda)}{\sigma_{i}}\Bigg)^{2}\Bigg\}\label{likelihood}
\end{equation}
where $p_{j}=\{\Upsilon_{*\text{D},j},\Upsilon_{*\text{B},j},M_{200,j}\}$,
$N$ is the number of observational points for each galaxy, $\sigma_{i}$
is the data error, $V_{\text{obs},j}(r_{i})$ is the observed circular
velocity of the $j$-th galaxy at the radius $r_{i}$ and $V_{c}(r_{i},p_{j},\beta,\lambda)$
is the total rotation curve, which was expressed in equation (\ref{circular}). 
We use the values of $V_{\text{obs},j}(r_{i})$ as provided by 
the SPARC catalogue. 
To constrain $\beta$ and $\lambda$ with respect to some set of galaxies
it is necessary to consider an overall likelihood $\mathcal{L}$.
Since the observational data measurements of the galaxies are independent,
the overall likelihood function can be computed just multiplying the
distributions for each galaxy. Hence, the full likelihood function
is given by 
\begin{equation}
\mathcal{L}(\textbf{p},\beta,\lambda)=\prod_{j=1}^{N_{g}}\mathcal{L}_{j}(p_{j},\beta,\lambda)\;,\label{fulllikelihood}
\end{equation}
where $\textbf{p}=\{p_{1},...,p_{N_g}\}$ and $N_{g}$ is the total
number of galaxies.

According to the Bayes theorem, the posterior distribution is proportional
to the prior times the likelihood. We assume here uniform prior for
each parameter. Namely, the stellar mass-to-light ratios, for disk
and bulge, are restricted $0.3<\Upsilon_{*\text{D}}<0.8$ and $0.3<\Upsilon_{*\text{B}}<0.8$,
in agreement with stellar population model analysis\cite{2014ApJ...788..144M,2014PASA...31...36S}.
A wide range for the other parameters is considered: $10^{9}<M_{200}/\text{M}_{\odot}<10^{14}$,
$-2<\beta<2$ and $\overline{\lambda}_{0}<\lambda/\text{kpc}<100$,  where $\overline{\lambda}_{0}$
is the mean value among the smallest observable radii when the $N_g$ galaxies are considered. The lower limit on $\lambda$  is imposed
to avoid undesired divergences when $\lambda\rightarrow0$.

There are several methods in literature for sampling the parameter
space, starting with the well-known Metropolis-Hastings\cite{1953JChPh..21.1087M}
algorithm. In this work we chose the affine-invariant ensemble sampler
proposed in ref.\cite{goodman2010ensemble}, which was implemented
by the stable and well tested open-source Python package \texttt{emcee}\cite{2013PASP..125..306F}.
According ref.\cite{goodman2010ensemble} this affine-invariant sampler
offers several advantages over traditional MCMC sampling methods e.g.
high performance and a hand-tuning of few parameters compared to traditional
samplers.

\section{Analysis and Results}

\label{section5}

The SPARC catalogue contains 175 galaxies, hence a complete analysis
would require 384 free parameters: $175\times(\Upsilon_{\text{*D}},M_{200})+ 32\times\Upsilon_{\text{*B}}$ (many galaxies do not show a bulge) plus $ \beta,\lambda$. Even for the affine-invariant ensemble
sampler, this number of parameters is quite high and the sampling
of the likelihood becomes computationally too expensive. Thus, we
decided to analyse 4 random sets of 10 galaxies each. For two sets (B and D) we have then 25 free parameters each, while for the other two sets we have 23 parameters each.
The calculation of one set demands roughly one day of computation on a machine with 4 CPUs and 16 gigabytes of RAM. 

\begin{figure}[tbp]
	\centering 
	\includegraphics[width=7cm]{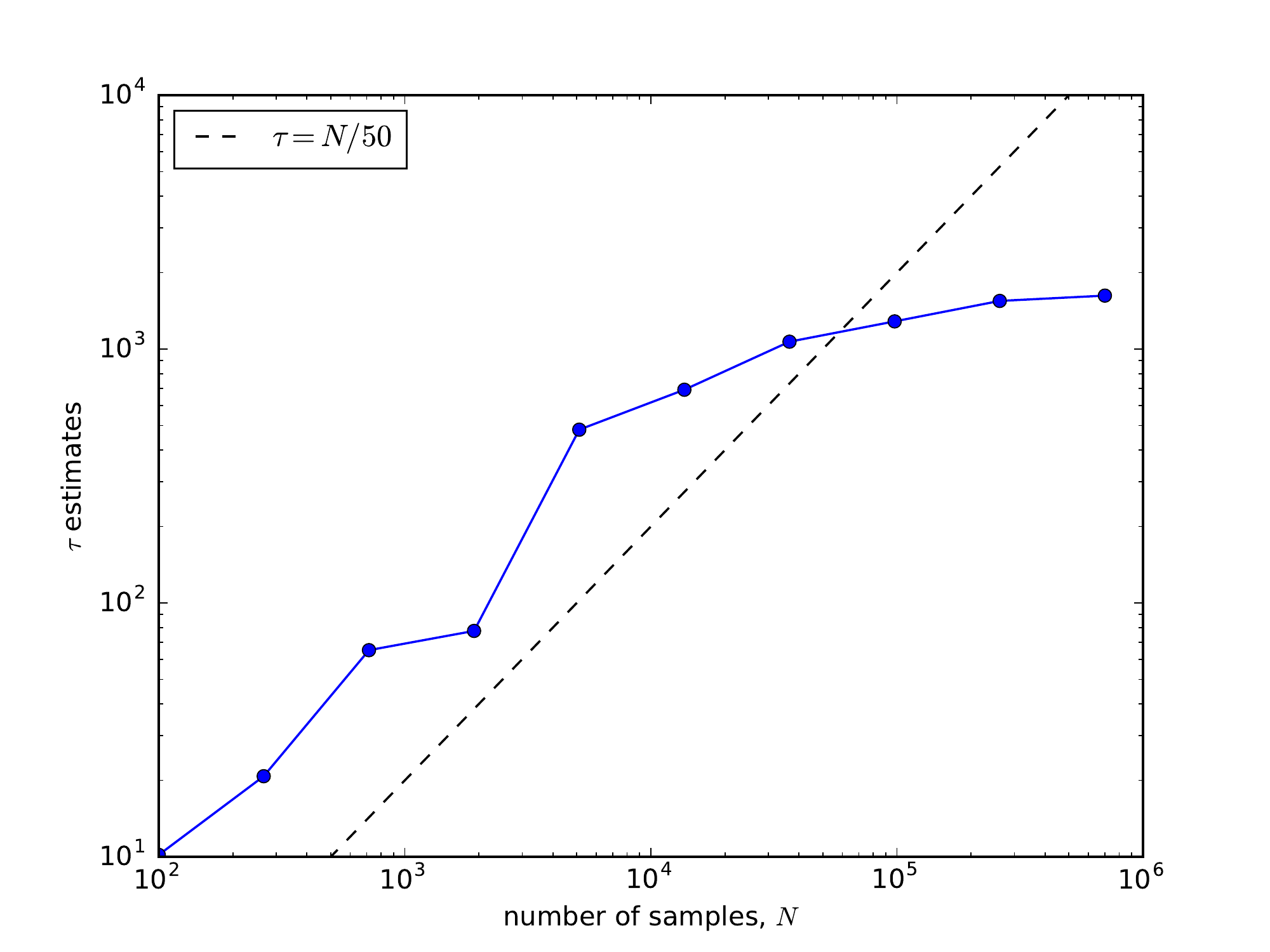} \includegraphics[width=7cm]{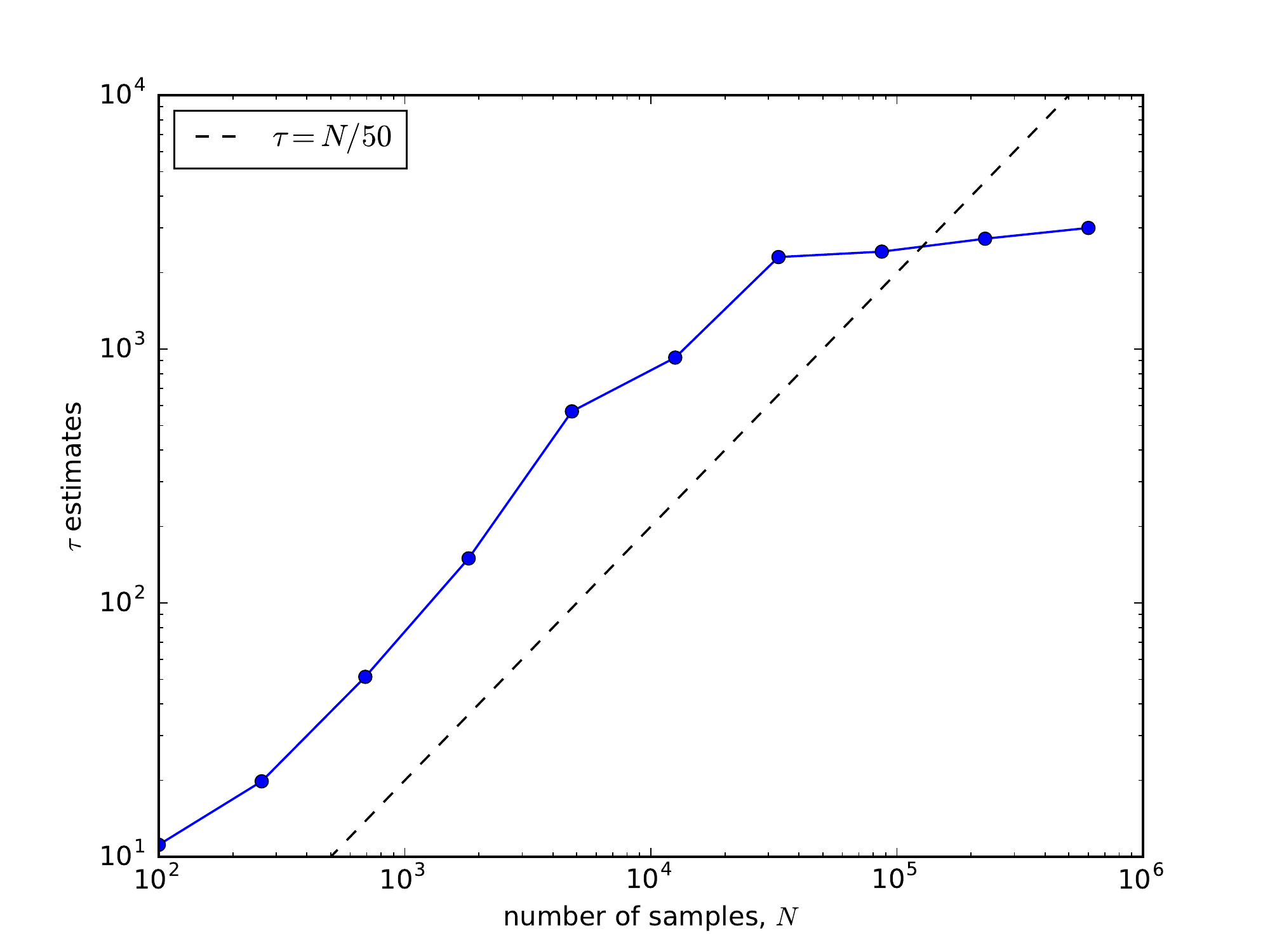}\\
	\includegraphics[width=7cm]{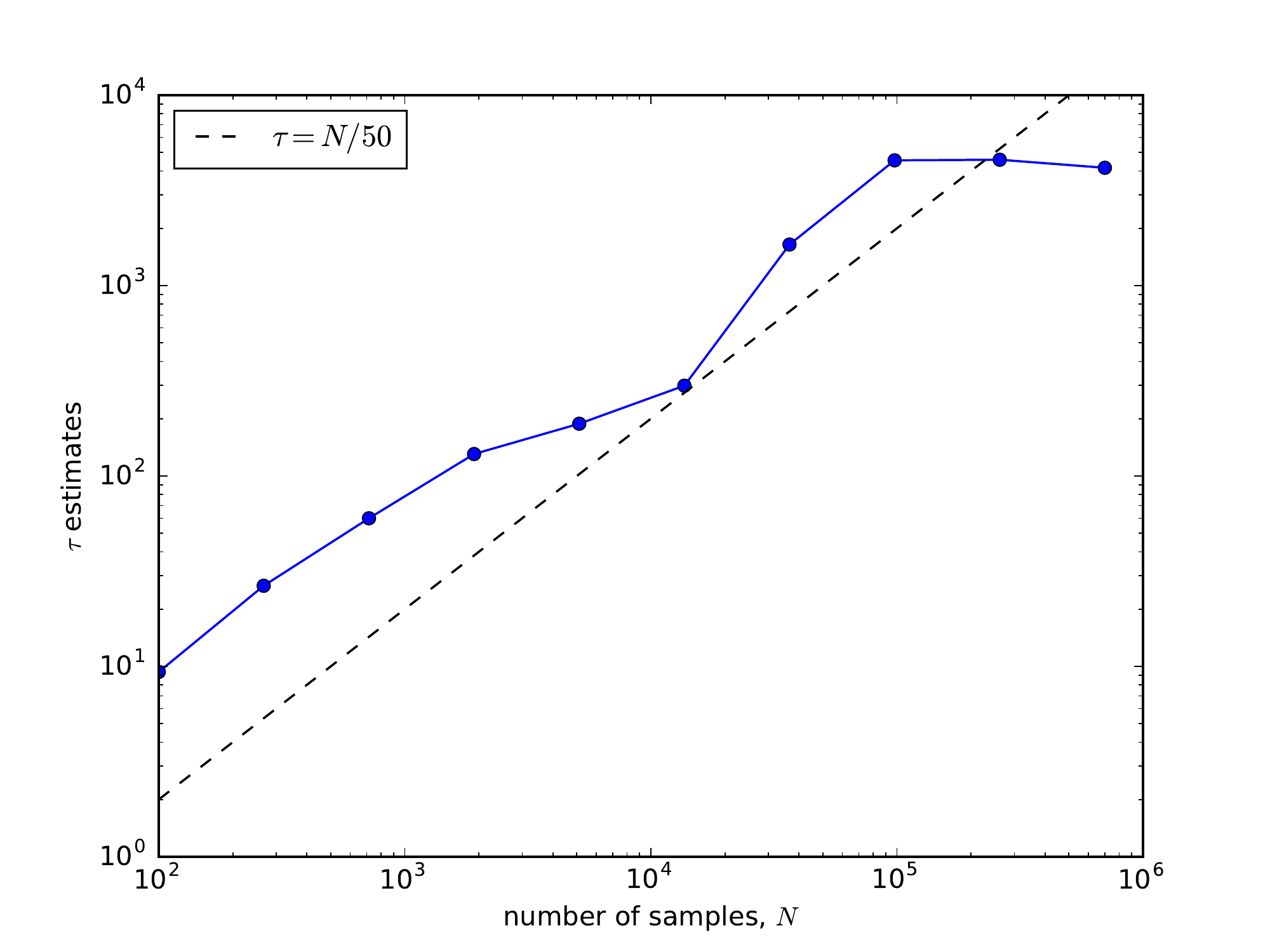} \includegraphics[width=7cm]{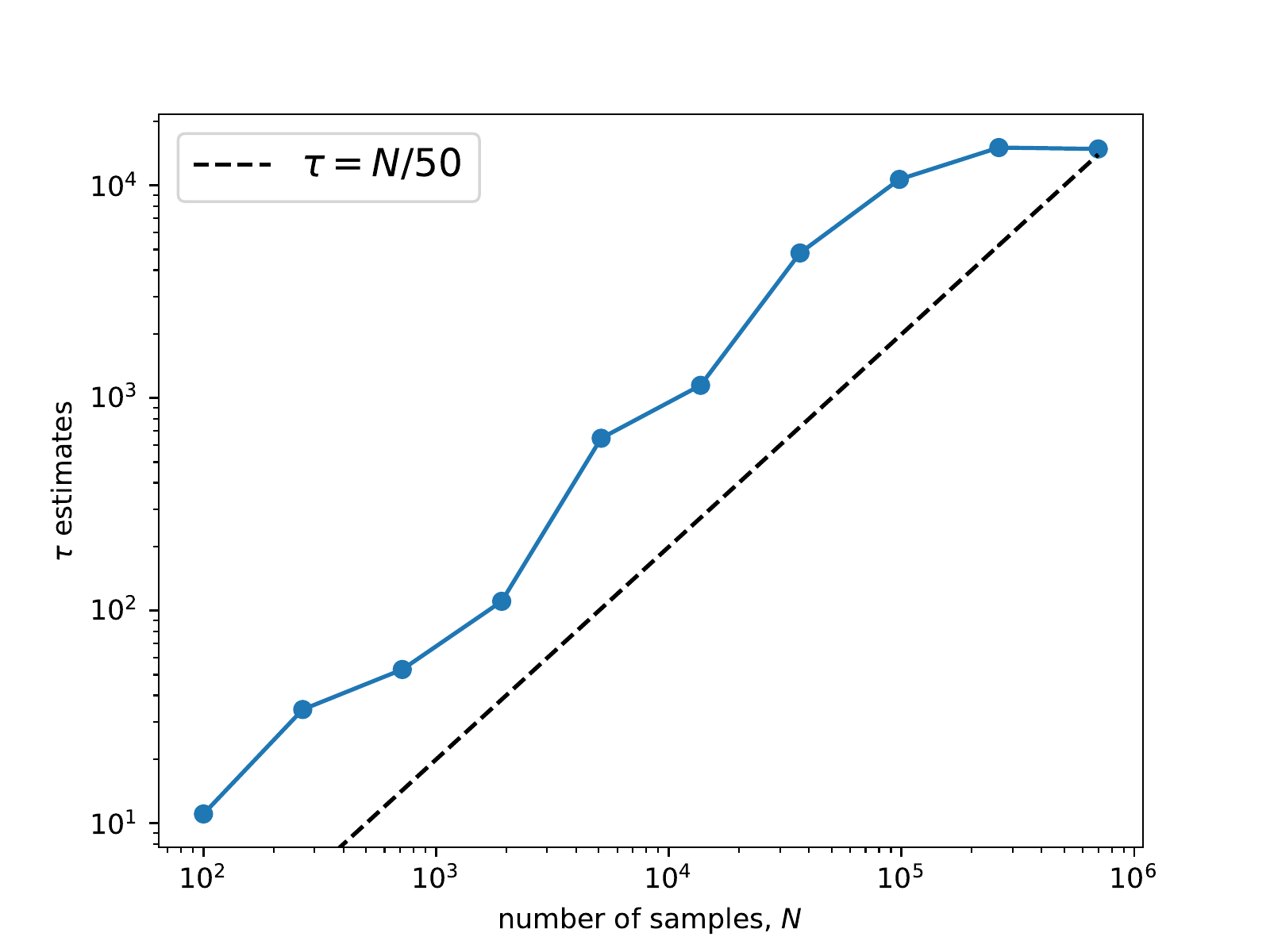}
	\caption{\label{autocorrelation}The autocorrelation time analysis (blue solid line)
		for each set of galaxies respectively containing 10 objects each. When $\tau$ reaches the dotted line  the convergence of the chains is achieved, see ref.\cite{2013PASP..125..306F}.}
\end{figure}

A critical issue for every MCMC sampler is the question of the convergence
of the chains\cite{cowles1996markov}. Here,
we will follow the \texttt{emcee} developers recommendation in ref.\cite{2013PASP..125..306F}
to use the autocorrelation time $\tau$ as a diagnostics of the MCMC
performance. The autocorrelation estimations is very well detailed in ref.\cite{sokal1997monte} so that here we will just present some essential points in order to clarify the convergence diagnostic.

Let us define the normalized autocorrelation function, $\rho_f$, 
\begin{equation}
\rho_f(\tau)= \frac{C_f(\tau)}{C_f(0)}\;,
\end{equation} 
with
\begin{equation}
C_f(\tau)=\frac{1}{M-\tau}\sum_{t=1}^{M-\tau}\left[f(X(t+\tau))-\langle f\rangle\right]\left[f(X(t))-\langle f\rangle\right]\;,
\end{equation}
where $\langle f\rangle=\frac{1}{M}\sum_{t=1}^{M}f(X(t))$, $X(t)$ is the sampled random variable of the parameter space, and $M$ is the total length of the chain. $C_f$ is the autocovariance function of a stochastic process: it  measures the covariance between samples separated by a time lag $\tau$. If at certain value of $\tau$, namely $\hat{\tau}$, we have $C_f(\hat{\tau})\rightarrow0$, we can say that independent results are obtained. Thus, $\hat\tau$ gives a threshold of how many samples of the posterior are minimally necessary for producing independent samples. The $\tau$ estimation, $\tau_{\text{est}}$, is given by 
\begin{equation}
\tau_{\text{est}}(N)=1+2\sum_{\tau=1}^{N}\rho_f(\tau)\;, 
\end{equation} 
where $N$ starts at $N\ll M$. We have plotted $\tau_{\text{est}}$ for the sets that we considered in this work in figure \ref{autocorrelation}.

\begin{table}[tbp]
		\centering
		\small
		\begin{tabular}{|l|c|c|c|c|c|c|c|c|c|c|c|c|}
			\hline 
			Set  & $a_{f}$  & \multicolumn{2}{|c|}{Best-fit values} &$k_\text{Y}$& $\chi_{\text{red,tot}}^{2}$ &$k_\text{sg}$& $\chi_{\text{red,tot}}^{2}|_{\beta = 0}$&$N$&$\Delta\text{BIC}$&$2\log B_{12}$&CL\\
			\hline
			   &           & $\beta$                 & $\lambda(\text{kpc})$   &  &     &   &     &    &     &     &$\sigma$\\\hline
			A  &   0.16    & $0.34_{-0.10}^{+0.12}$  & $10.27_{-3.82}^{+2.89}$ &25& 0.88&23 & 1.11&206 &32.18&31.84&$5.29$ \\\hline
			B  &   0.14    & $0.30\pm0.08$           & $7.42_{-3.99}^{+2.94}$  &23& 0.80& 21& 0.96&180 &17.12&23.21&$4.44$\\\hline
			C  &   0.12    & $0.28_{-0.08}^{+0.09}$  & $8.18_{-6.31}^{+5.39}$  &25& 0.83& 23& 1.04&163 &20.52&12.41&$3.09$ \\\hline
			D  &   0.15    & $0.54_{-0.10}^{+0.11}$  & $4.15_{-0.95}^{+0.81}$  &23& 0.78& 21& 1.02&196 &32.92&20.38&$4.12$\\\hline
	  Combined &    -      &$0.34\pm 0.04$           &$5.61\pm0.91$            &90&0.82 & 88&1.03 &745 &91.61&87.83&8.26\\
	    	\hline
	    \end{tabular}%
	    \caption{\label{tabelabetalambda} The acceptance fraction $a_{f}$, the maximum likelihood estimation for $(\beta,\lambda)$, 
		the total goodness of fit $\chi_{\text{red,tot}}^{2}$ and the one calculated fixing $\beta = 0$, 
		for each set of 10 galaxies and for the combination of the data sets. $k_\text{Y}$ ($k_\text{sg}$) is the number of free parameters 
		for the Yukawa model (for standard gravity), while $N$ is the number of data points. 
		We also 
		report the values for the $\Delta$BIC, for $2\log B_{12}$, and the confidence level (CL), see text for more details.}
\end{table}

Thus, we ran chains increasing the number of iterations until  finally it is possible to perceive a plateau for $\tau$  as we show in figure \ref{autocorrelation}, which indicates that $\hat\tau$ has been found. After estimating $\hat{\tau}$, we discard the number of iterations $N_{\text{disc}}\sim\hat{\tau}$ (burn-in) and  compute the posterior on the parameters. The \texttt{emcee} developers suggest that a number of iterations $M>50\hat{\tau}$ is  enough to achieve the convergence of the chains. We performed a second MCMC sampling to check this assumption, but now considering a shorter chain with the number of iterations $M=70\hat{\tau}$ with $N_{\text{disc}}=\hat{\tau}$ and we obtained the same results for the posterior. Hence, indeed when the chain length crosses the dotted line, $N=50\tau$, we can achieve the convergence.

Another quantity to monitor is the acceptance fraction $a_{f}$ which
is the fraction of proposed steps that are accepted in the sampling
process. In our analysis we obtain an acceptance that is between 
0.1 and 0.2, as can be seen in table \ref{tabelabetalambda}.

The main results are displayed on table \ref{tabelabetalambda}. We show the
best-fit values for the parameters $(\beta,\lambda)$ and their $1\sigma$
error bars, the acceptance fraction $a_{f}$ and the overall goodness
of fit $\chi_{\text{red,tot}}^{2}$, i.e. considering all galaxies
of the same set together. 
It is possible to see an improvement of $\chi^2$ from the standard $\beta=0$ model to the Yukawa model for each set and also in the combination of all sets. 
The analysis with all sets combined takes into account that the total data, i.e. summing all data points, is 745. The total number of parameters is 90: 
$40\times (\Upsilon_\text{*D}, M_{200})+8\times\Upsilon_\text{*B}$ plus $\beta,\lambda$, for a total   of 655 degrees of freedom.

In table \ref{tabelaresultado1} we arranged
the best-fit values and their $1\sigma$ error bars for the galaxy-specific
parameters $\Upsilon_{*,\text{D}},\Upsilon_{*,\text{B}},\text{and }M_{200}$
and also the individual goodness of fit $\chi_{\text{red}}^{2}$. In table \ref{tabelaresultado2} we have the same but for the $\beta=0$ case.

\begin{figure}[tbp]
	\centering 
	\includegraphics[width=10cm]{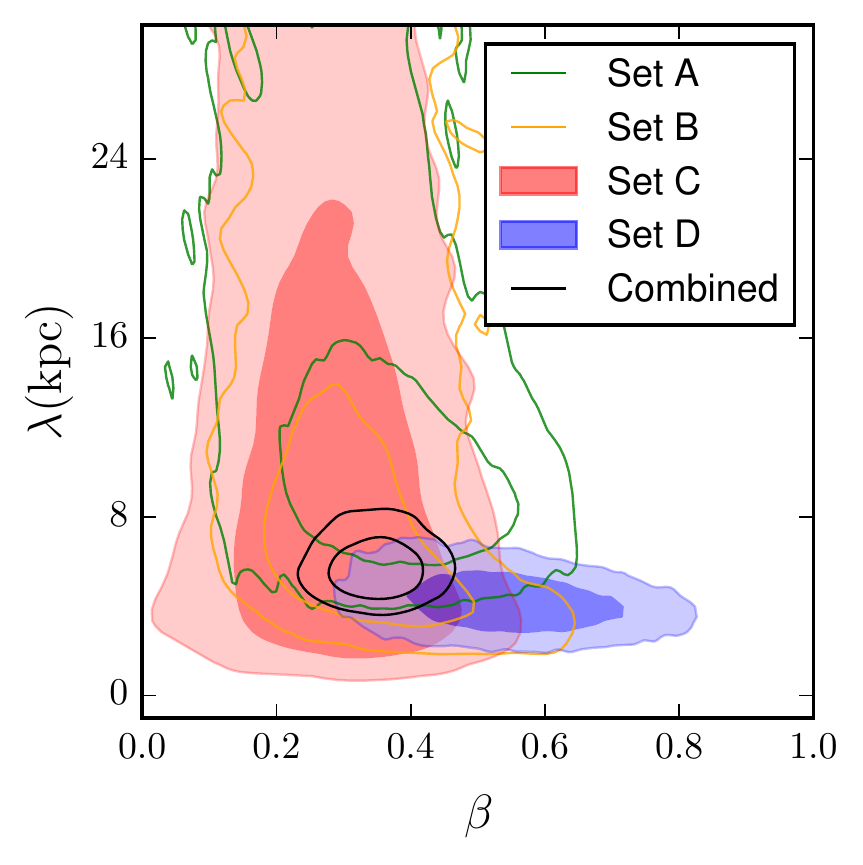}\\
	\includegraphics[width=6cm]{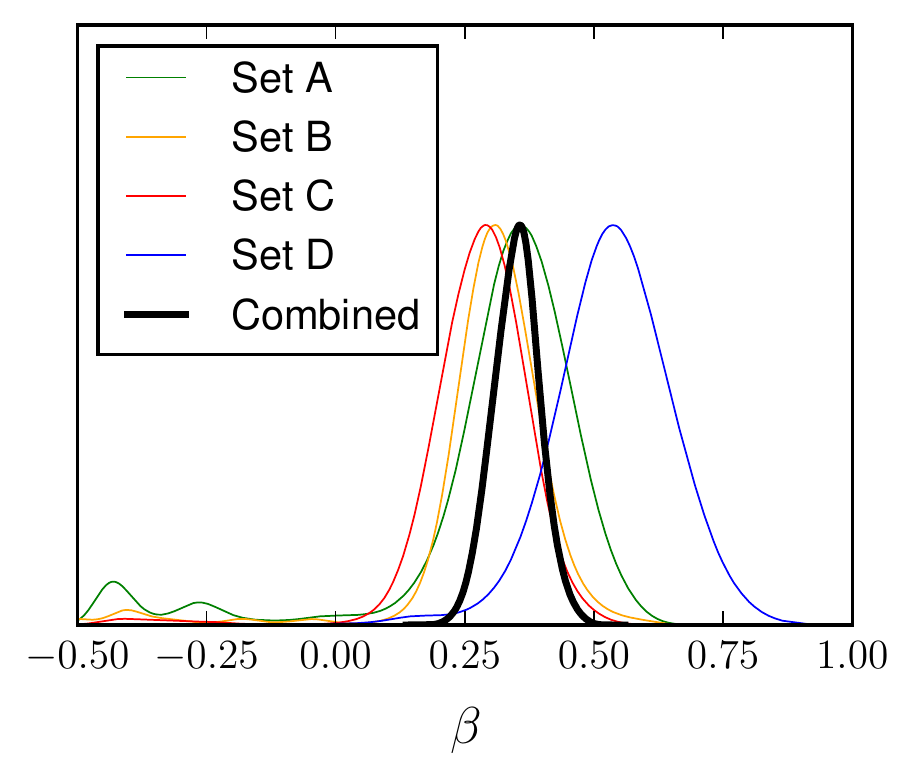} \includegraphics[width=5.7cm]{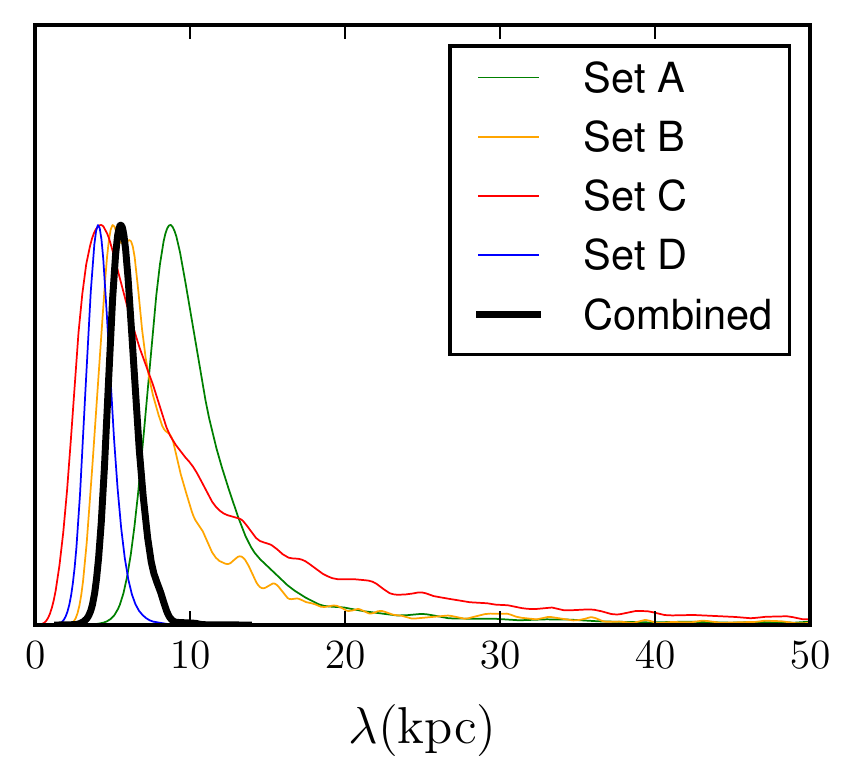}
	\caption{\label{scatterplot}The marginalized distribution of the $(\beta,\lambda)$
		and the one-dimensional posterior distribution according the set of
		10 galaxies each and the combined analysis.}
\end{figure}

\begin{figure}[tbp]
	 \centering
		\includegraphics[width=5cm]{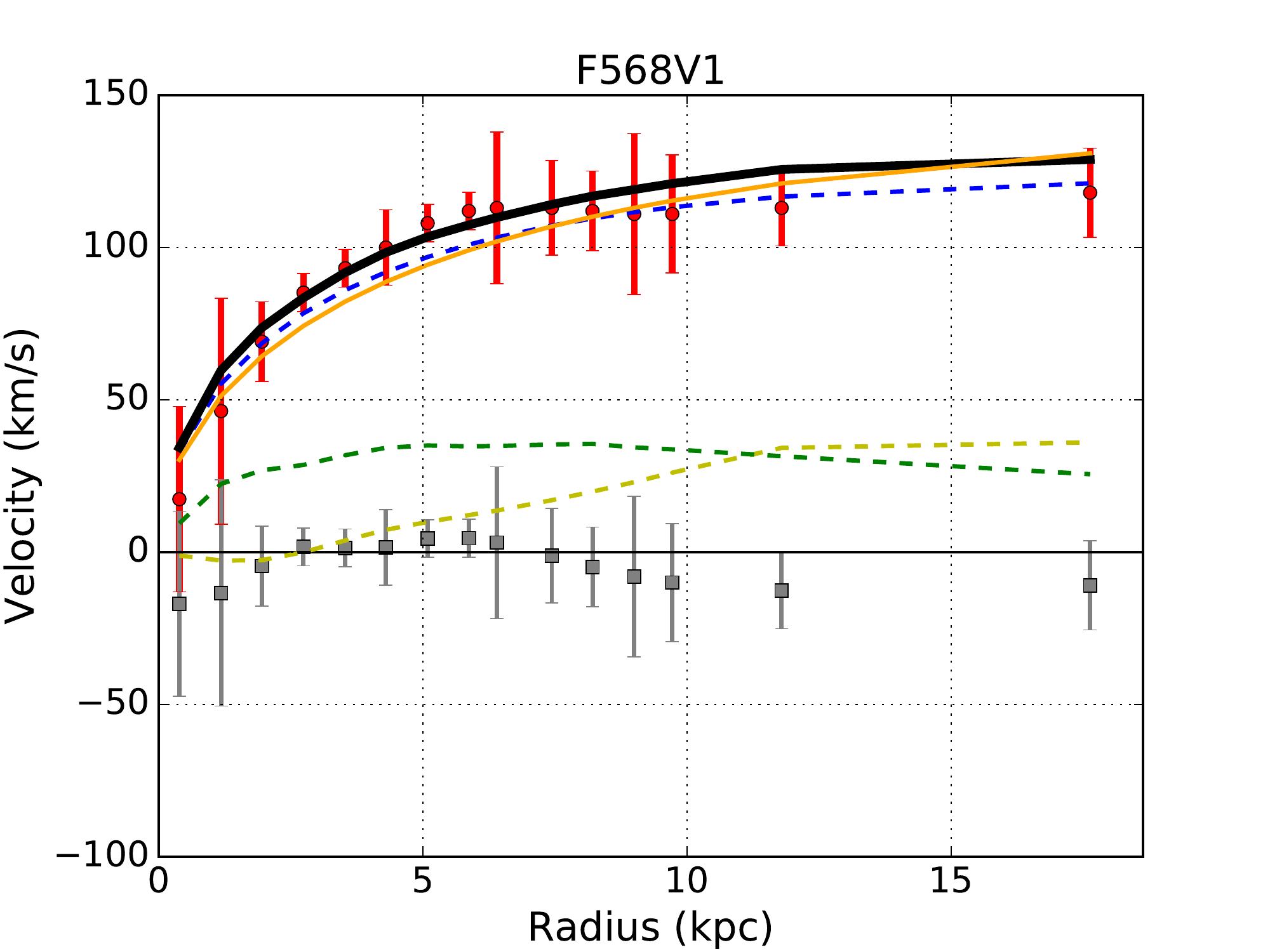} 
		\includegraphics[width=5cm]{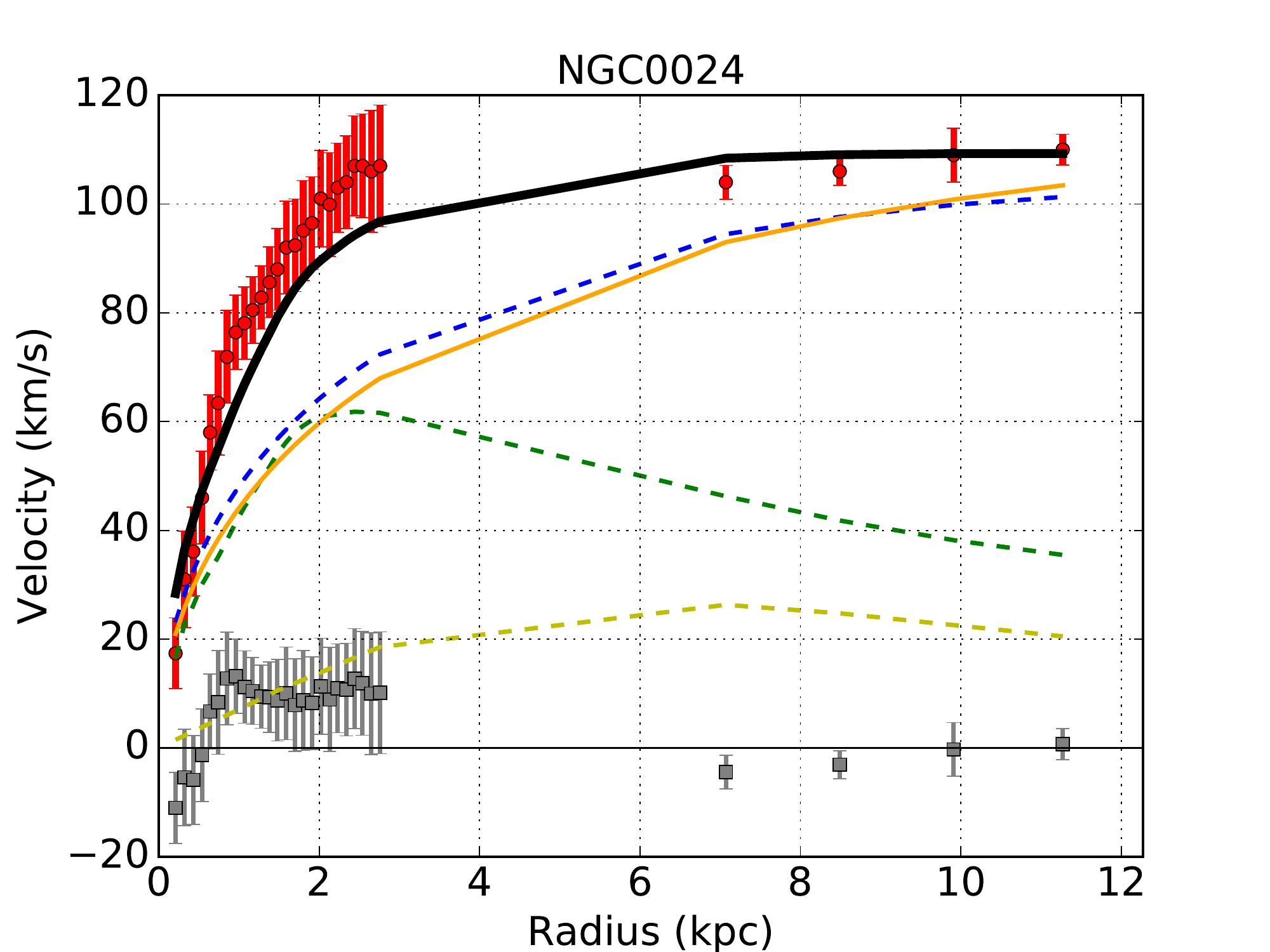} \\ 
		\includegraphics[width=5cm]{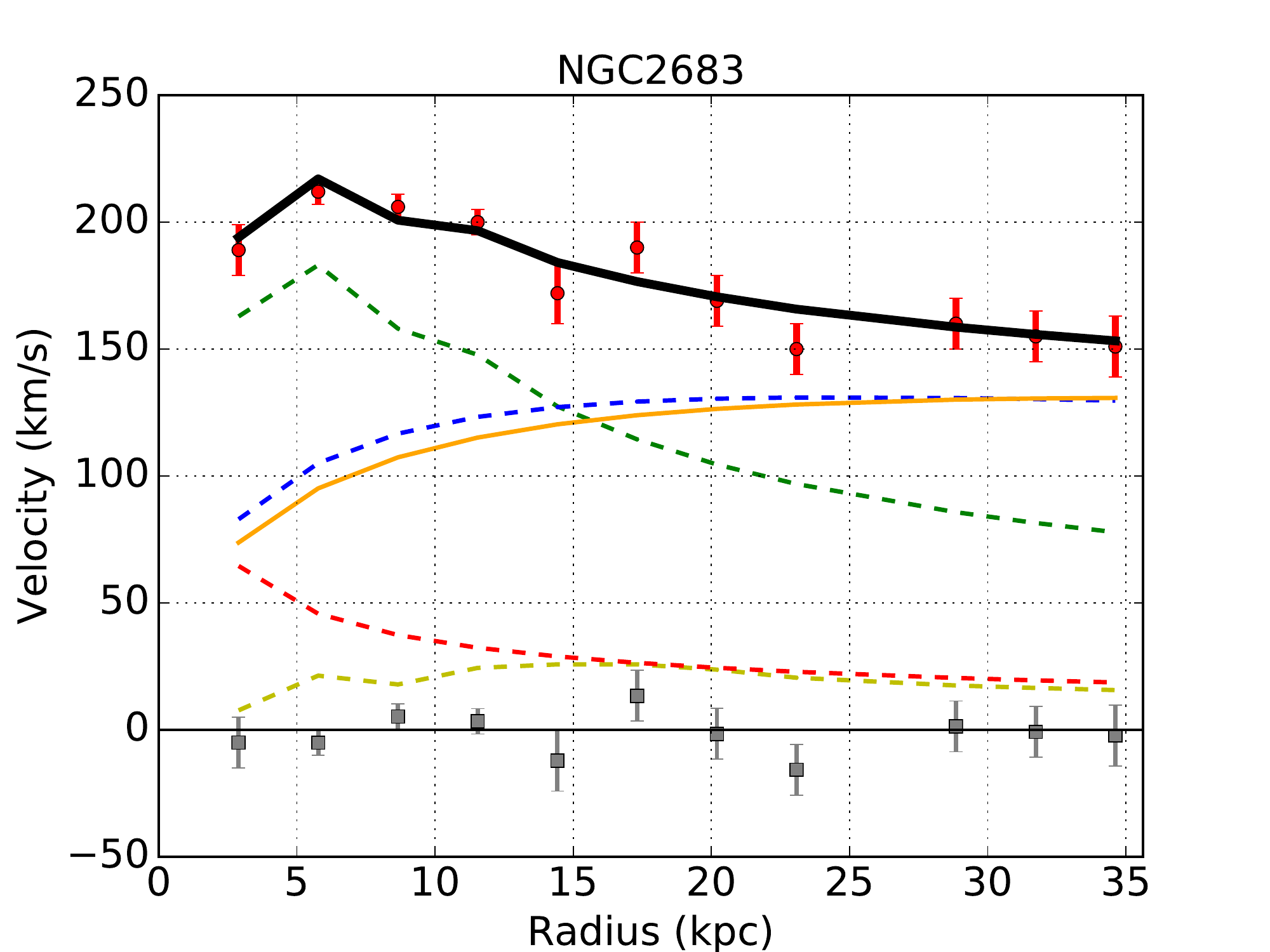} 
		\includegraphics[width=5cm]{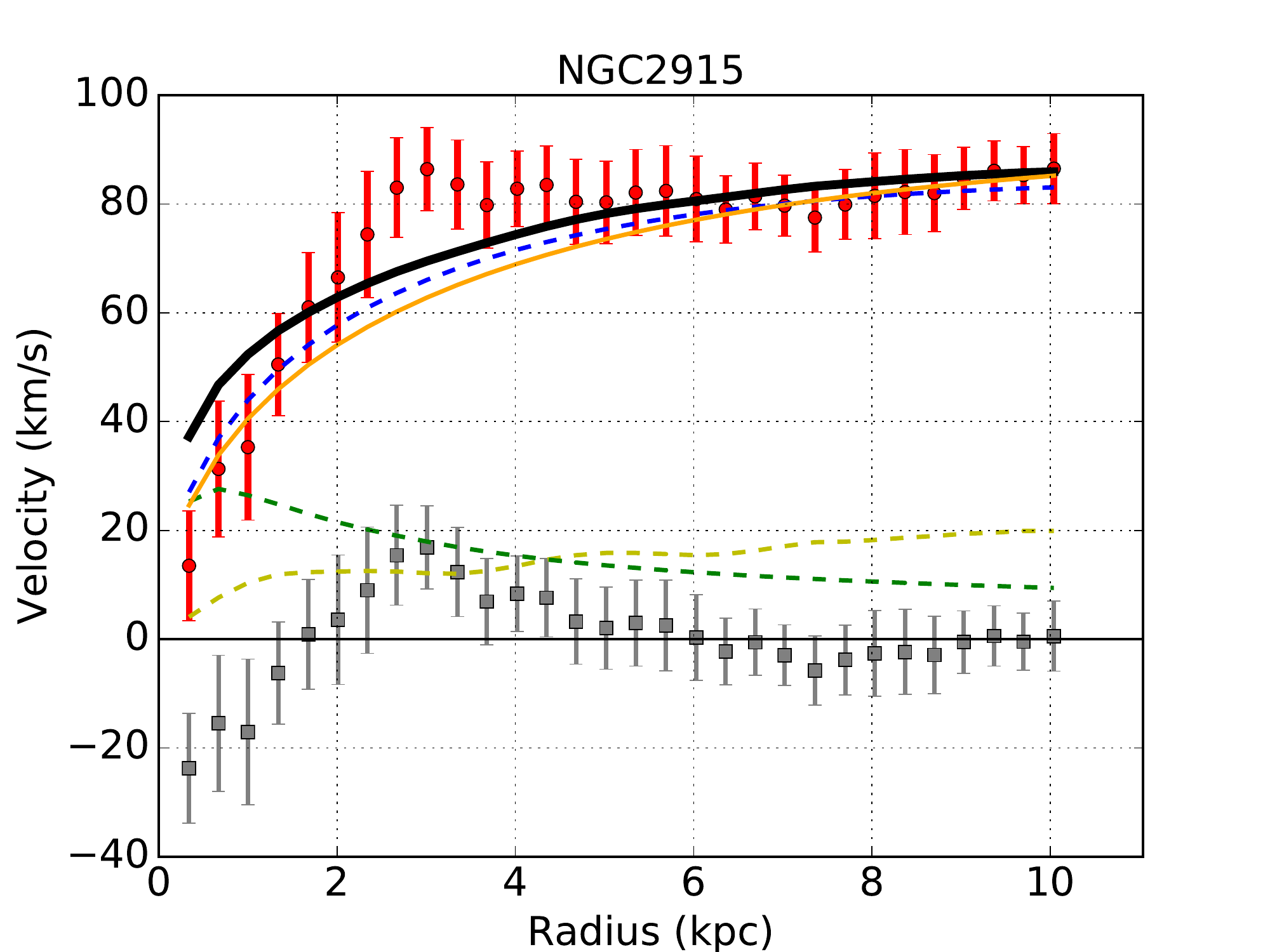} \\ 
		\includegraphics[width=5cm]{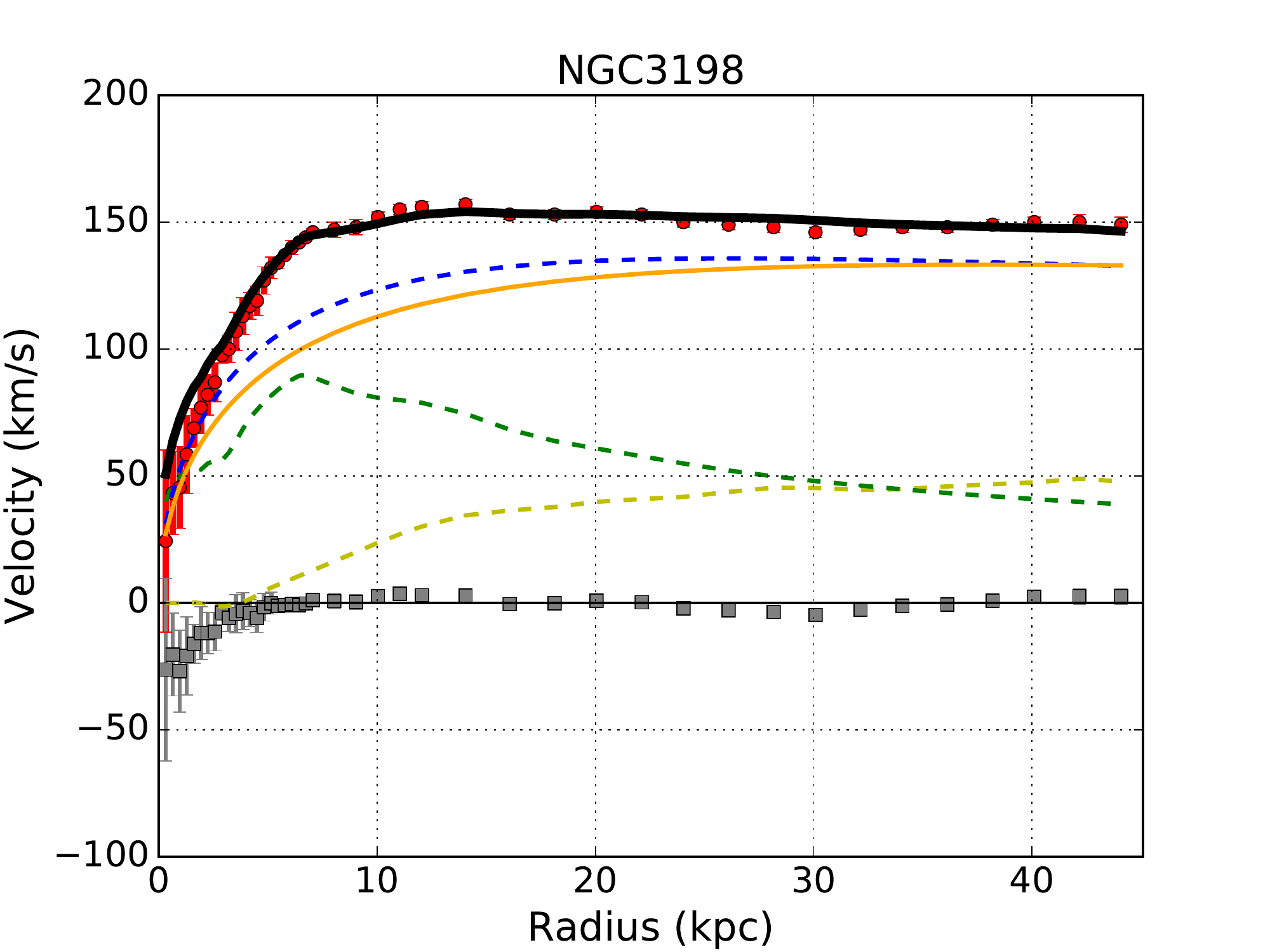} 
		\includegraphics[width=5cm]{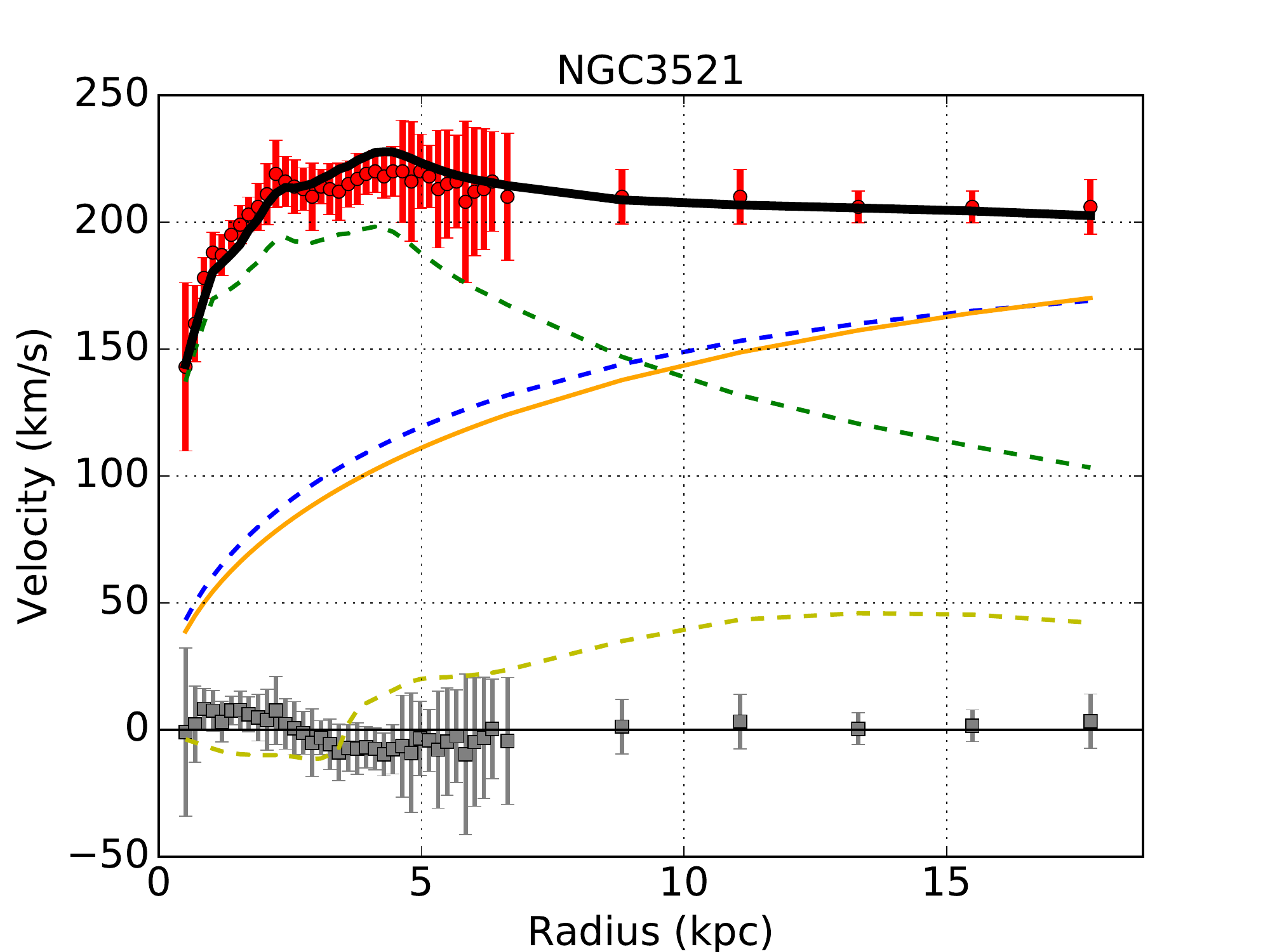} \\ 
		\includegraphics[width=5cm]{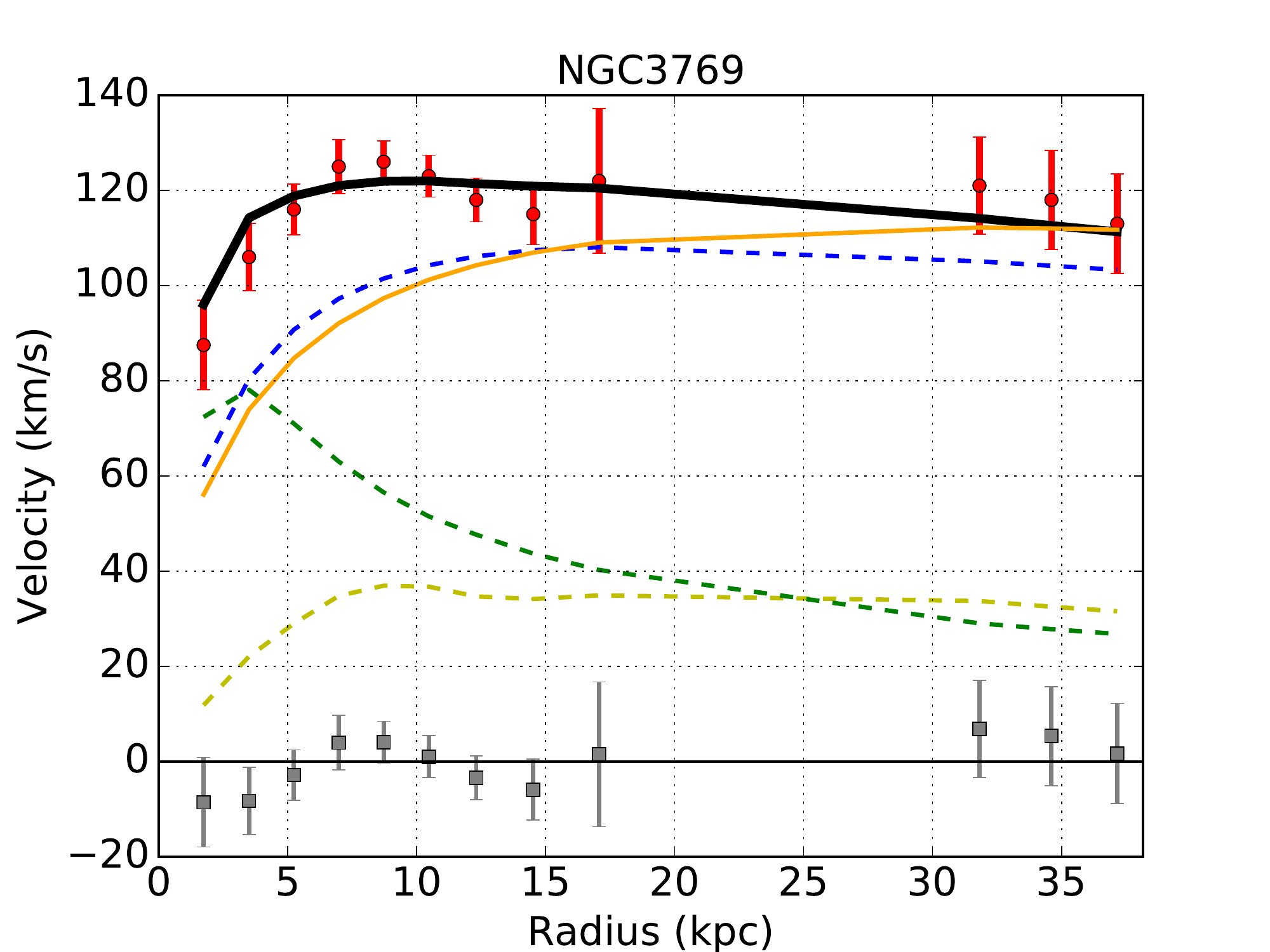}  
		\includegraphics[width=5cm]{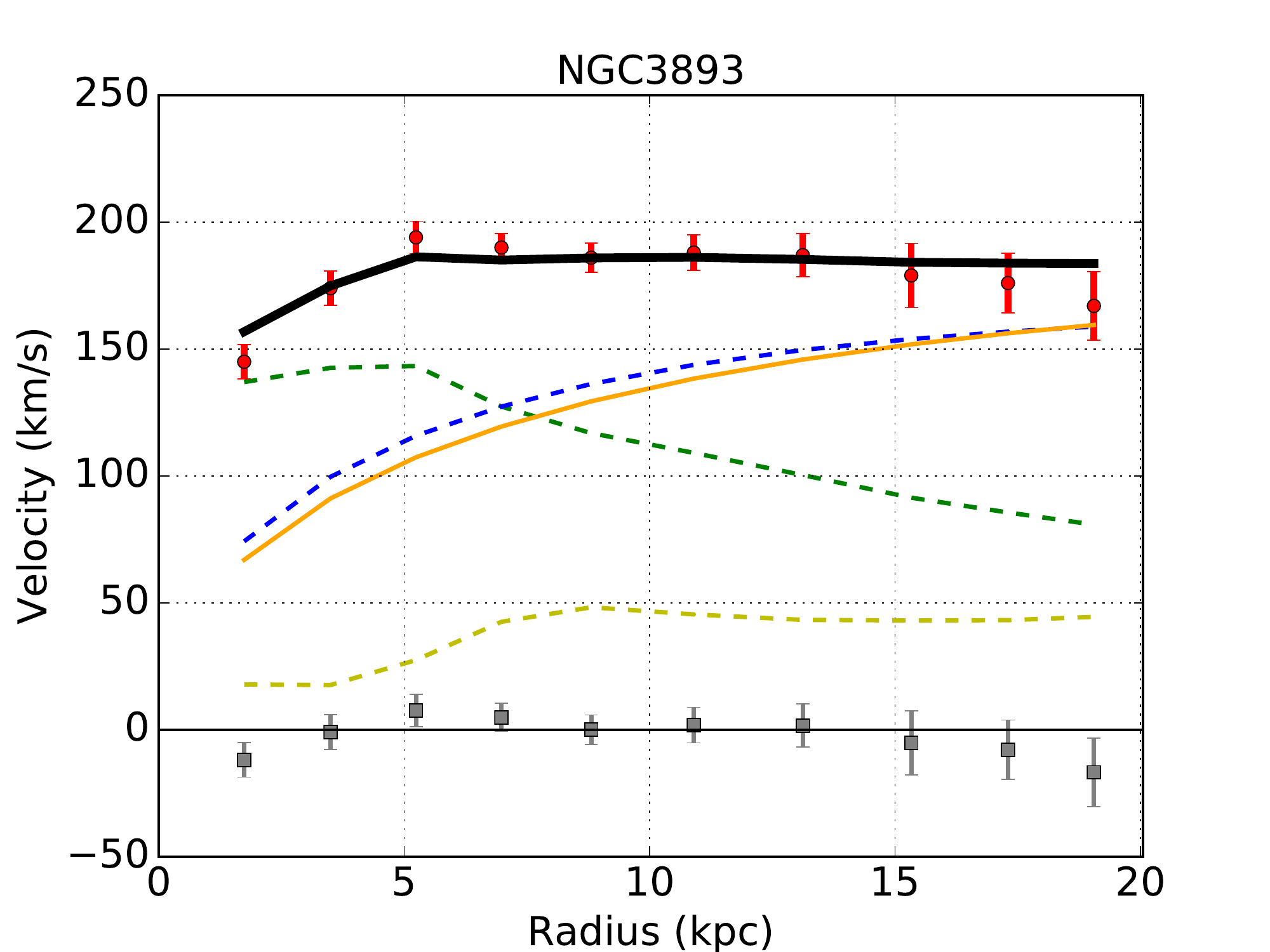} \\ 
		\includegraphics[width=5cm]{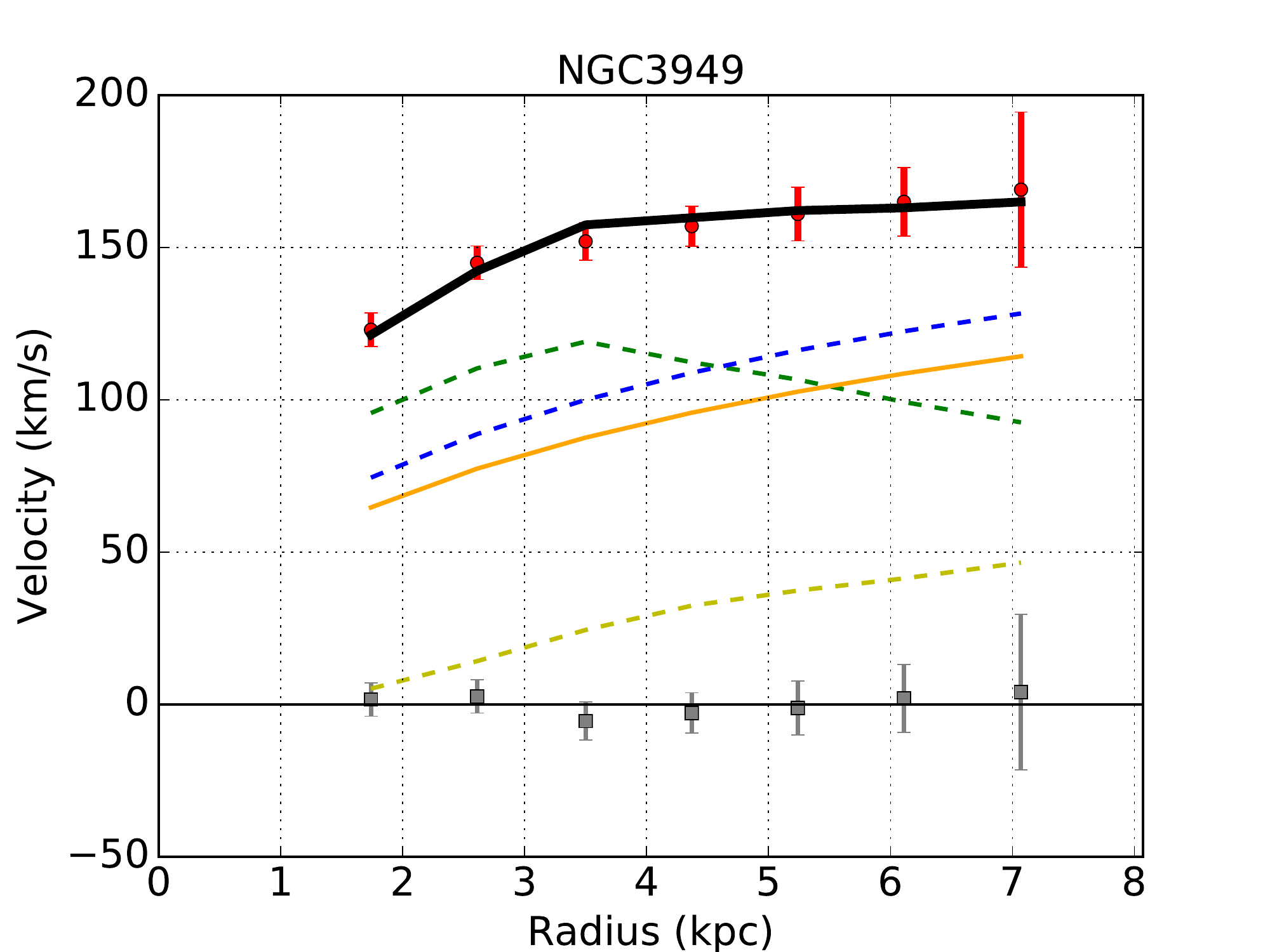} 
		\includegraphics[width=5cm]{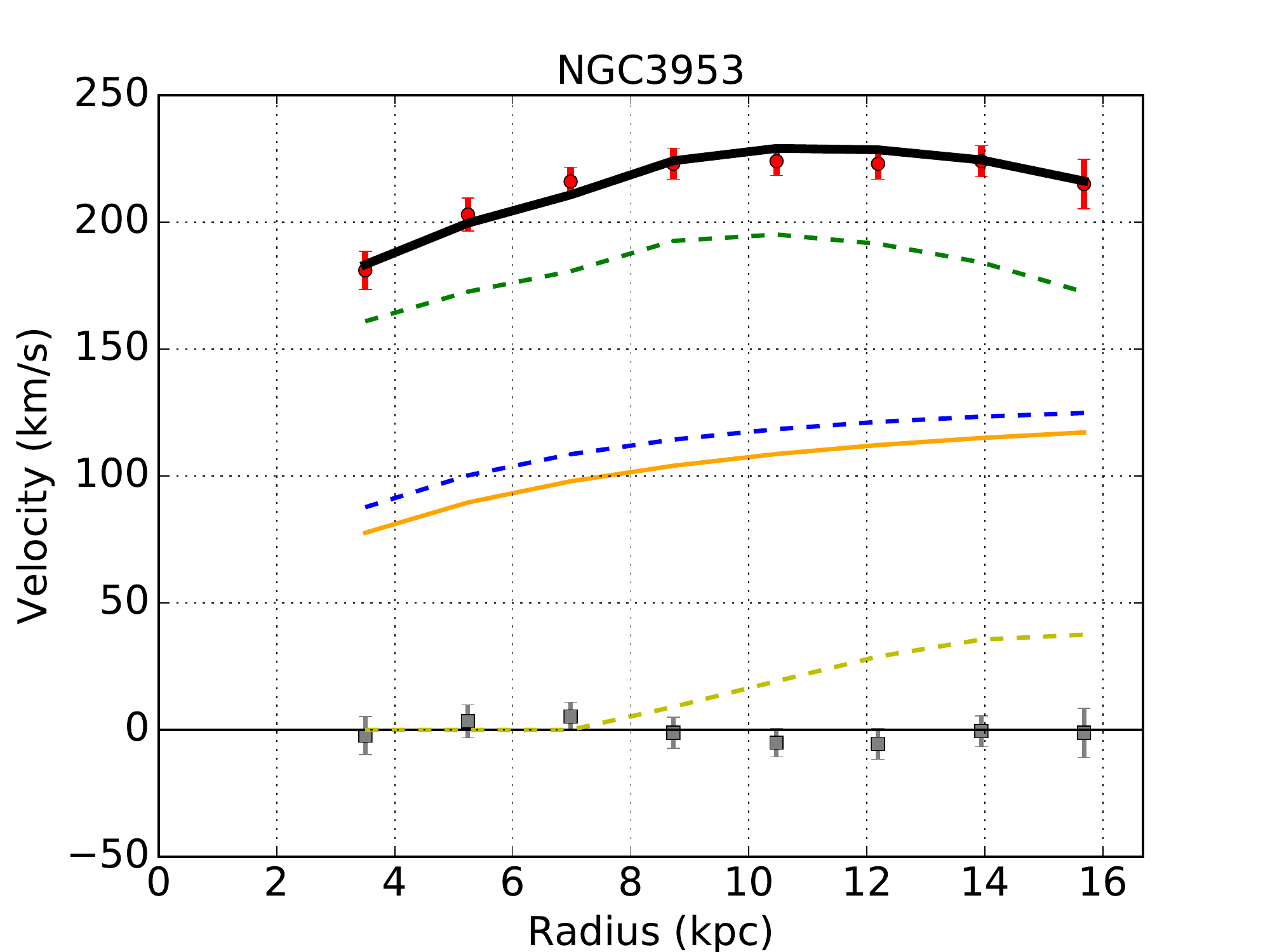}\\ 
		\caption{\label{rotationcurveA}The rotation curves and their components: gas (dashed
			yellow line), disk (dashed green line which corresponds to $\sqrt{\Upsilon_{*\text{D}}V_{\text{disk}}^{2}}$), bulge (dashed red line which corresponds to $\sqrt{\Upsilon_{*\text{B}}V_{\text{bulge}}^{2}}$) and
			dark matter with Yukawa-like corrections (dashed blue line). The black
			solid line is the overall best-fit (see equation \ref{circular}) and the
			values for the parameters are displayed on tables \ref{tabelaresultado1}
			and \ref{tabelabetalambda}, the orange solid line is the dark matter component for $\beta=0$. The red dots with error bars are the
			observational data taken from SPARC catalogue and the grey ones are
			the residual of the fit. We have plotted the results for the 
			set A.}	
\end{figure}

\begin{figure}[tbp]
		\centering 
		\includegraphics[width=5.5cm]{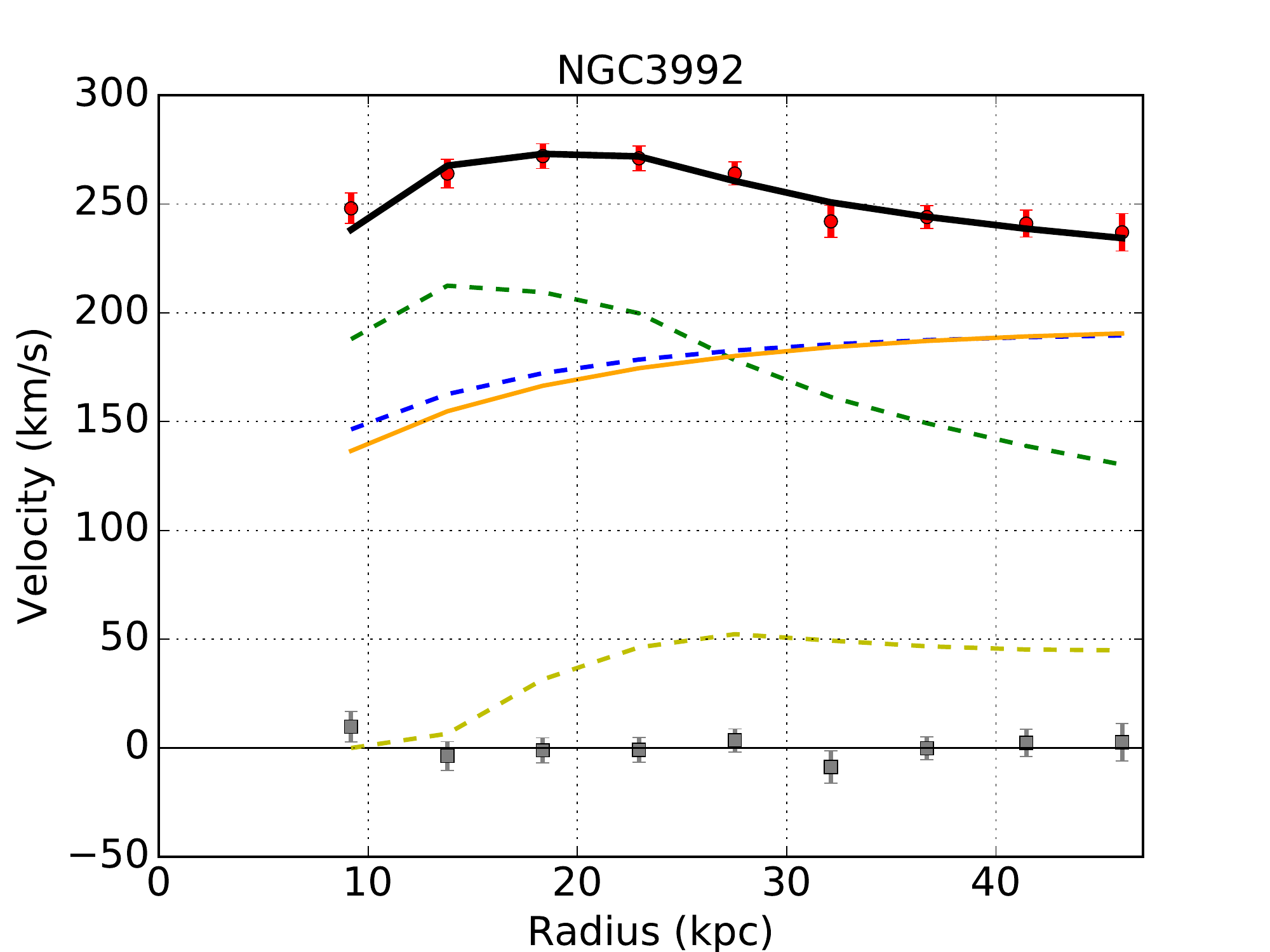} 
		\includegraphics[width=5.5cm]{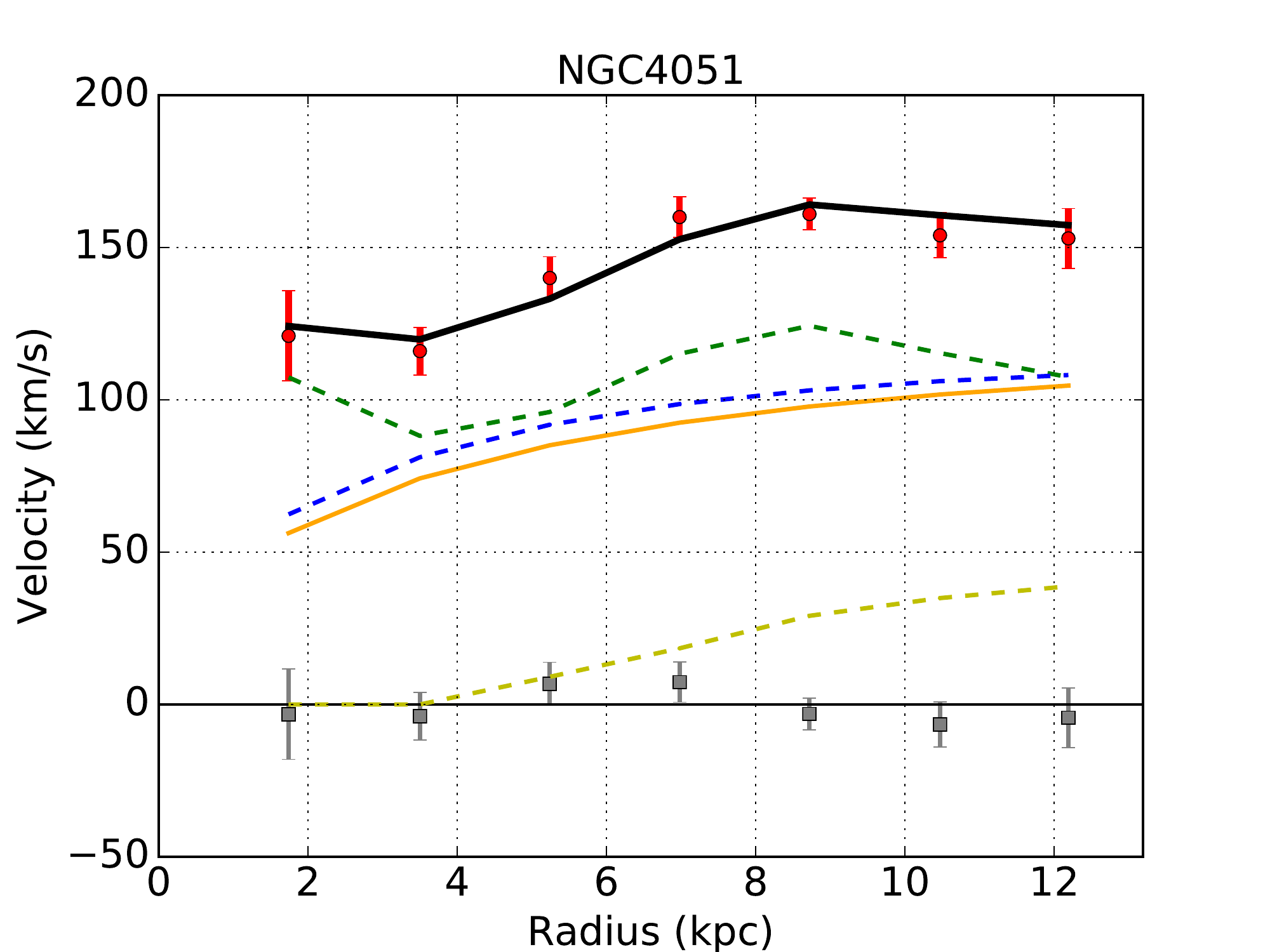} \\ 
		\includegraphics[width=5.5cm]{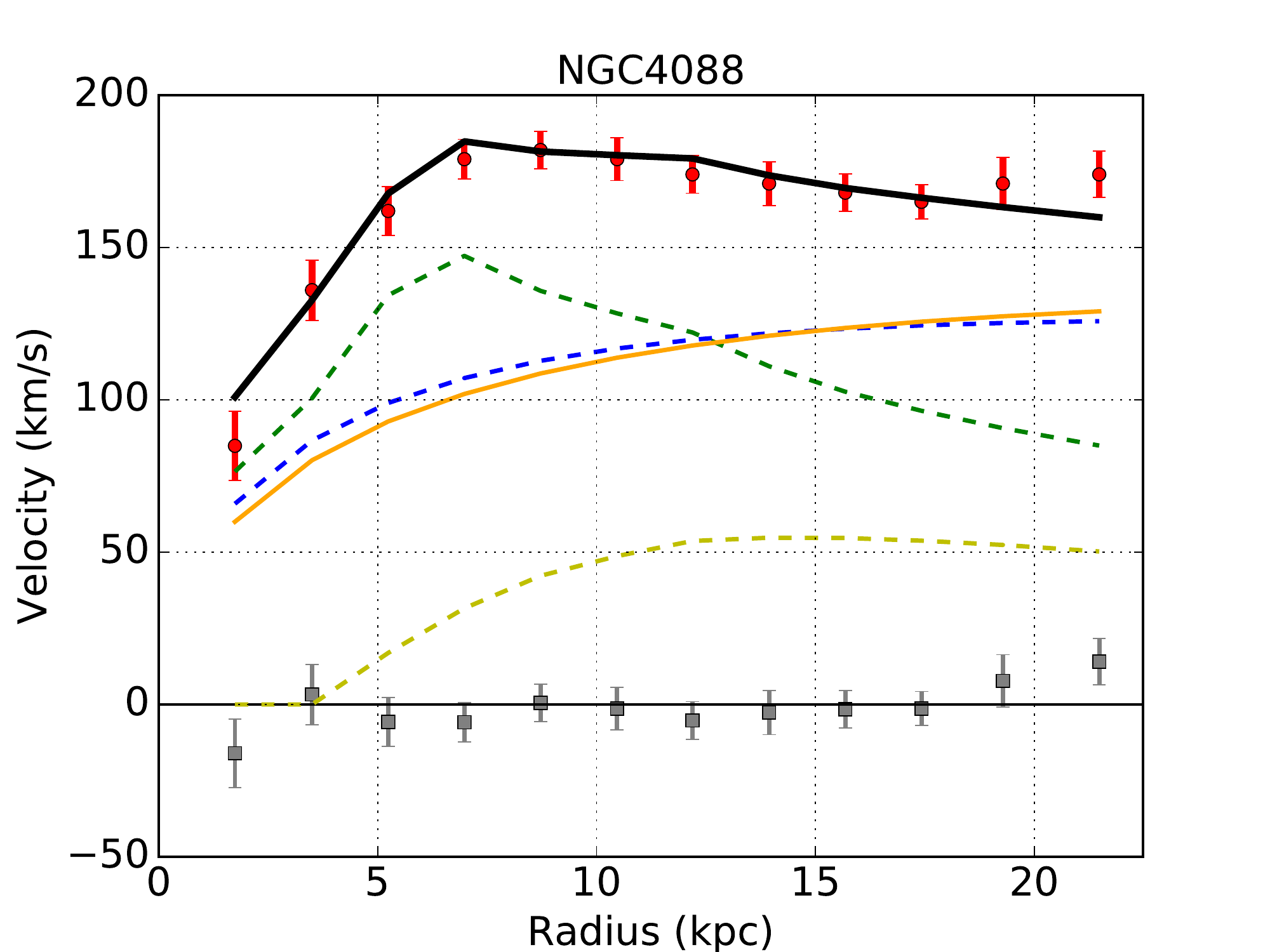} 
		\includegraphics[width=5.5cm]{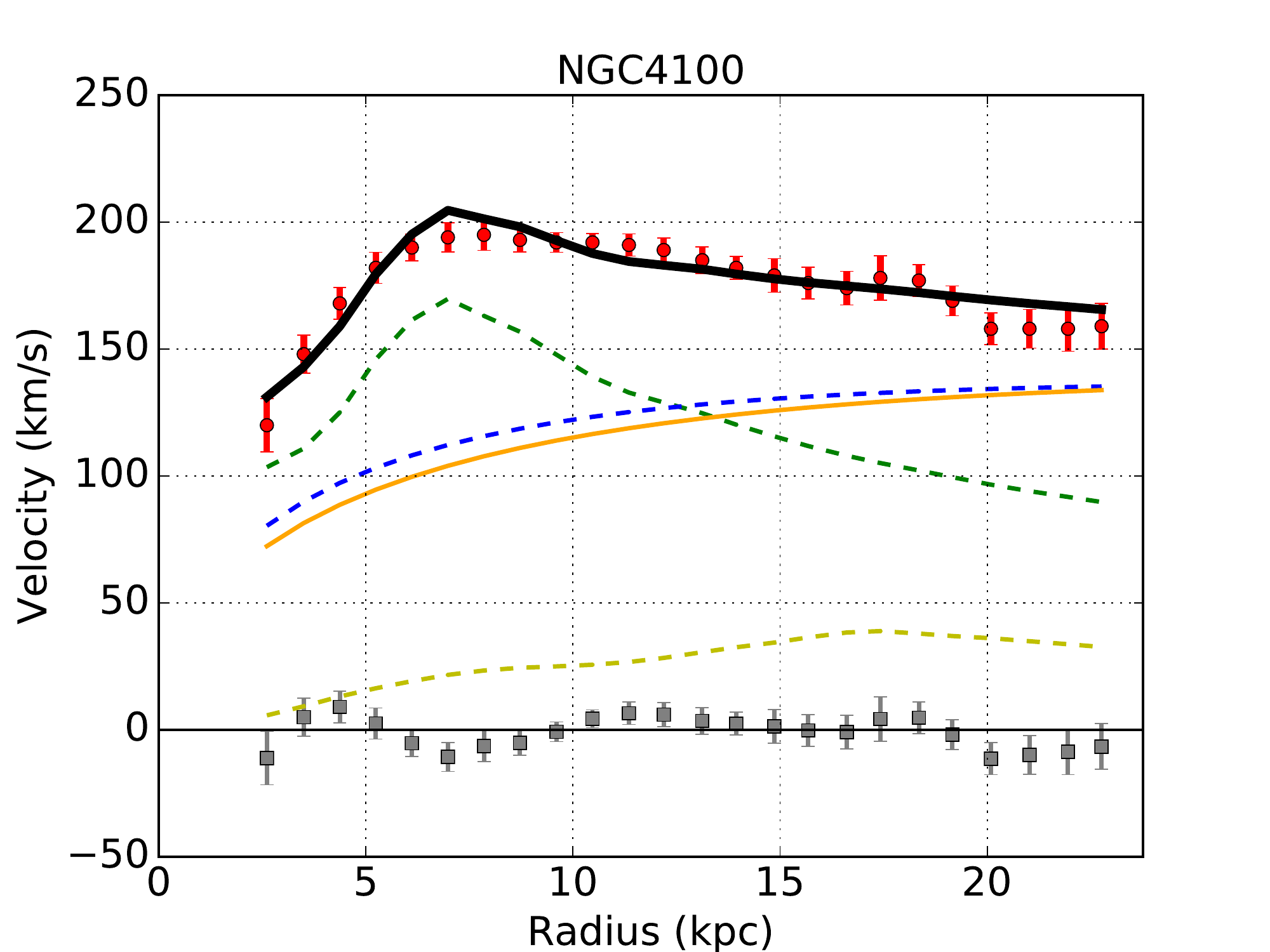} \\ 
		\includegraphics[width=5.5cm]{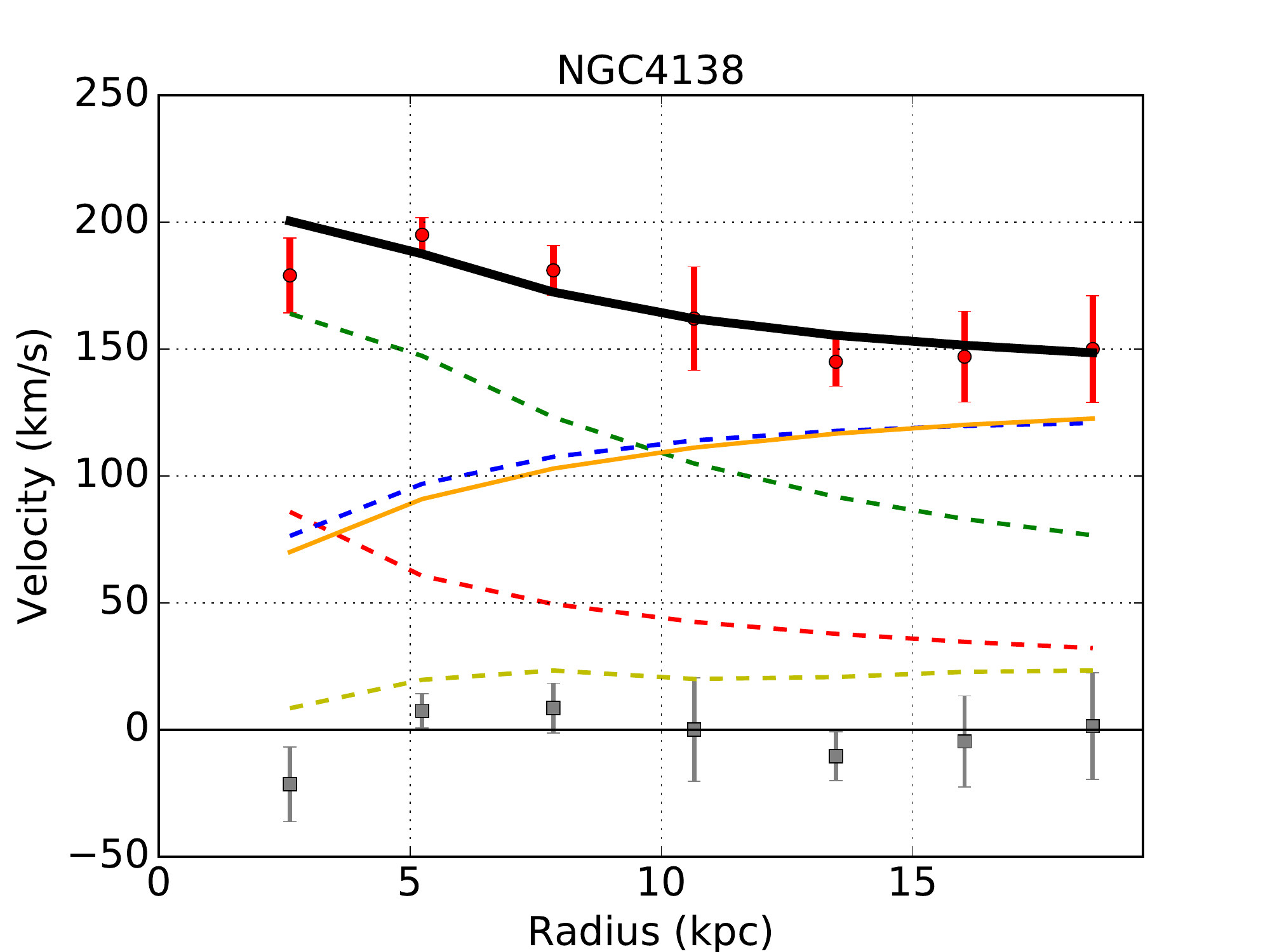} 
		\includegraphics[width=5.5cm]{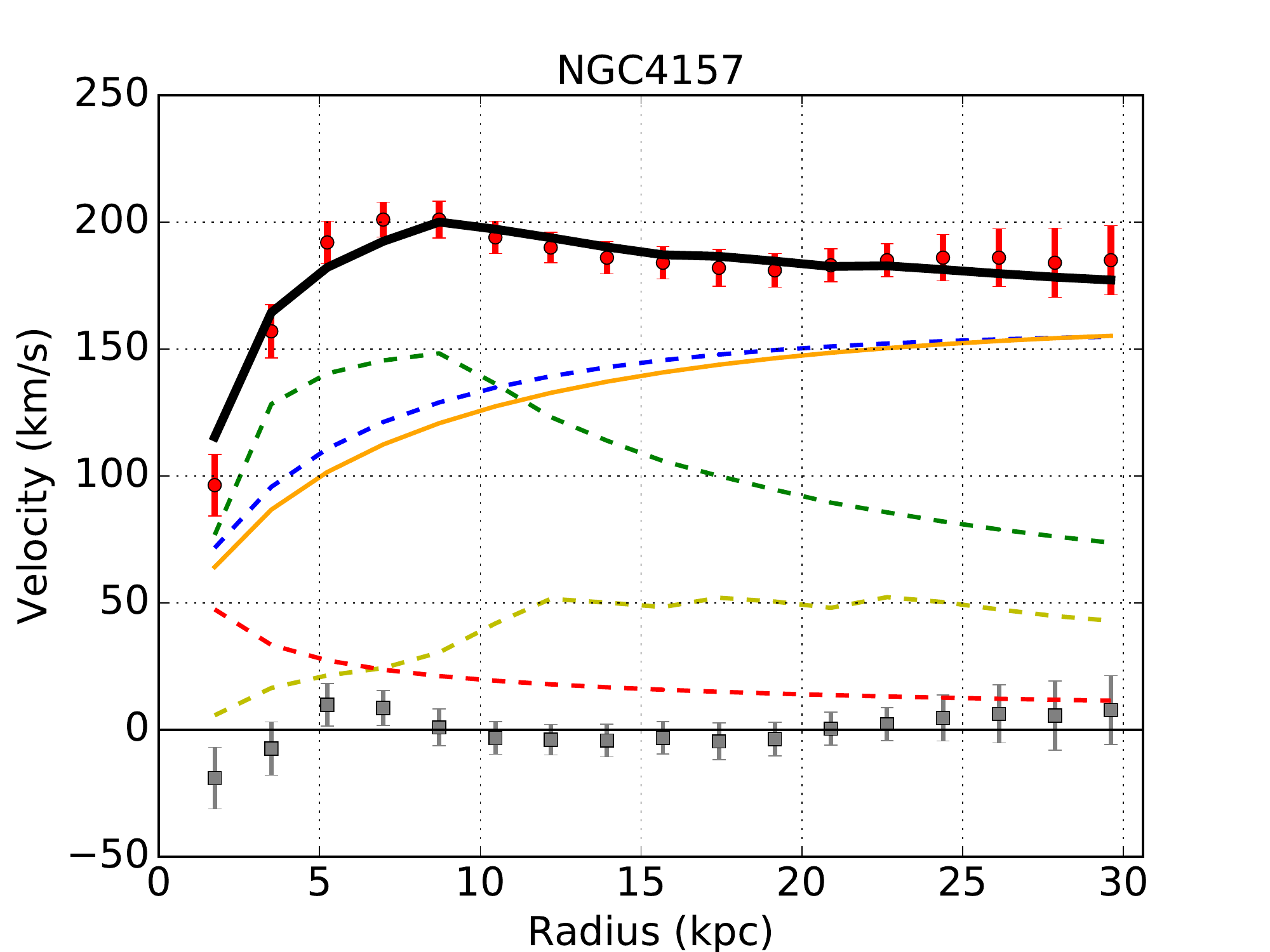} \\ 
		\includegraphics[width=5.5cm]{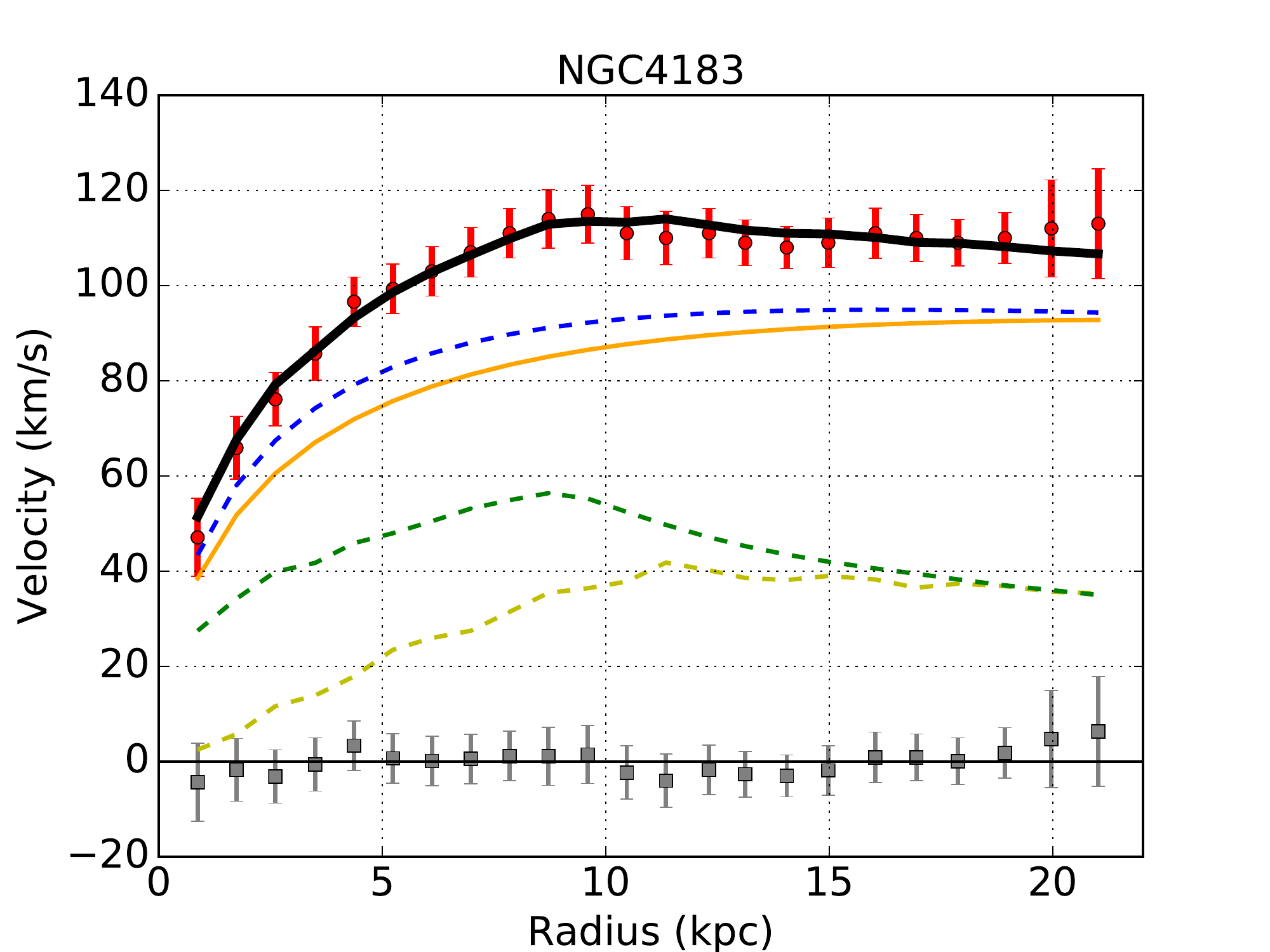} 
		\includegraphics[width=5.5cm]{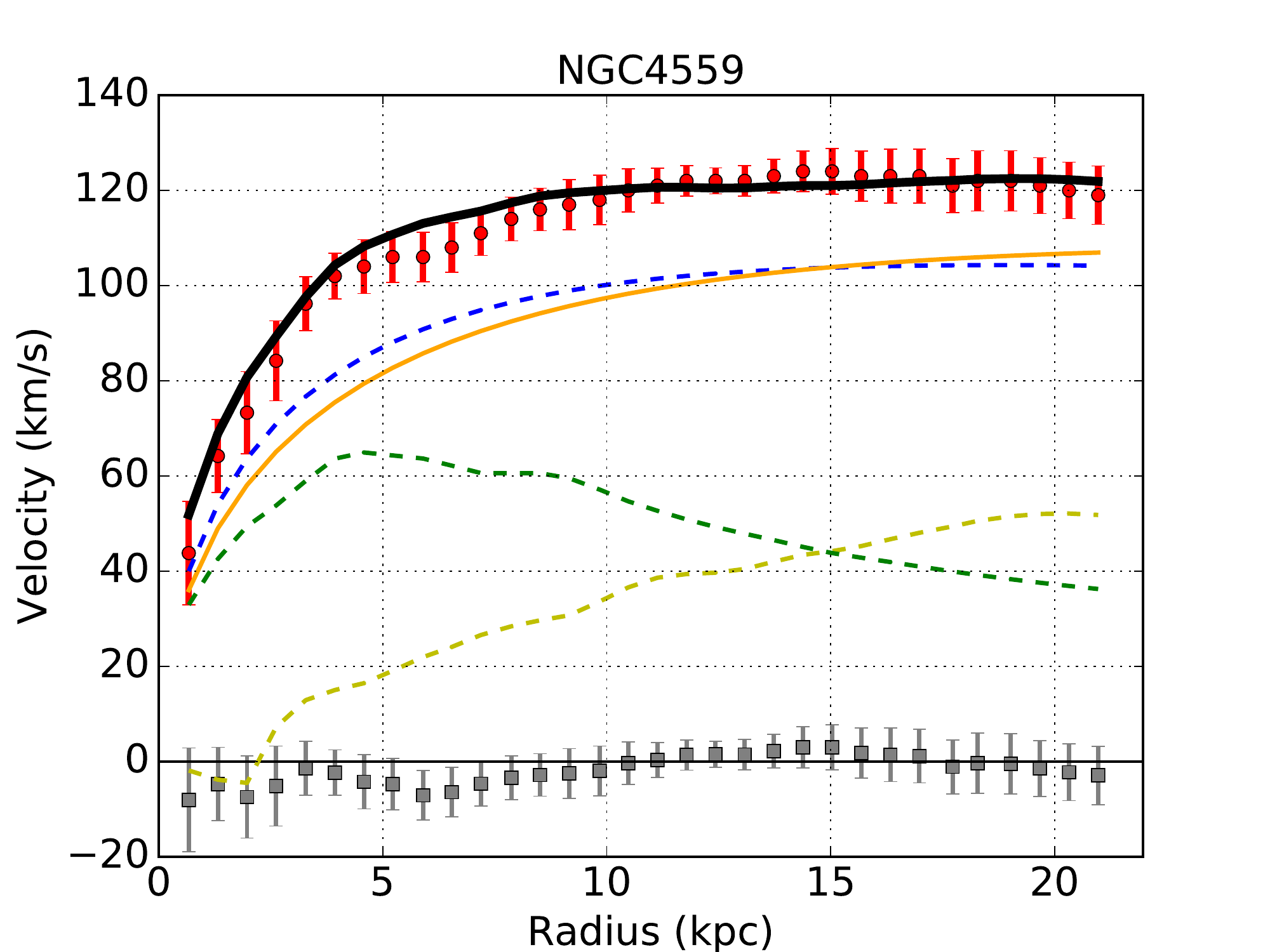} \\ 
		\includegraphics[width=5.5cm]{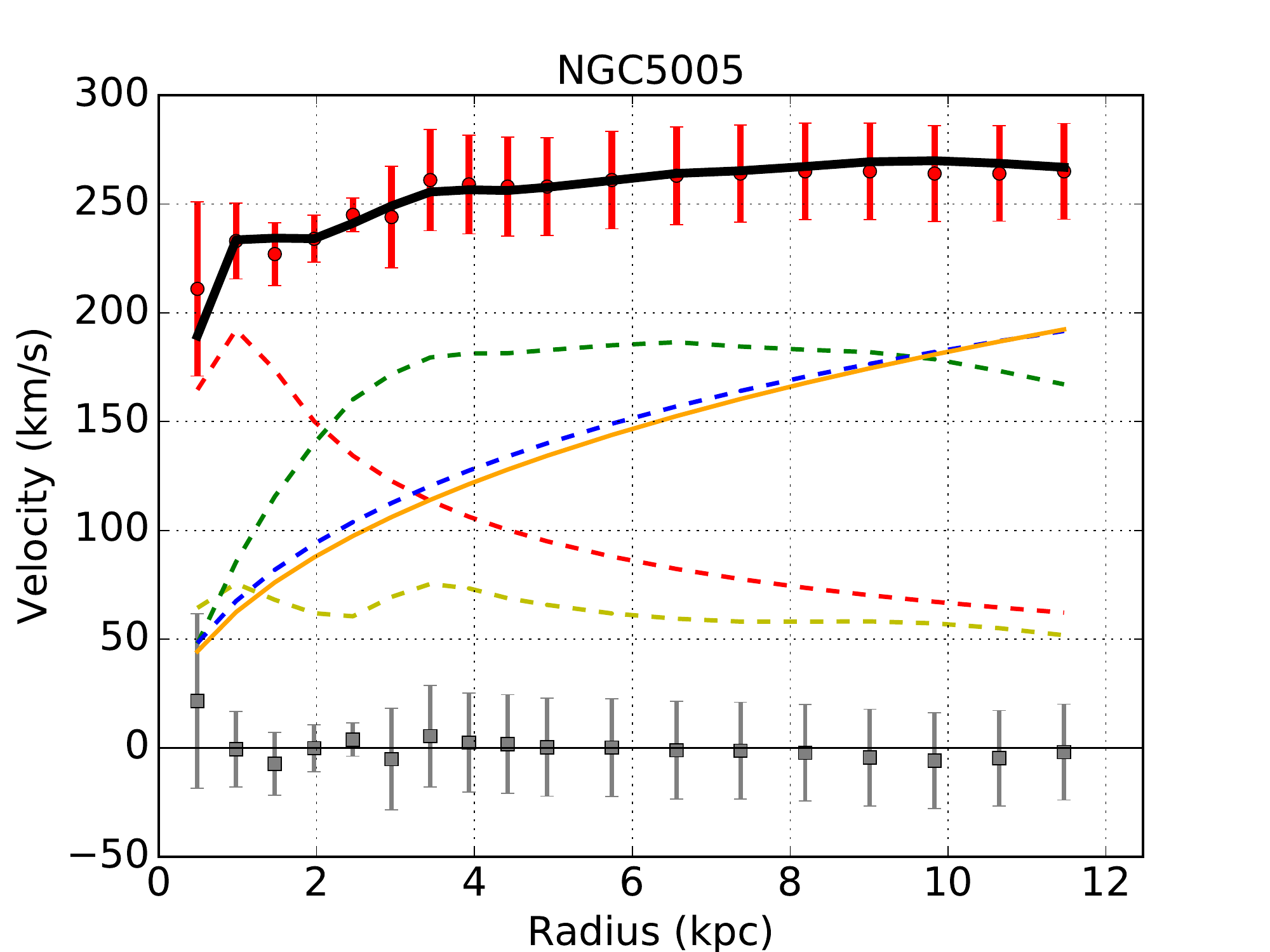} 
		\includegraphics[width=5.5cm]{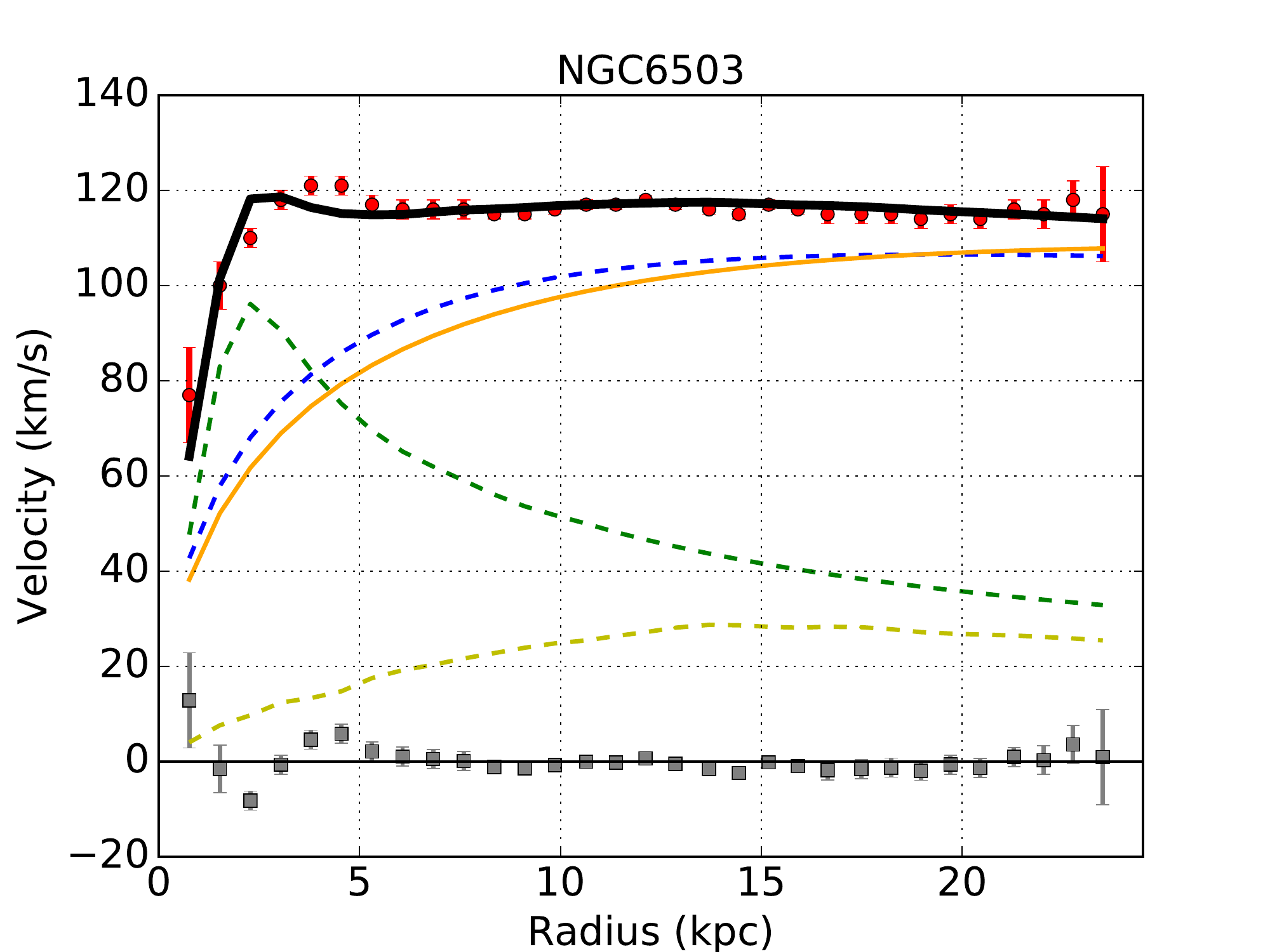} \\ 
		\caption{\label{rotationcurveB}Same as figure \ref{rotationcurveA}, but for set B. }
\end{figure}

\begin{figure}[tbp]
		\centering 
		\includegraphics[width=5.5cm]{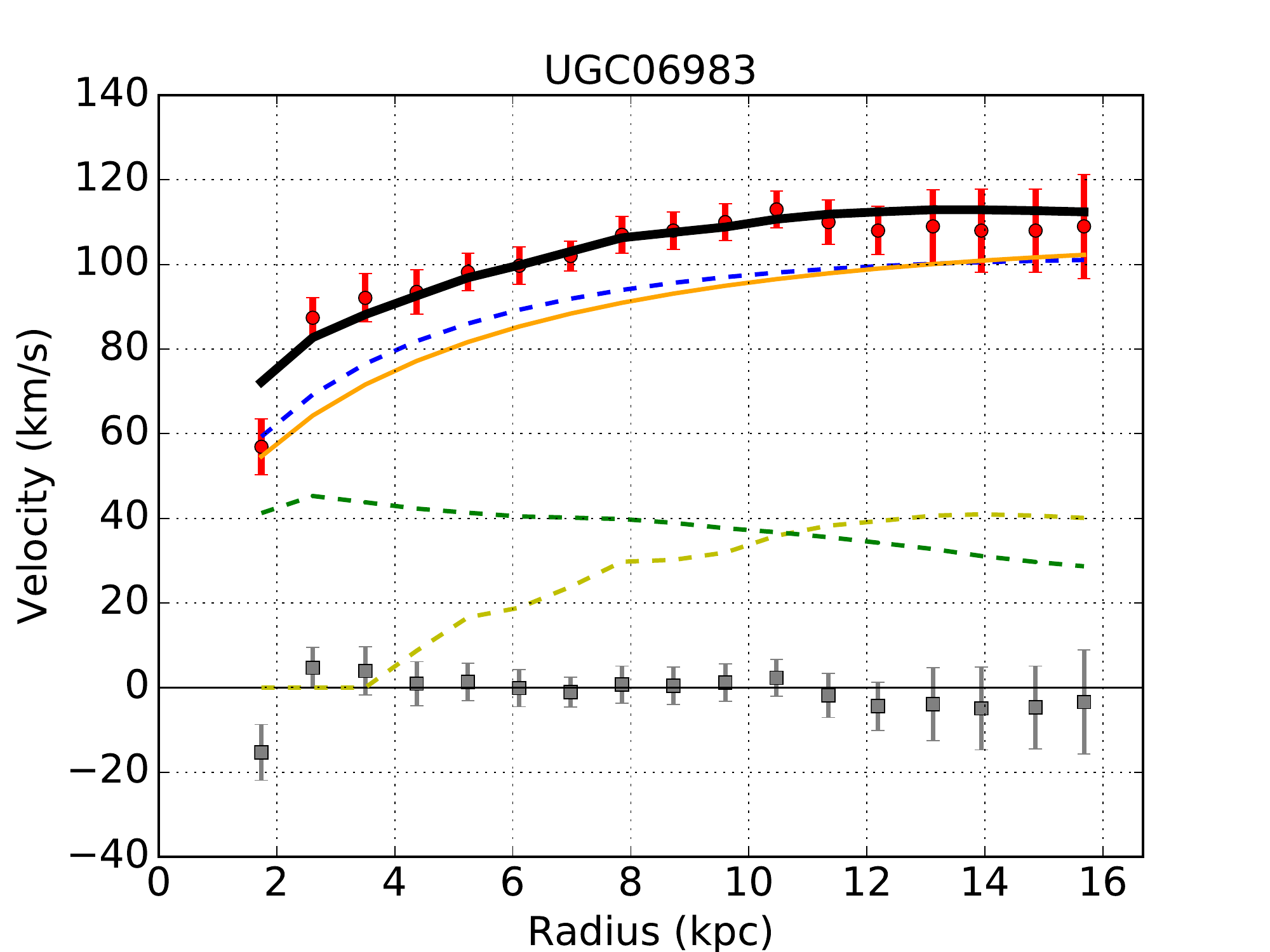} 
		\includegraphics[width=5.5cm]{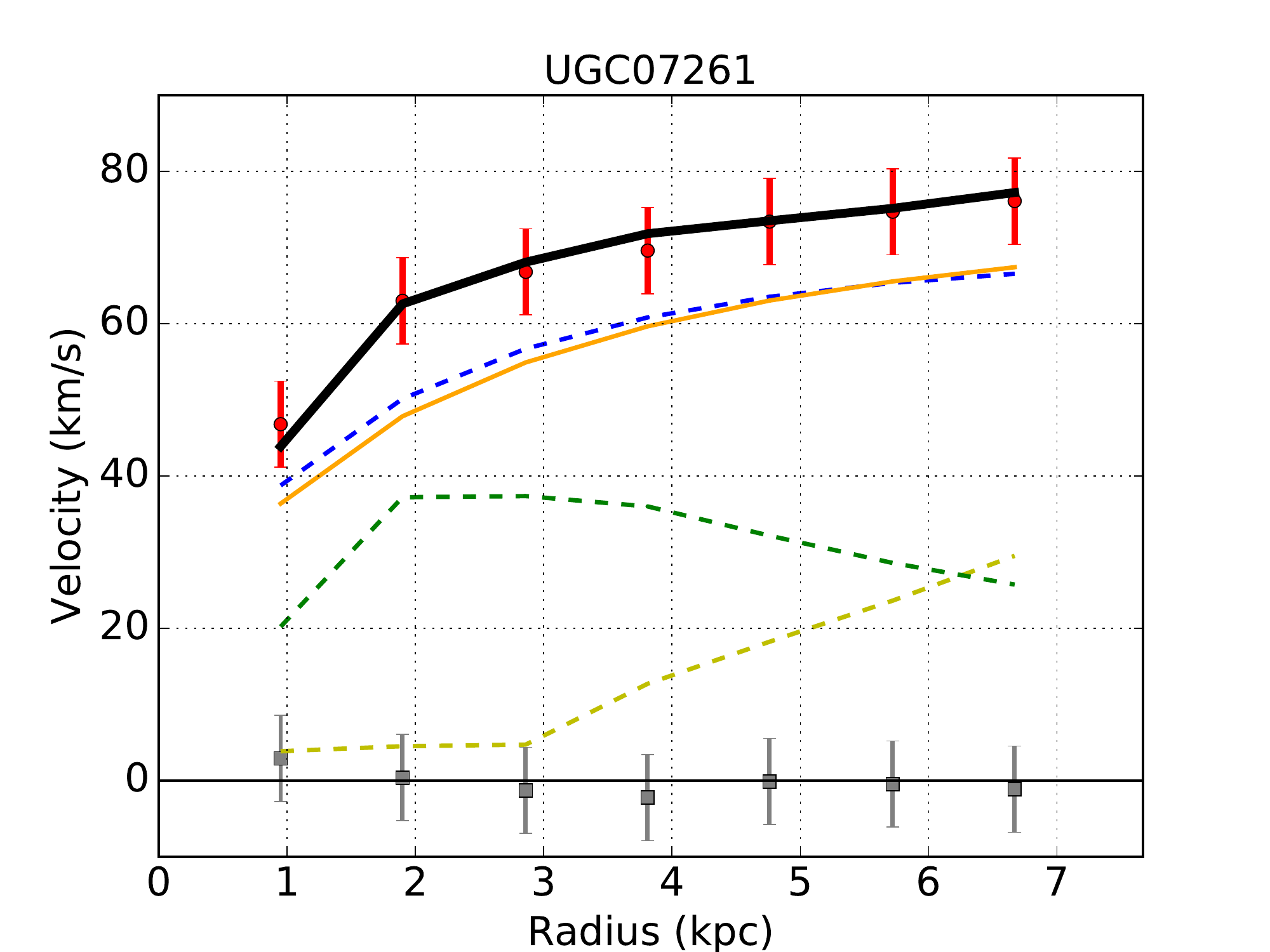} \\ 
		\includegraphics[width=5.5cm]{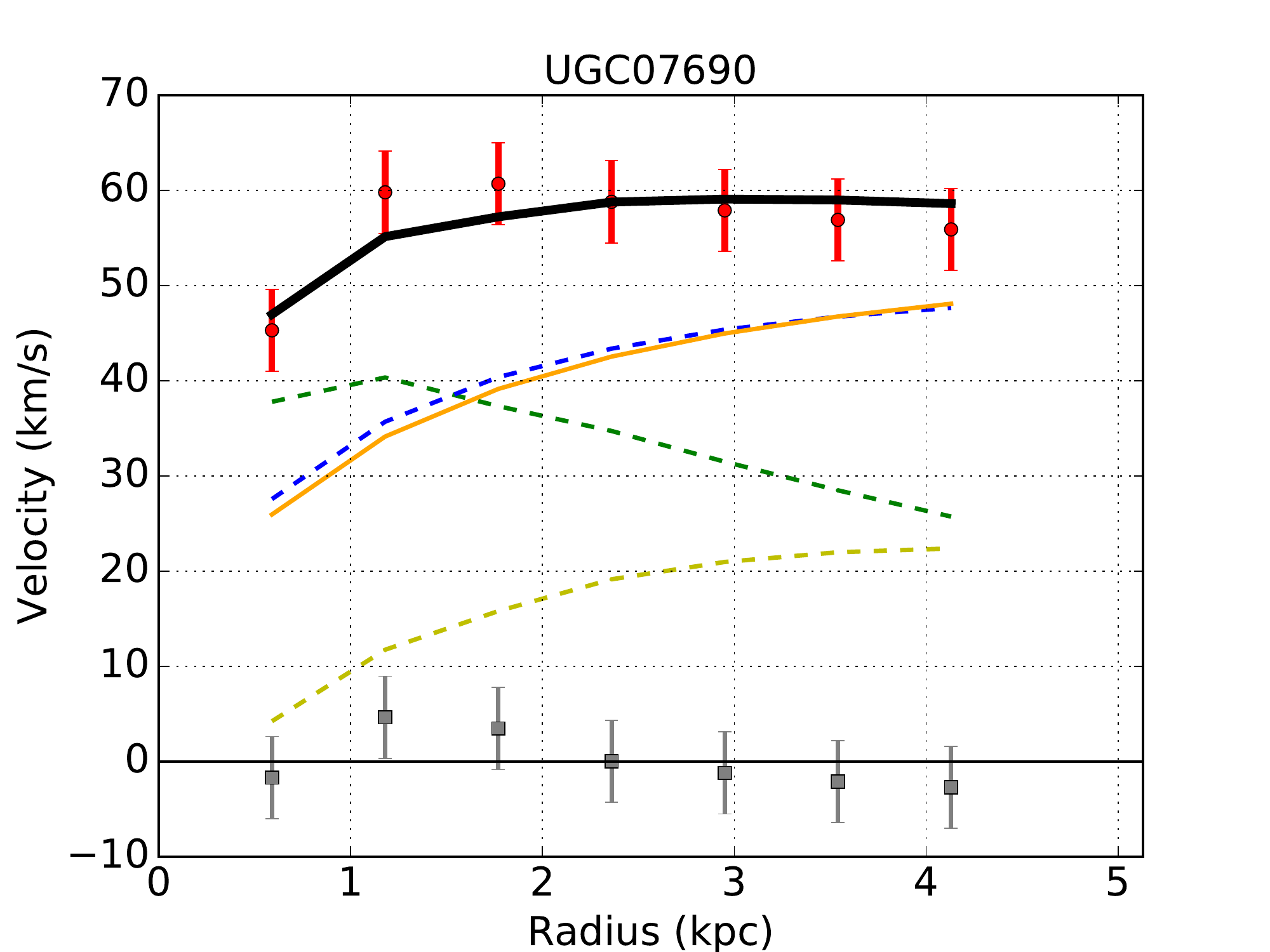} 
		\includegraphics[width=5.5cm]{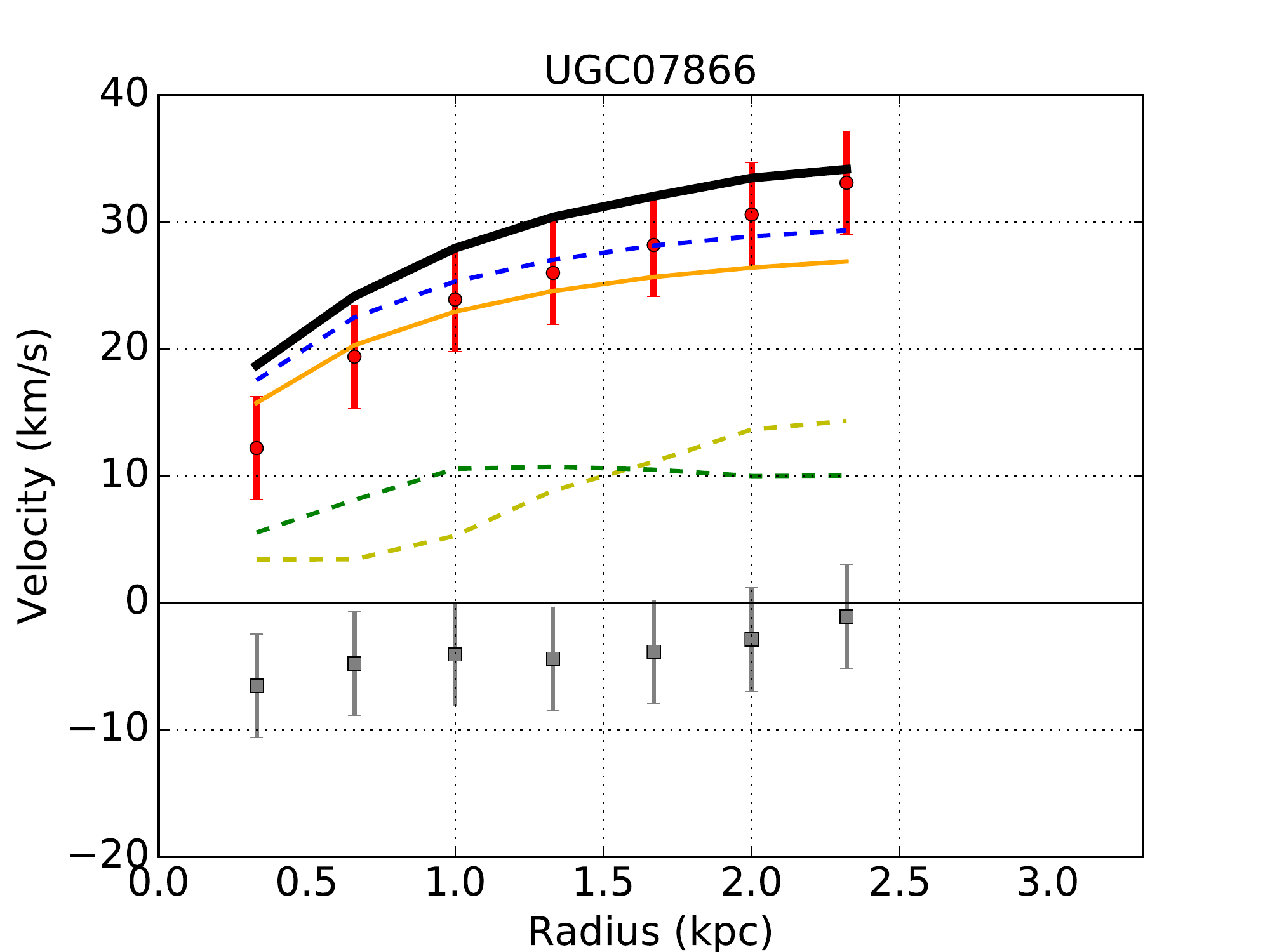} \\ 
		\includegraphics[width=5.5cm]{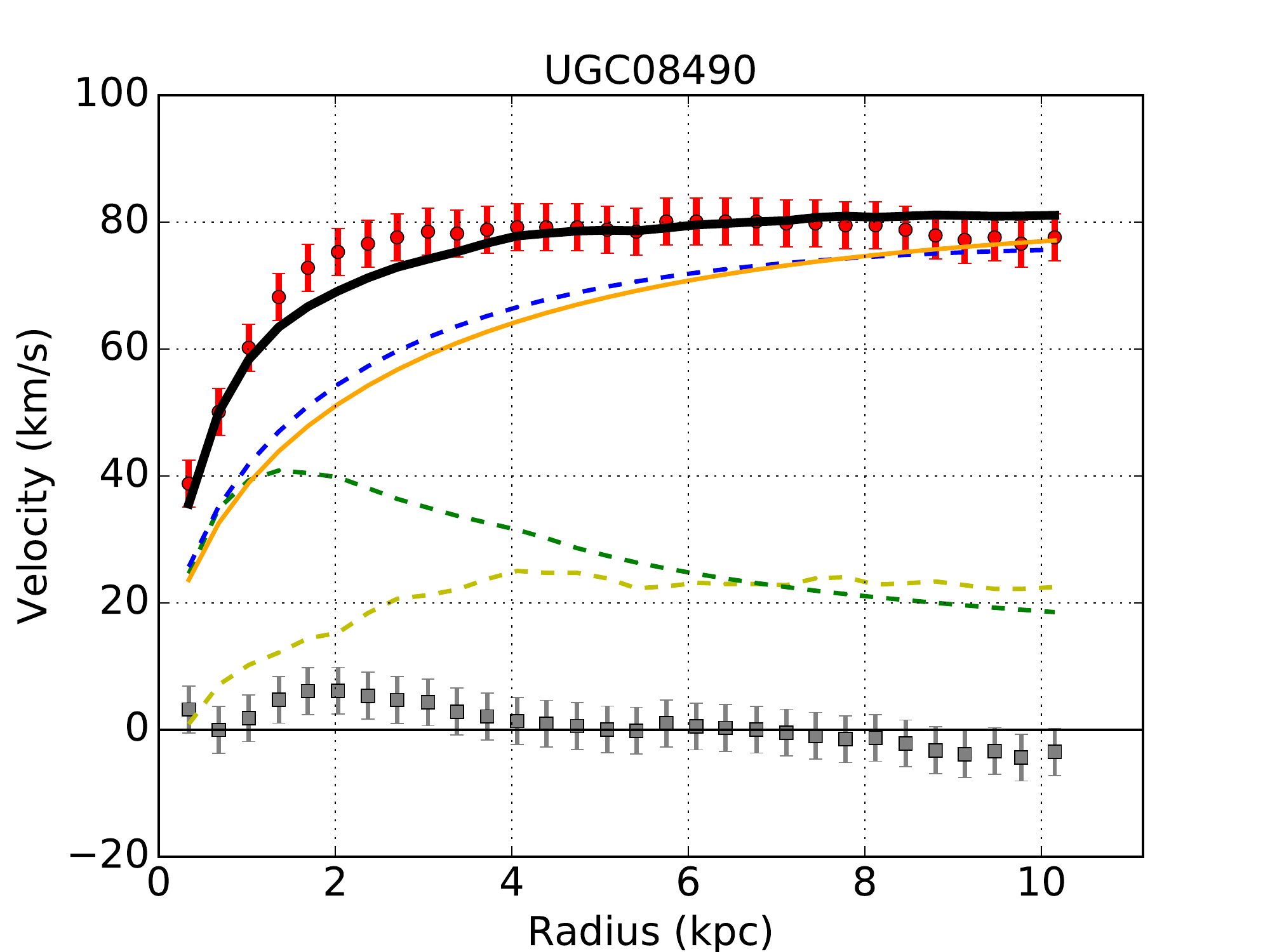} 
		\includegraphics[width=5.5cm]{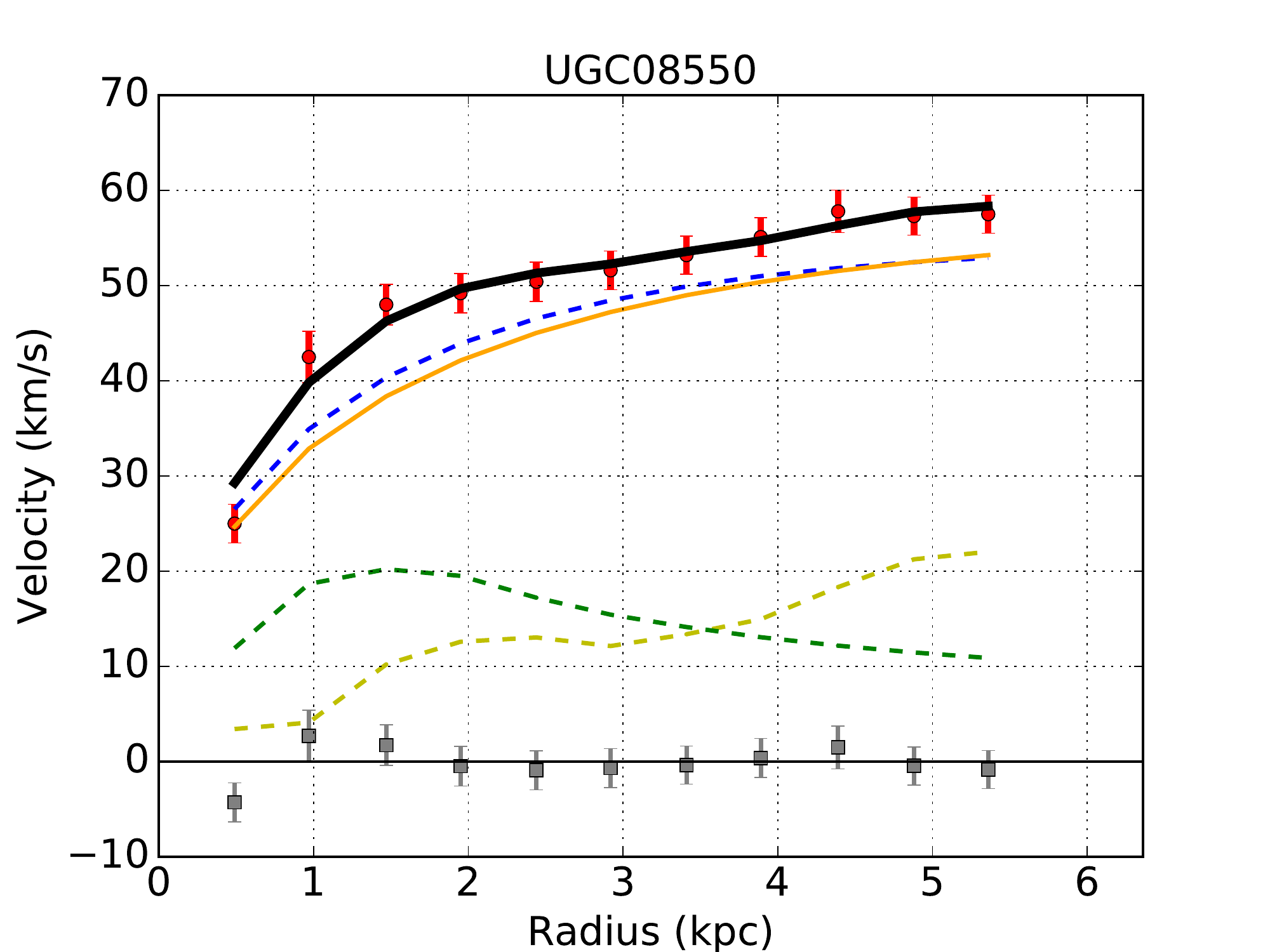} \\ 
		\includegraphics[width=5.5cm]{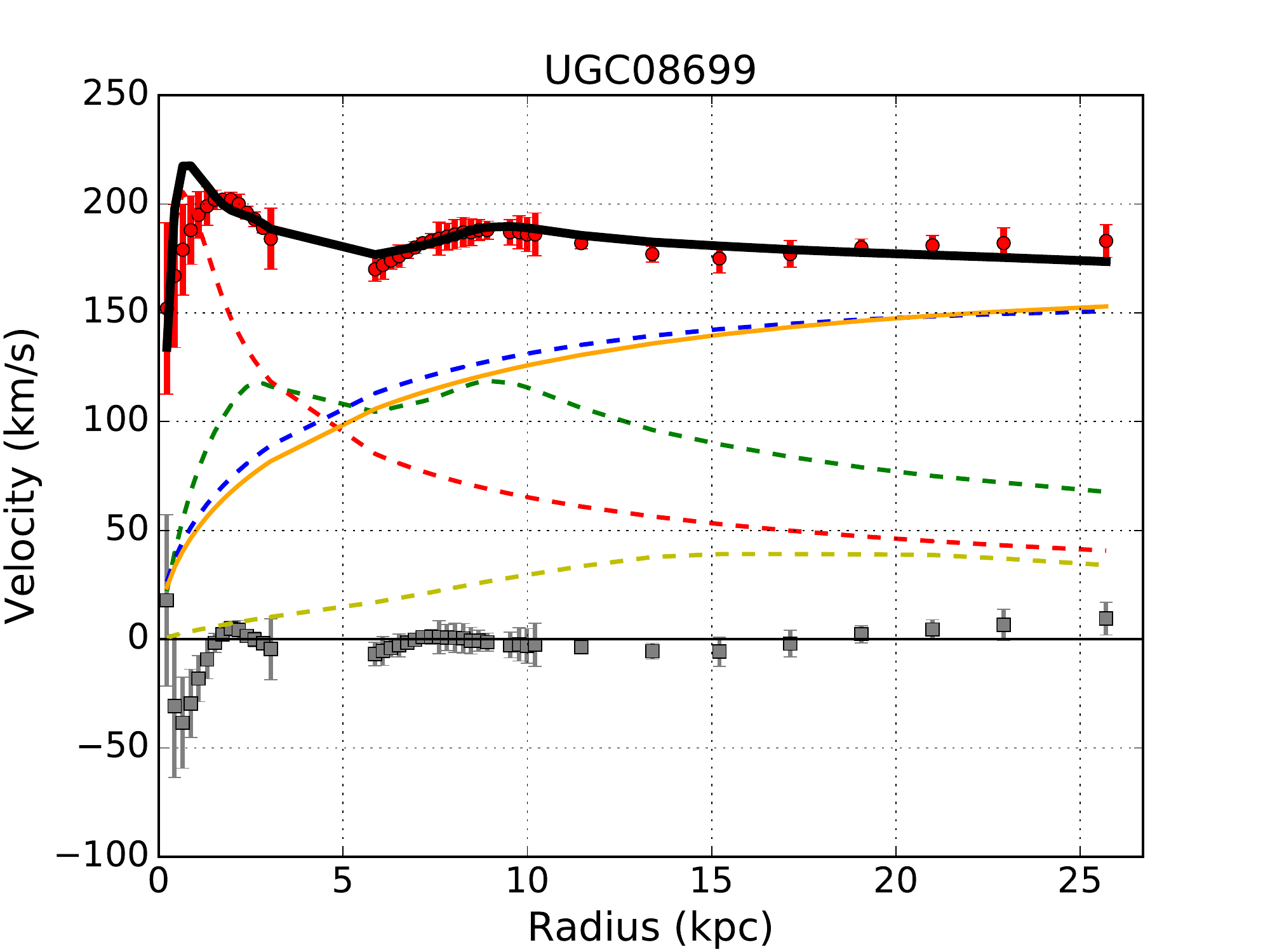} 
		\includegraphics[width=5.5cm]{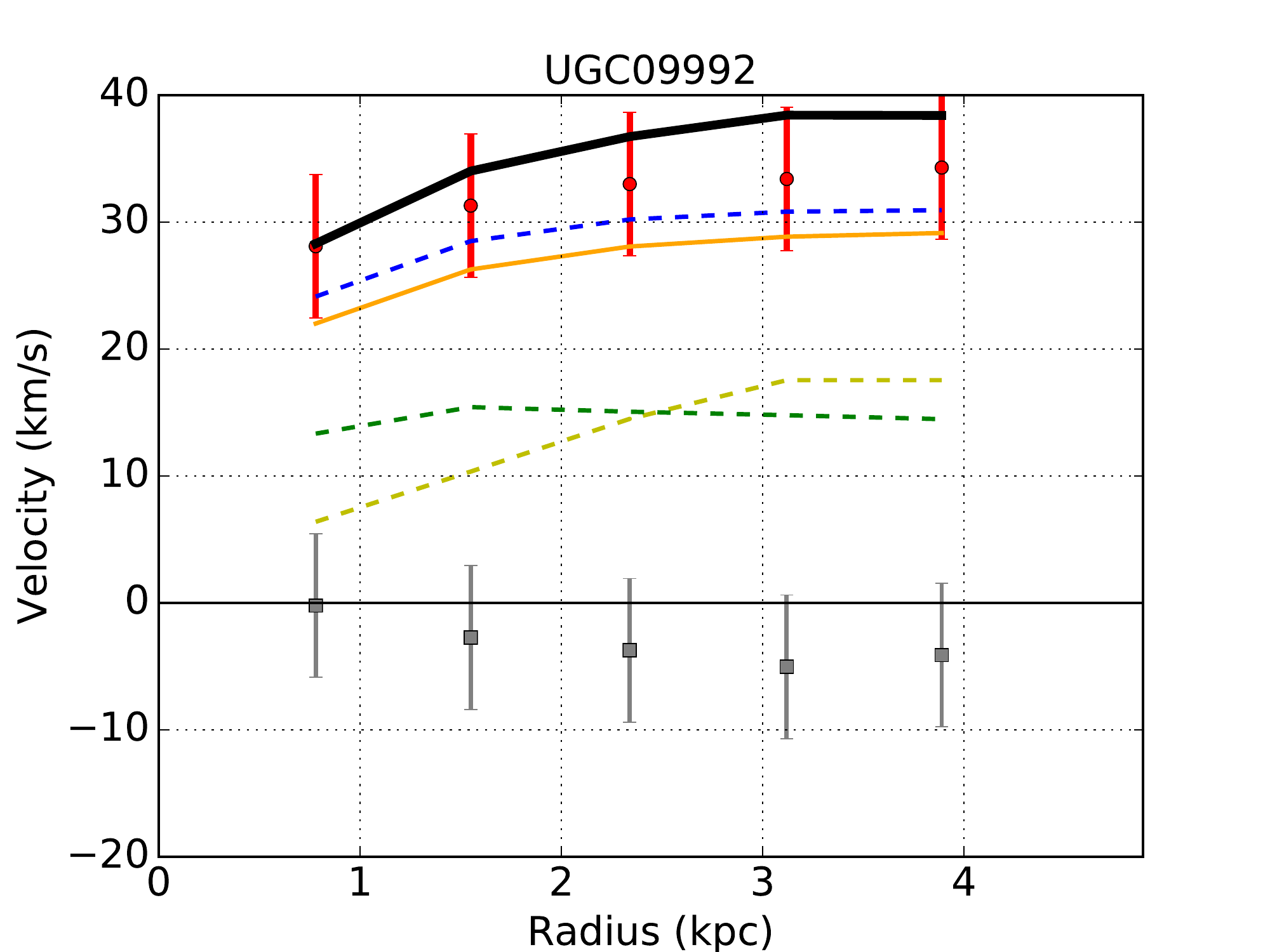} \\ 
		\includegraphics[width=5.5cm]{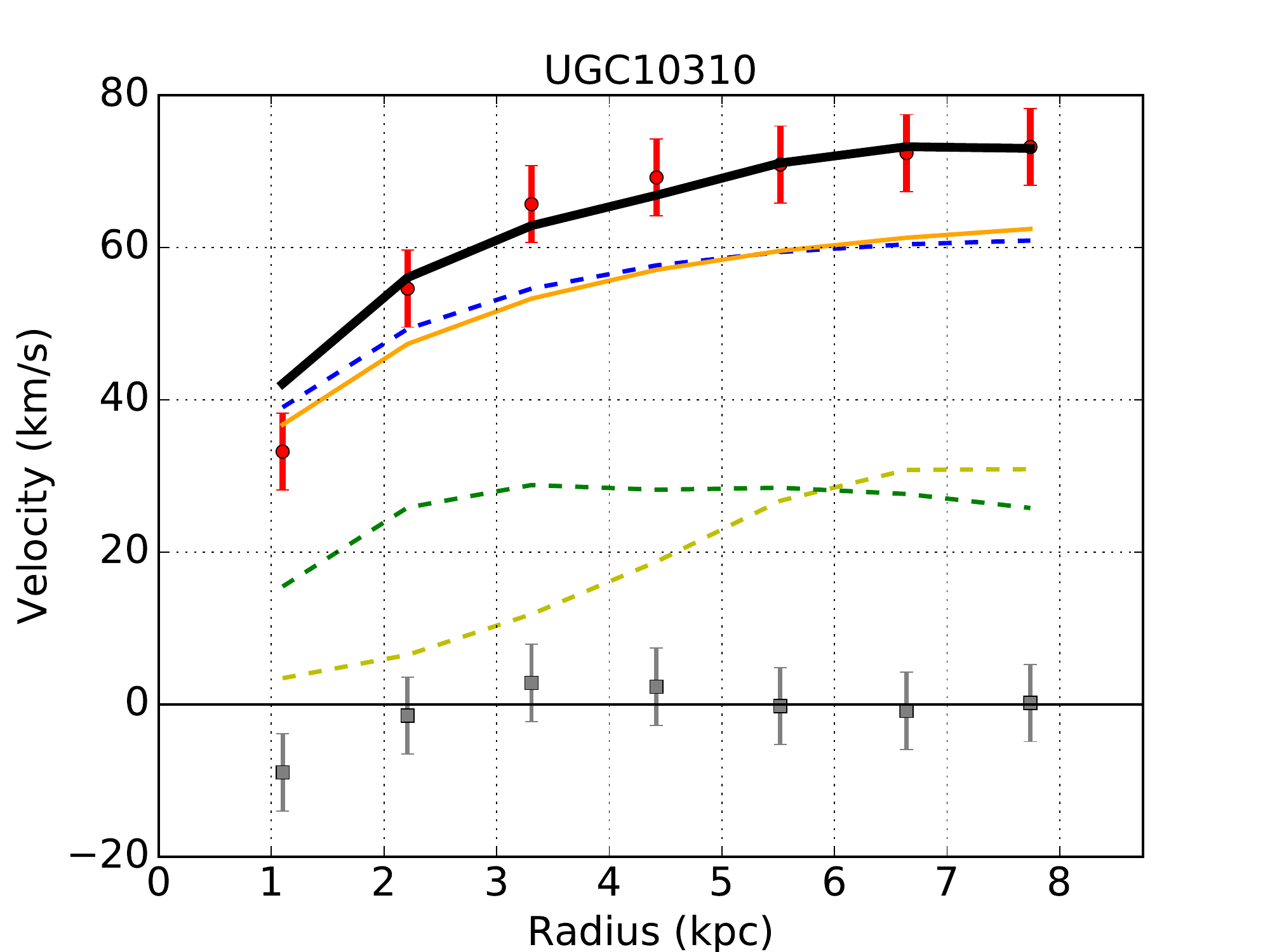} 
		\includegraphics[width=5.5cm]{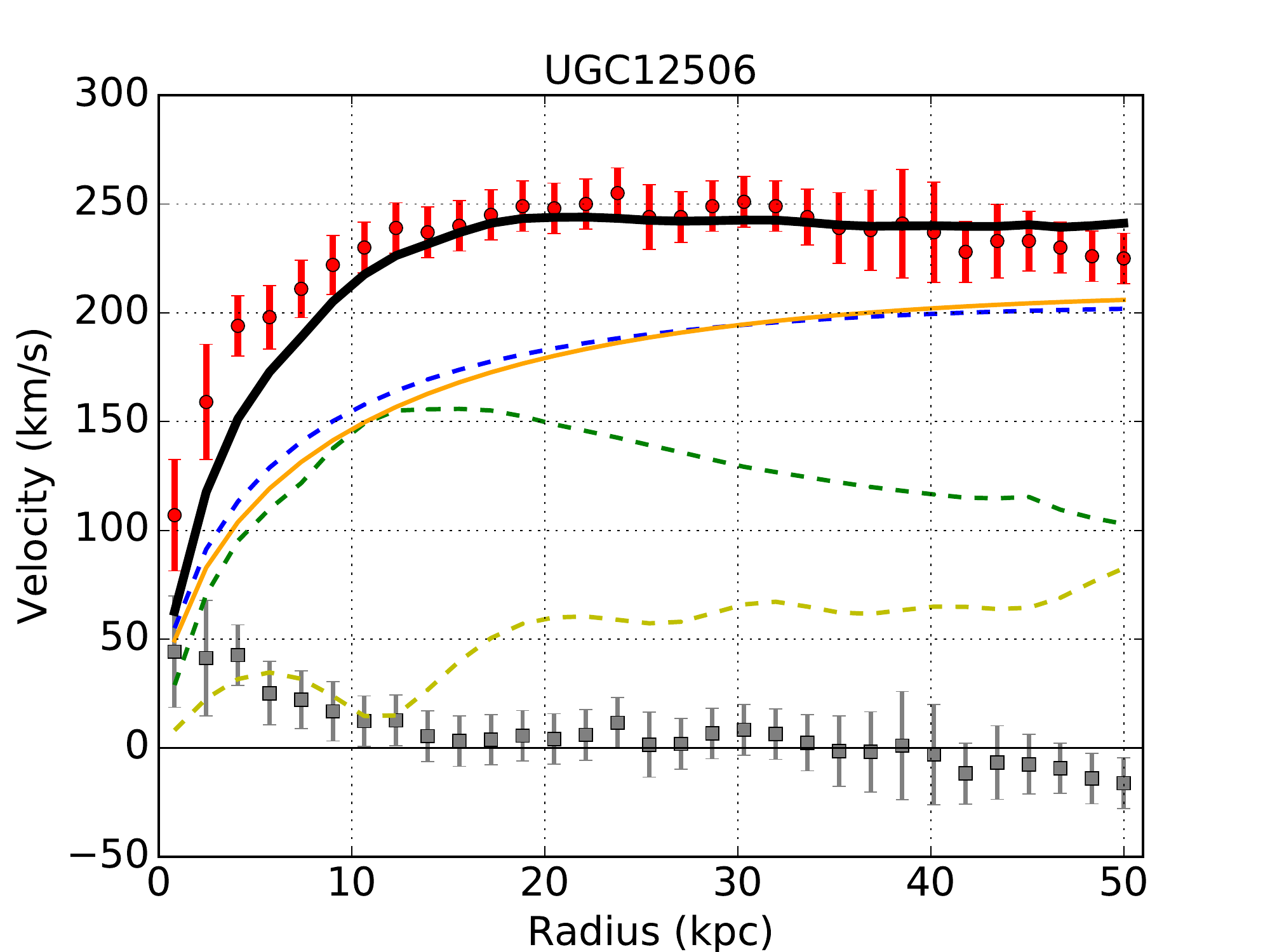} \\ 
		\caption{\label{rotationcurveC}Same as figure \ref{rotationcurveA}, but for set C. }
\end{figure}

\begin{figure}[tbp]
		\centering 
		\includegraphics[width=5.5cm]{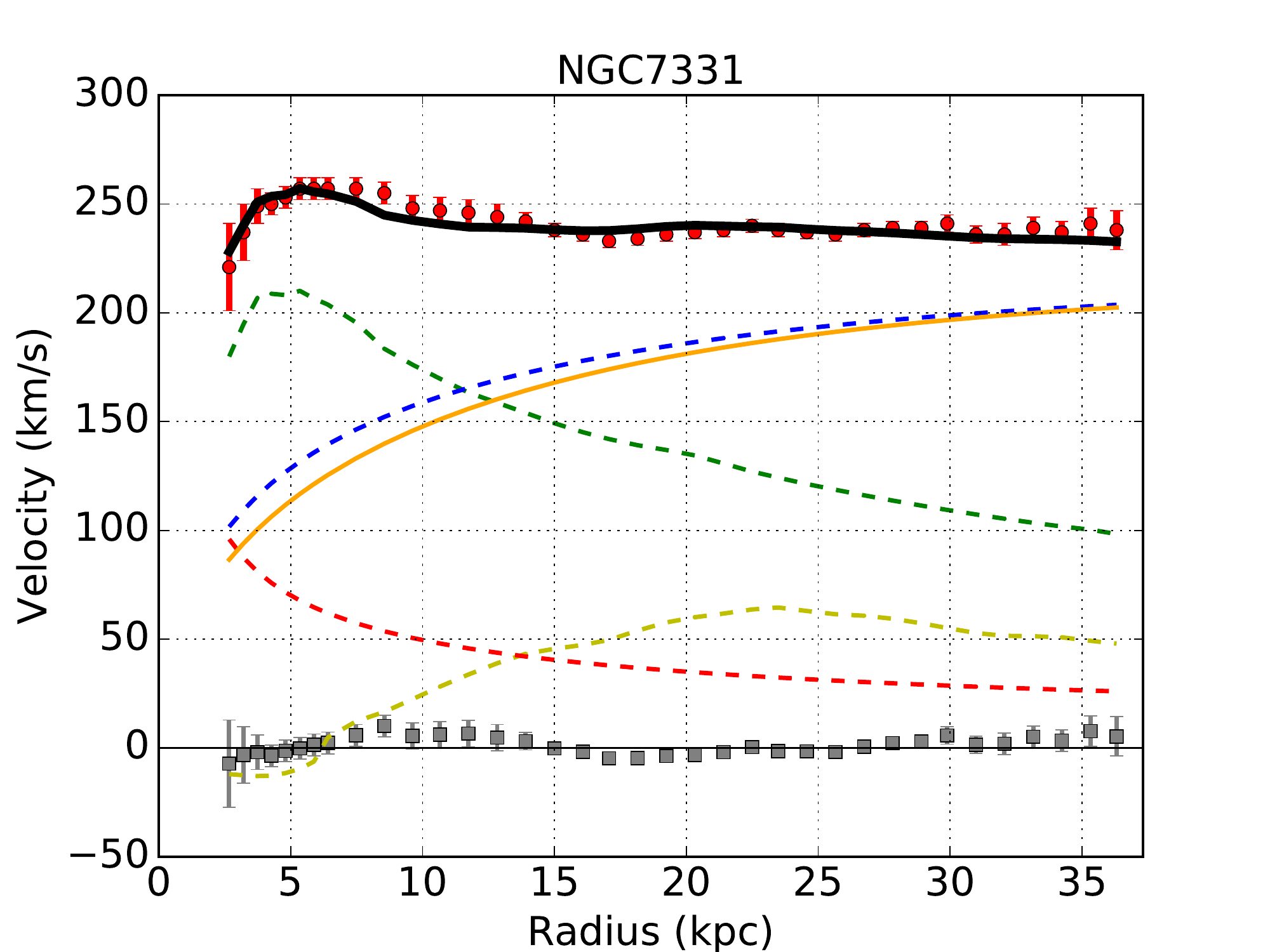} 
		\includegraphics[width=5.5cm]{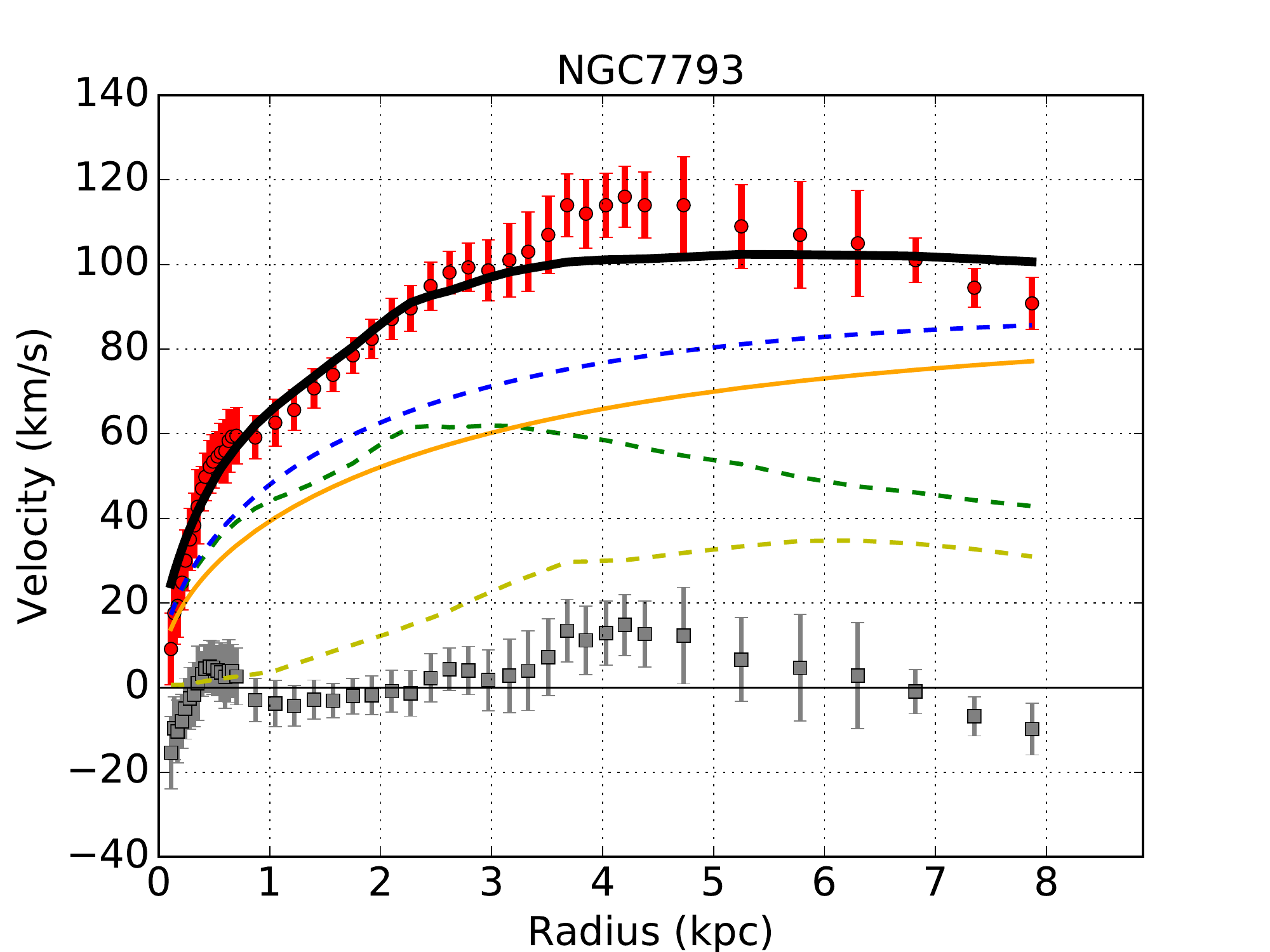} \\ 
		\includegraphics[width=5.5cm]{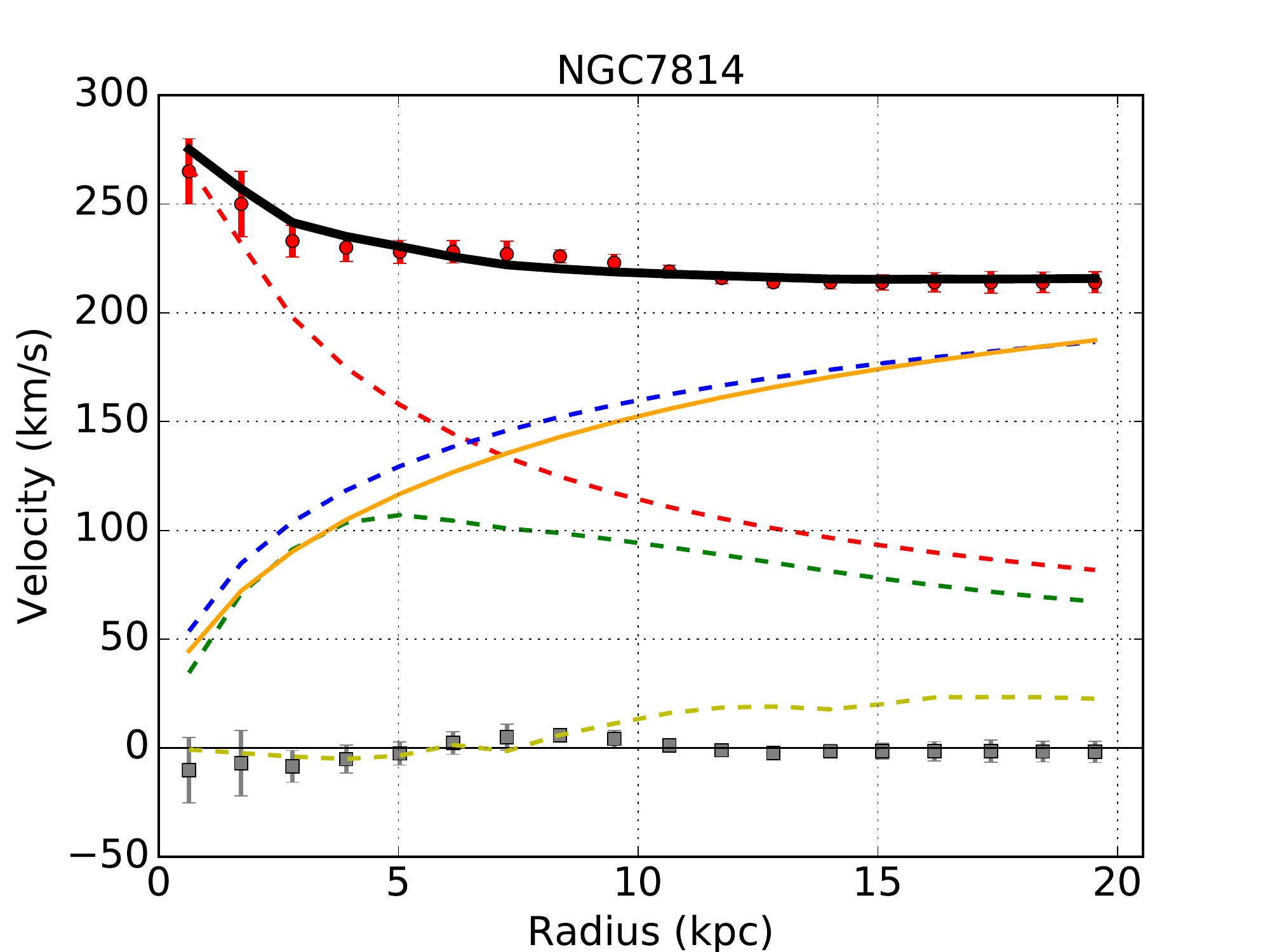} 
		\includegraphics[width=5.5cm]{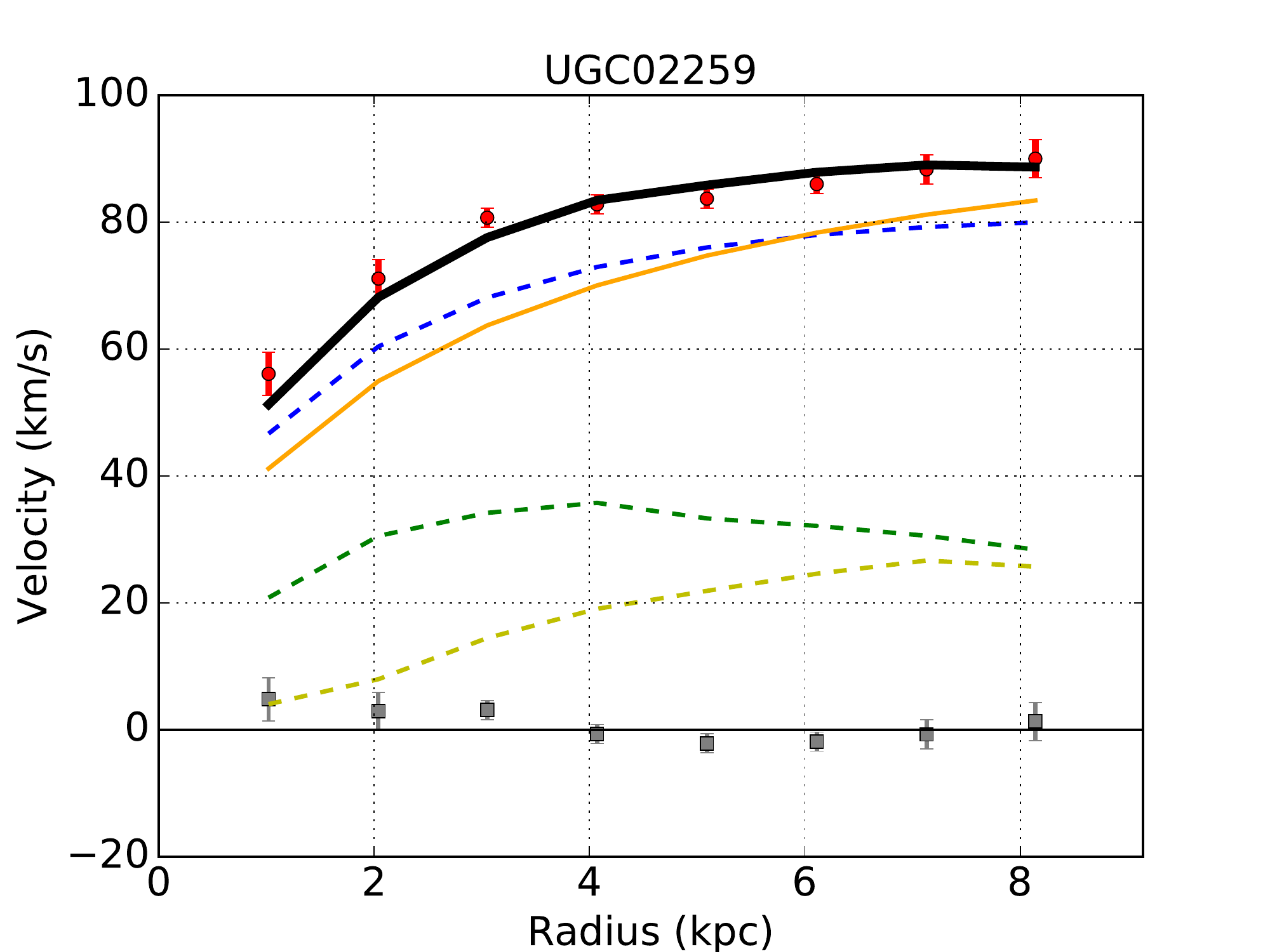} \\ 
		\includegraphics[width=5.5cm]{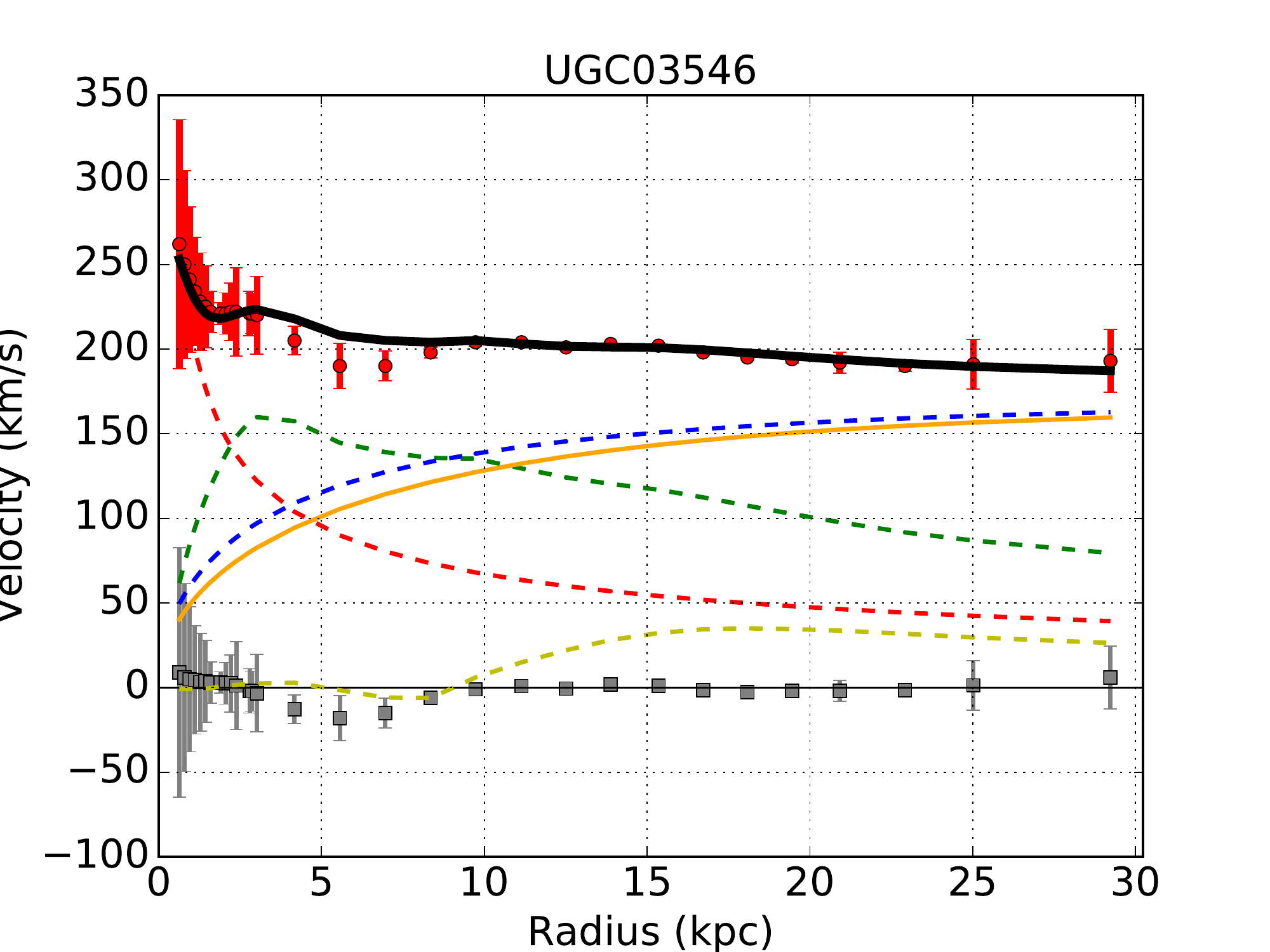} 
		\includegraphics[width=5.5cm]{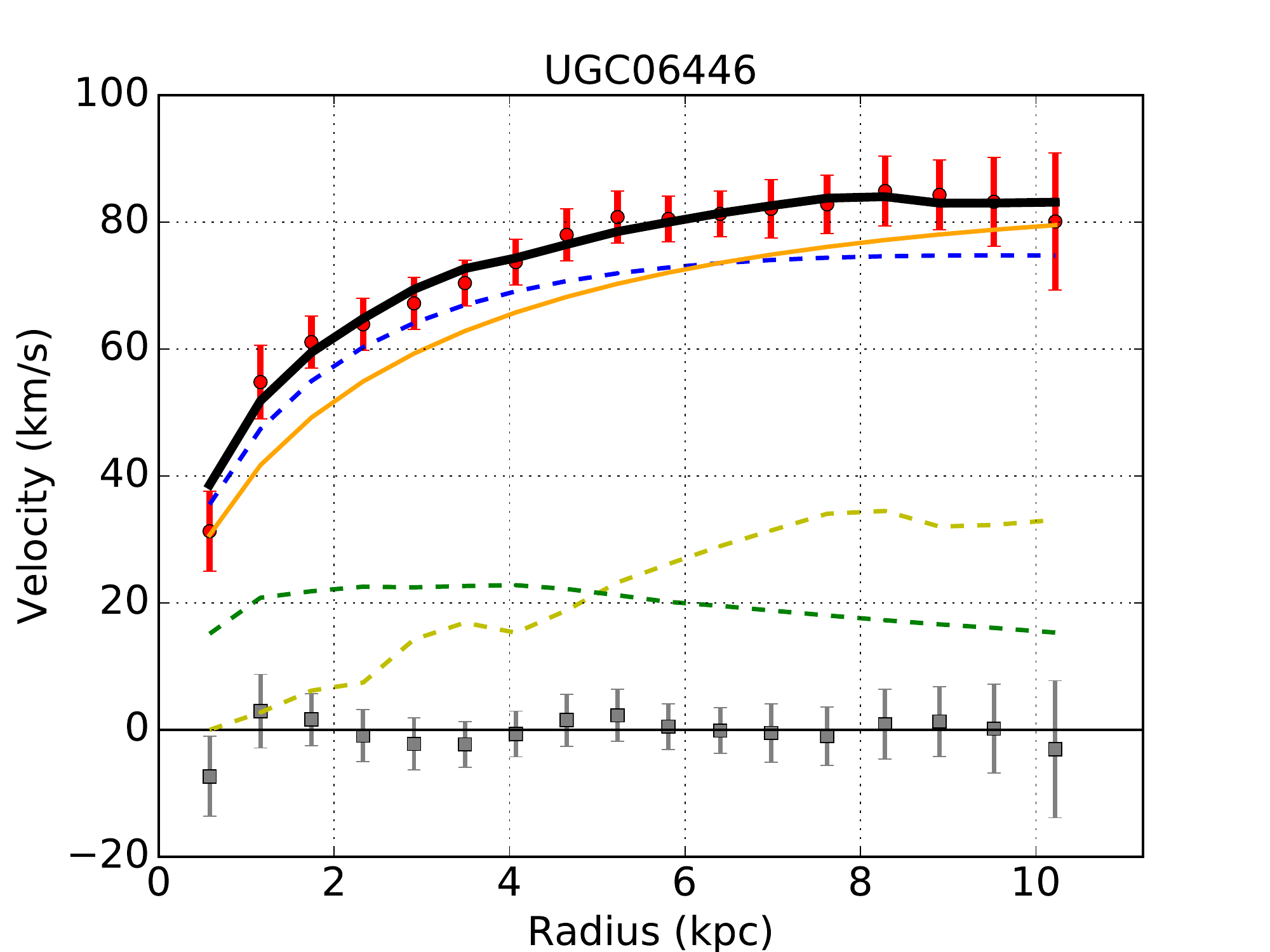} \\ 
		\includegraphics[width=5.5cm]{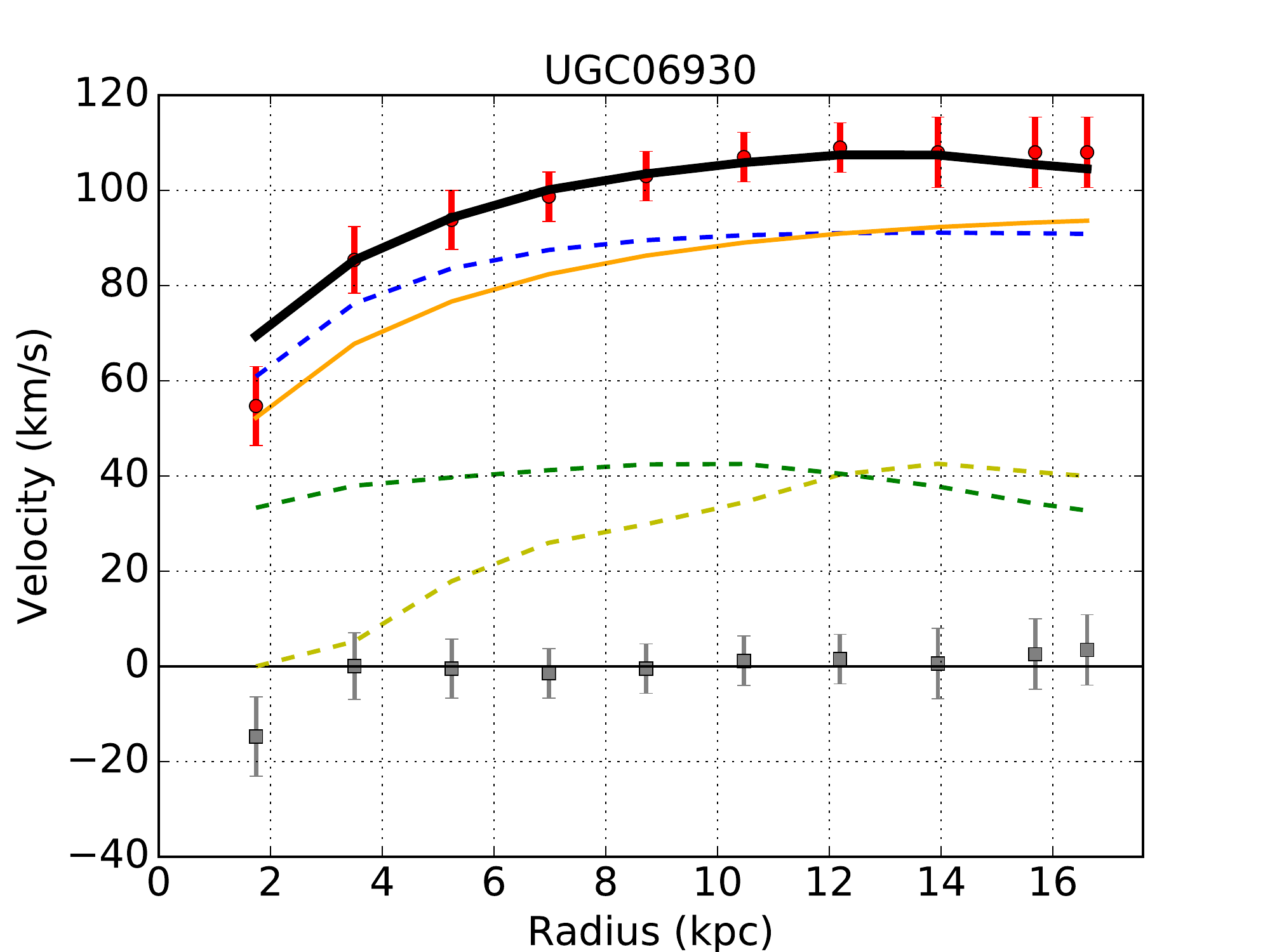} 
		\includegraphics[width=5.5cm]{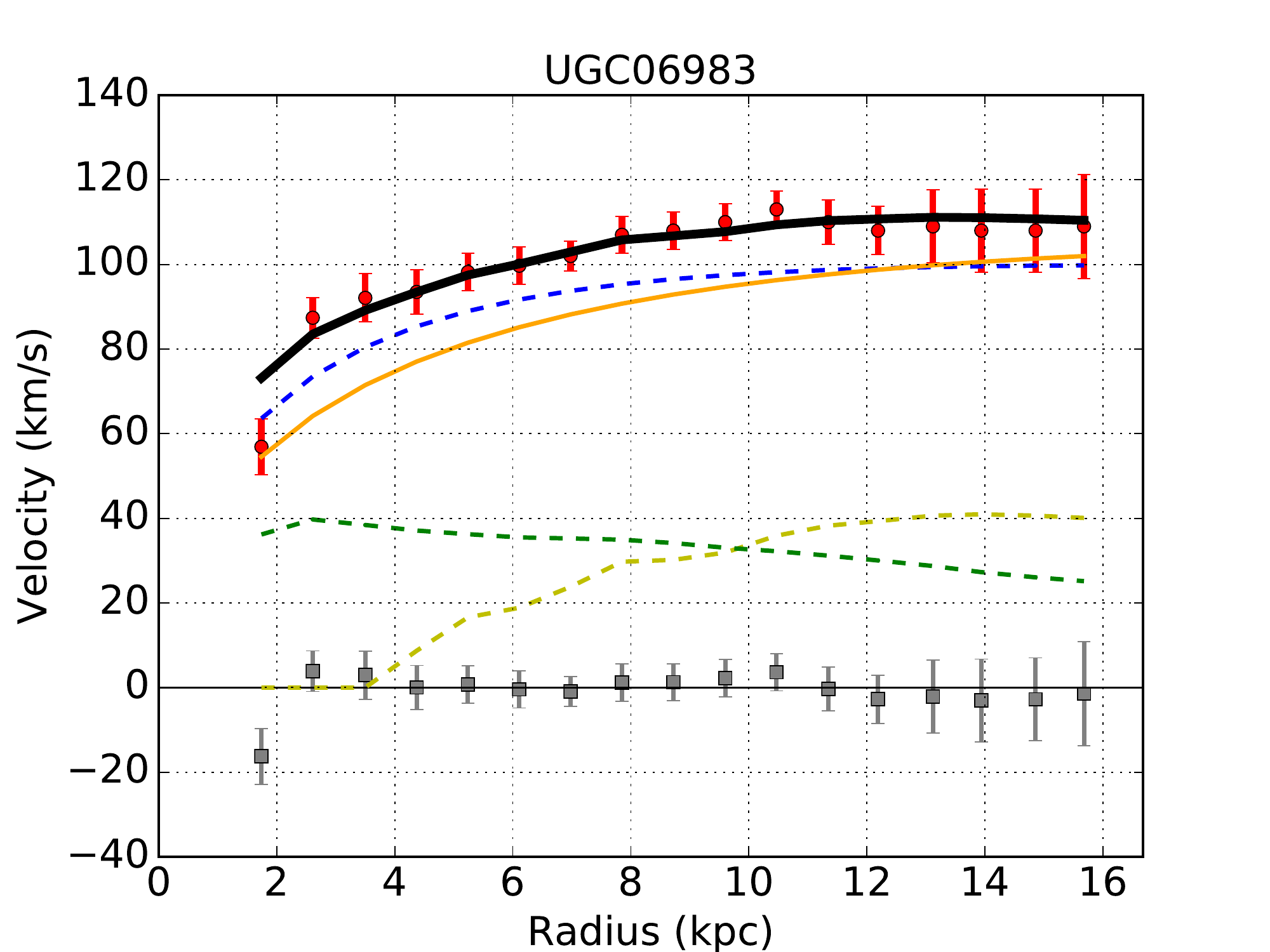} \\ 
		\includegraphics[width=5.5cm]{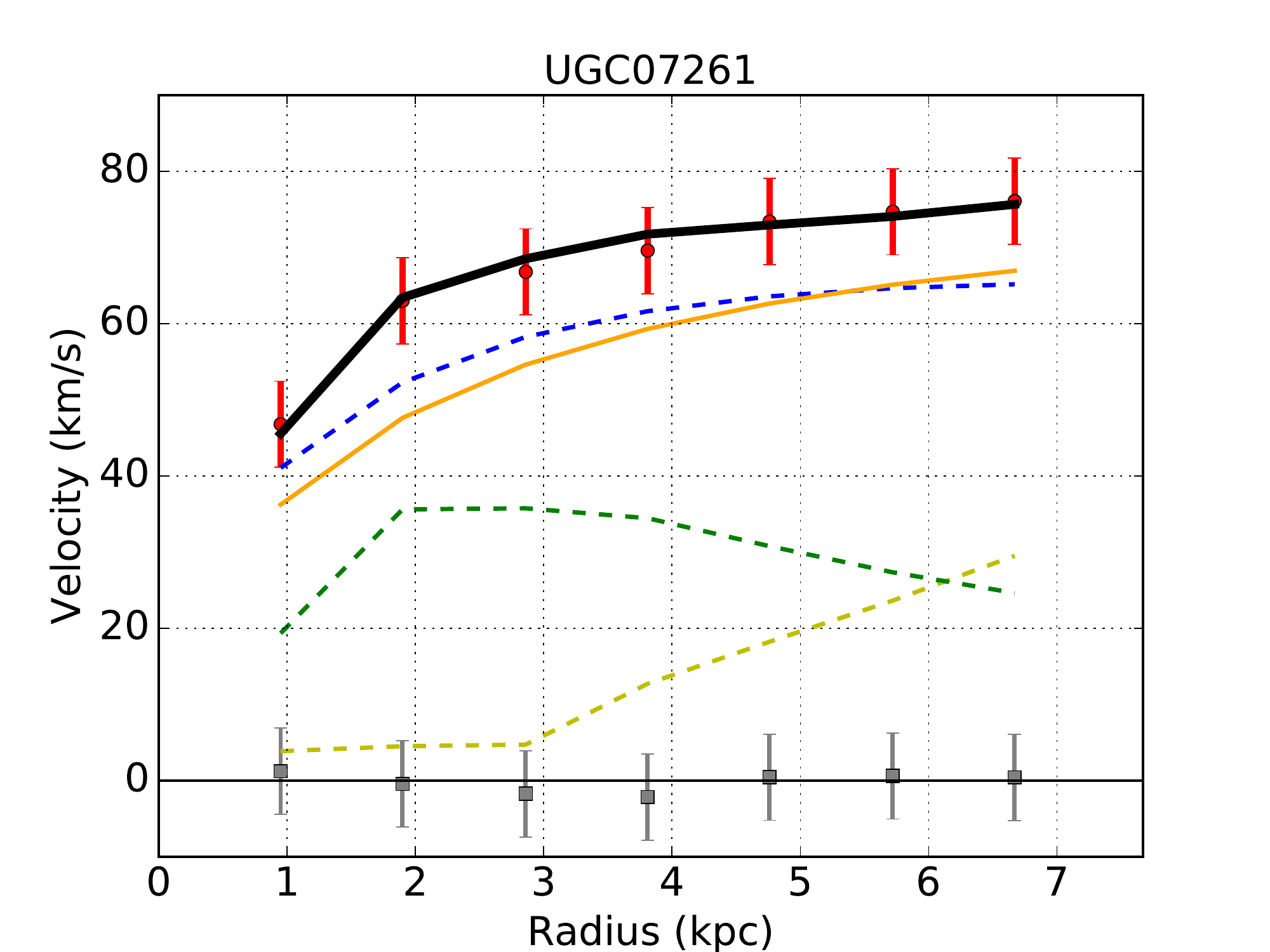} 
		\includegraphics[width=5.5cm]{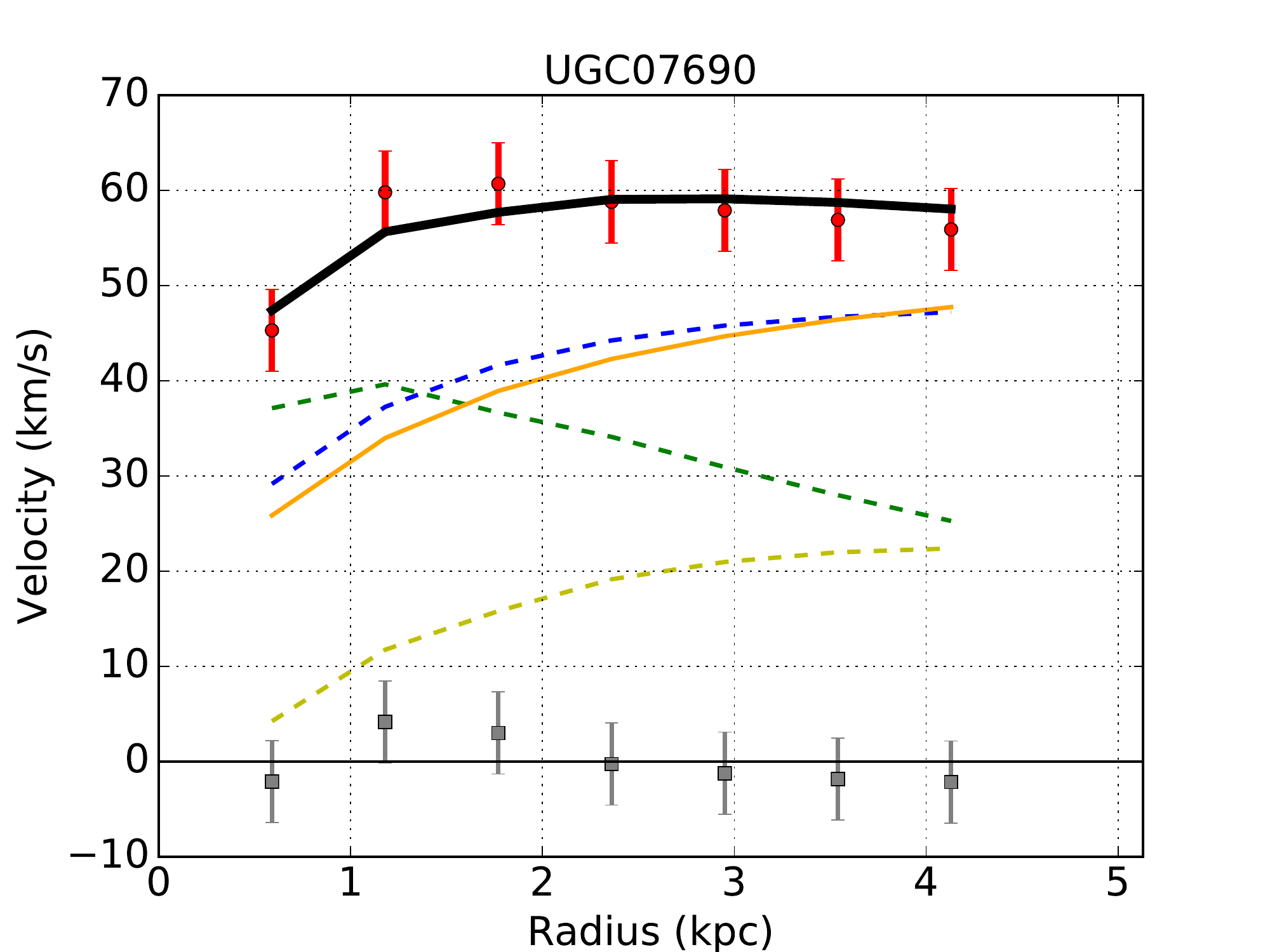} \\ 
		\caption{\label{rotationcurveD}Same as figure \ref{rotationcurveA}, but for set D. }
\end{figure}

In figure \ref{scatterplot} we  plotted the $1\sigma$ and $2\sigma$
contours of the marginalized distribution for the parameters ($\beta,\lambda$)
and their one-dimensional distributions for each set. Finally, the
individual rotation curves, evaluated using the equation (\ref{circular})
with the best-fit values of tables \ref{tabelaresultado1} and \ref{tabelabetalambda},
are plotted in figures \ref{rotationcurveA},\ref{rotationcurveB},\ref{rotationcurveC} and \ref{rotationcurveD}.

We have also computed the combined posterior for the parameters $\beta$ and $\lambda$  multiplying the marginalized ones of each set, the combined result is shown in figure \ref{scatterplot}. The best-fit and 
1$\sigma$ ranges for the parameters $\beta$ and $\lambda$ for the combined analysis are also displayed on table \ref{tabelabetalambda}. 

\begin{figure}[tbp]
	\centering 
	\includegraphics[width=15cm]{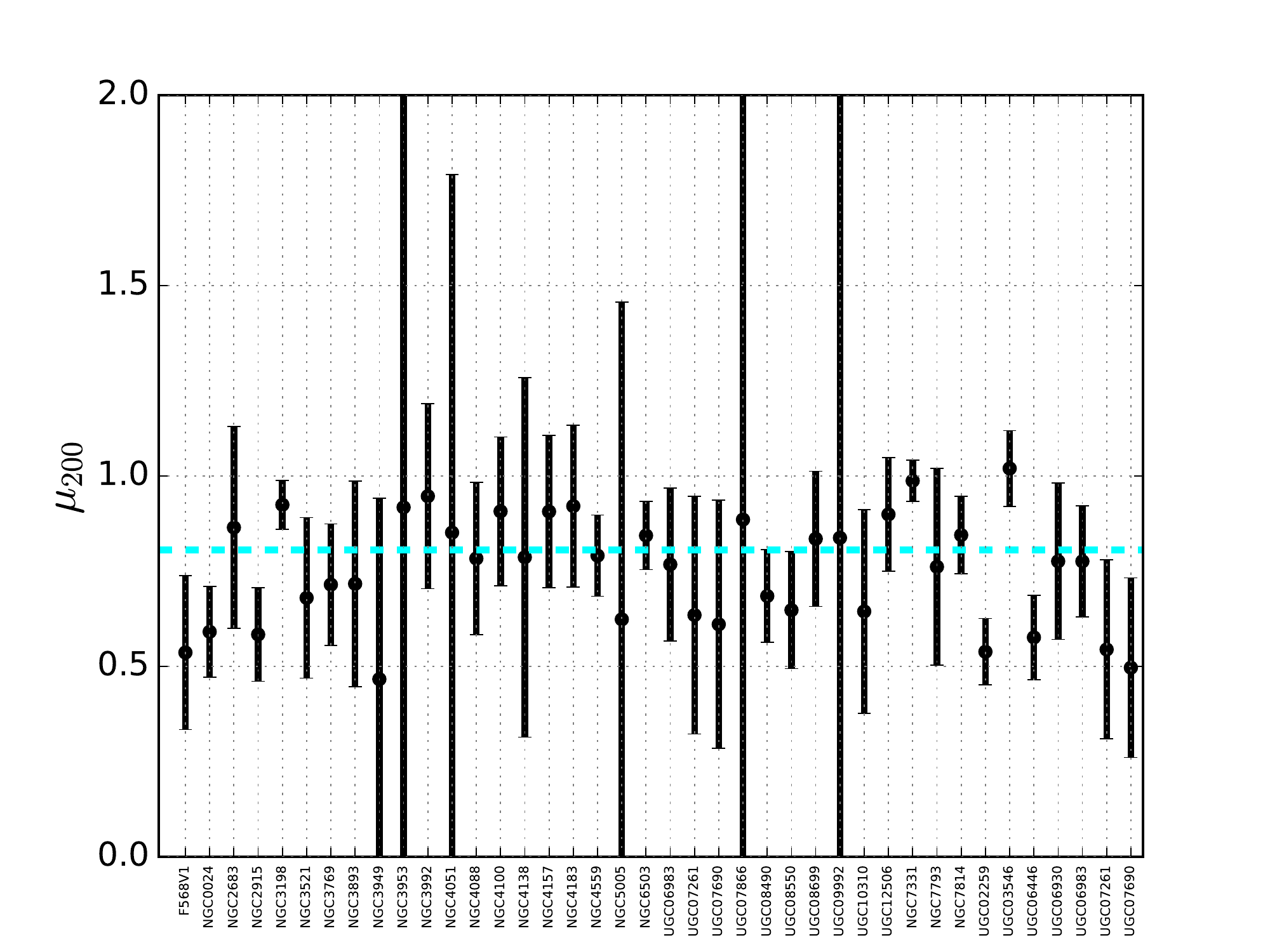}\\
	\caption{\label{deltaM200}The ratio $\mu_{200}$ (black dots) and the respective error bars. The cyan dotted line is the average value of $\mu_{200}$.}
\end{figure}

According to our results, we obtained an attractive Yukawa interaction, thus a decrease of the amount of dark matter necessary for a galaxy  to reproduce the behaviour of rotation curves data, is expected. 
We quantify this decrease with the  quantity $\mu_{200}\equiv \frac{M_{200}}{M_{200(\beta=0)}}$,  plotted  in figure \ref{deltaM200}. In order to propagate the errors in $\mu_{200}$ we assumed as the standard deviation the largest value of the asymmetric error bars. 
We also quantify the ratios for the other parameters $\Upsilon_\text{*D}$ and $\Upsilon_\text{*B}$, namely $\gamma_\text{*D}\equiv\frac{\Upsilon_\text{*D}}{\Upsilon_\text{*D}(\beta=0)}$ and 
$\gamma_\text{*B}\equiv\frac{\Upsilon_\text{*B}}{\Upsilon_\text{*B}(\beta=0)}$. The errors on $\Upsilon_\text{*D}$ and $\Upsilon_\text{*B}$ are also asymmetric, hence for 
$\gamma_\text{*D}$ and $\gamma_\text{*B}$  we propagate the errors as we did for $\mu_{200}$.	
The average value of $\mu_{200}$ is $\langle\mu_{200}\rangle=0.80\pm0.02$, corresponding to a $20\%$ of reduction of dark matter due the fifth force. 
We also obtained the average values $\langle\gamma_\text{*D}\rangle=0.96\pm0.01$ and $\langle\gamma_\text{*B}\rangle=0.96\pm0.04$.

It must be noted that the simple way of counting degrees of freedom adopted above is actually misleading. In fact, the observational data are not compared just to a theoretical model
depending on free parameters; rather, they are compared to the sum of a theoretical model (the NFW profile) plus the {\it observed} baryonic component, possibly rescaled by $\Upsilon_{*,\text{D}},\Upsilon_{*,\text{B}}$. Also for this reason, we turn now to the Bayesian evidence to assess the relative probability of the two models, with and without the Yukawa correction. In the Bayesian evidence ratio, in fact, only the difference of degrees of freedom between model matters, and this difference arises only because of the Yukawa theoretical model.
The evidence is given by
\begin{equation}
E=\int \mathcal{L(\mathbf{p},\beta,\lambda)}\mathcal{P}(\mathbf{p},\beta,\lambda)d\beta d\lambda d\mathbf{p}\;,
\end{equation} 
where $\mathcal{P}$ is the prior distribution.The Bayes ratio between the model 1 ($\beta\neq0$) and  model 2 ($\beta=0$) is defined as
\begin{equation}
B_{12}=\frac{\int \mathcal{L}_1(\mathbf{p},\beta,\lambda)\mathcal{P}_1(\mathbf{p},\beta,\lambda)d\beta d\lambda d\mathbf{p}}{\int \mathcal{L}_2(\mathbf{p})\mathcal{P}_2(\mathbf{p}) d\mathbf{p}}\;.
\end{equation}

Approximating the likelihood and the priors as Gaussian, we can compute analytically the evidence as
\begin{equation}
E=\mathcal{L}_\text{max}\sqrt{\frac{\det\mathbf{P}}{\det\mathbf{Q}}}\exp\left[-\frac{1}{2}(\hat{\theta}_\alpha F_{\alpha\beta}\hat{\theta}_\beta+\bar{\theta}_\alpha P_{\alpha\beta}\bar{\theta}_\beta-\tilde{\theta}_\alpha Q_{\alpha\beta}\tilde{\theta}_\beta)\right]\;,
\end{equation}
where $-2\ln\mathcal{L}_\text{max}=\chi^2_\text{min}$, $\theta_\alpha=\{\mathbf{p},\beta,\lambda\}$ for our general case; $\hat{\theta}_\alpha$ are the best fit values for the parameters and $\bar{\theta}_\alpha$ are the prior means. The matrix $\mathbf{Q}$ is $\mathbf{Q}=\mathbf{F}+\mathbf{P}$ and $\tilde{\theta}_\alpha=(\mathbf{Q}^{-1})_{\alpha\beta}[F_{\beta\sigma}\hat{\theta}_\sigma+P_{\beta\sigma}\bar{\theta}_\sigma]$, where $\mathbf{F}$ is the Fisher matrix and $\mathbf{P}$ is the inverse of the covariance matrix of the prior. When the prior is weak, which is our case, the expression above can be simplified and the Bayes ratio becomes
\begin{equation}
B_{12}=e^{-\frac{1}{2}(\chi^2_\text{min,1}-\chi^2_\text{min,2})}\sqrt{\frac{\det\mathbf{P}_1\det\mathbf{F}_2}{\det\mathbf{P}_2\det\mathbf{F}_1}}\;.
\end{equation}   

If we assume diagonal matrices for $\mathbf{P}$  the determinant is just the product of the diagonal entries, i.e. the inverse of the squared errors. If the prior is flat, as in our case, we can take the variance of an uniform distribution as squared error. Since the models 1,2 share most of the parameters, in the ratio $\det\mathbf{P}_1/\det\mathbf{P}_2$ all terms except the $\beta,\lambda$ simplify. Hence we have
\begin{equation}
B_{12}=e^{-\frac{1}{2}(\chi^2_\text{min,1}-\chi^2_\text{min,2})}\frac{1}{p_\beta p_\lambda}\sqrt{\frac{\det\mathbf{F}_2}{\det\mathbf{F}_1}}\;,
\end{equation}
where $p_\beta,p_\lambda$ are the square root of the variance of the uniform distribution assumed for $\beta,\lambda$. 

The Bayes factor for the combined sets is computed considering the combination of the Fisher matrices of the sets. Since the correlation between the sets is zero, the combined Fisher matrix, $\mathbf{F}_\text{comb}$, is a block diagonal matrix where the diagonal entries are the Fisher matrices of each set.
Thus, it is straightforward to calculate the determinant of $\mathbf{F}_\text{comb}$ for models 1 and 2. We assumed the same structure for the combined $\mathbf{P}$, namely $\mathbf{P}_\text{comb}$, 
since the the minimum value $\lambda_0$ changes according to the set.  

Once we have $B_{12}$, then the probability $\mathcal{P}_{12}$ that the right model is 1 rather than 2 is
\begin{equation}
\mathcal{P}_{12}=\frac{B_{12}}{1+B_{12}}\;.
\end{equation}

As an aside, we also computed the Bayesian Information Criterion (BIC), which gives a very simple approximation to the evidence. The expression for BIC is given by \cite{schwarz1978estimating}
\begin{equation}
\text{BIC}= -2\ln \mathcal{L}_\text{max}+2k\ln N\;,
\end{equation}   
where $k$ is the number of free parameters and $N$ is the number of data points. The values of $k$ for both models, namely $k_\text{Y} $ for the Yukawa model and $k_\text{sg}$ 
for standard gravity, are reported in table \ref{tabelabetalambda} together with the number of data points $N$ used in each set and in the combined analysis. In our case, the likelihoods are Gaussian and  hence we have again $-2\ln\mathcal{L}_\text{max}=\chi^2_\text{min}$. 
The relative BIC ($\Delta\text{BIC}$) is defined as
\begin{equation}
\Delta\text{BIC}\equiv\text{BIC}|_{\beta=0}-\text{BIC}|_{\beta\neq0}\;.
\end{equation} 
The relative BIC  approximates then $2\ln B_{12}$ in the limit in which the variance of the parameters decreases
when going from prior to posterior by a factor $N$. 
The $\Delta\text{BIC}$  and the confidence level (CL) associated to $\mathcal{P}_{12} $ values for each set and for the combined analysis are displayed in table \ref{tabelabetalambda}. As expected, the BIC gives  a rough approximation to the Gaussian evidence. Both prefer the $\beta\not= 0$ model to an extremely high significance, more than 8$\sigma$ for the combined set.

\begin{table}[tbp]
	\centering
	\small
	\begin{tabular}{|l|c|c|c|c|c|c|}
		\hline
		Set &Galaxy &\multicolumn{3}{|c|}{Best-fit values}&$\chi^2_{\text{red}}$\\
		\hline
		&&$\Upsilon_{*\text{D}}$&$\Upsilon_{*\text{B}}$&$M_{200}(10^{11}M_{\odot})$&\\\hline
		A&F568V1 & $ 0.60 ^{+ 0.20 }_{- 0.10 }$ & -                              & $ 2.91 ^{+ 0.61 } _{- 0.82 }$ & 0.35\\\hline
		A&NGC0024 & $ 0.79 ^{+ 0.01 }_{- 0.01 }$ &-                              & $ 1.63 ^{+ 0.21 } _{- 0.28 }$ & 1.68\\\hline
		A&NGC2683 & $ 0.64 ^{+ 0.04 }_{- 0.04 }$ & $ 0.52^{+ 0.15 }_{- 0.21 }$   & $ 3.79 ^{+ 0.64 } _{- 0.81 }$ & 1.37\\\hline
		A&NGC2915 & $ 0.32 ^{+ 0.01 }_{- 0.02 }$ &-                              & $ 0.76 ^{+ 0.10 } _{- 0.14 }$ & 0.98\\\hline
		A&NGC3198 & $ 0.40 ^{+ 0.04 }_{- 0.05 }$ &-                              & $ 4.30 ^{+ 0.27 } _{- 0.29 }$ & 1.31\\\hline
		A&NGC3521 & $ 0.49 ^{+ 0.01 }_{- 0.02 }$ &-                              & $ 12.00 ^{+ 2.22 } _{- 2.81 }$ & 0.37\\\hline
		A&NGC3769 & $ 0.33 ^{+ 0.02 }_{- 0.03 }$ &-                              & $ 1.90 ^{+ 0.25 } _{- 0.32 }$ & 0.75\\\hline
		A&NGC3893 & $ 0.46 ^{+ 0.04 }_{- 0.04 }$ &-                              & $ 8.64 ^{+ 2.58 } _{- 2.22 }$ & 1.26\\\hline
		A&NGC3949 & $ 0.36 ^{+ 0.03 }_{- 0.05 }$ &-                              & $ 8.85 ^{+ 4.37 } _{- 5.81 }$ & 0.45\\\hline
		A&NGC3953 & $ 0.62 ^{+ 0.07 }_{- 0.07 }$ &-                              & $ 3.39 ^{+ 1.64 } _{- 2.82 }$ & 0.73\\\hline
		B&NGC3992 & $ 0.74 ^{+ 0.05 }_{- 0.03 }$ &-                              & $ 14.47 ^{+ 2.06 } _{- 2.11 }$ & 0.88\\\hline
		B&NGC4051 & $ 0.40 ^{+ 0.05 }_{- 0.10 }$ &-                              & $ 2.32 ^{+ 1.23 } _{- 1.64 }$ & 1.27\\\hline
		B&NGC4088 & $ 0.31 ^{+ 0.01 }_{- 0.01 }$ &-                              & $ 3.62 ^{+ 0.65 } _{- 0.76 }$ & 1.09\\\hline
		B&NGC4100 & $ 0.67 ^{+ 0.03 }_{- 0.03 }$ &-                              & $ 4.70 ^{+ 0.76 } _{- 0.83 }$ &1.20\\\hline
		B&NGC4138 & $ 0.69 ^{+ 0.09 }_{- 0.05 }$ & $ 0.53^{+ 0.10 }_{- 0.21 }$   & $ 3.18 ^{+ 1.00 } _{- 1.44 }$ & 2.67\\\hline
		B&NGC4157 & $ 0.35 ^{+ 0.02 }_{- 0.03 }$ & $ 0.45^{+ 0.09 }_{- 0.15 }$   & $ 7.44 ^{+ 1.24 } _{- 1.38 }$ & 0.76\\\hline
		B&NGC4183 & $ 0.49 ^{+ 0.09 }_{- 0.14 }$ &-                              & $ 1.32 ^{+ 0.22 } _{- 0.25 }$ & 0.19\\\hline
		B&NGC4559 & $ 0.31 ^{+ 0.01 }_{- 0.01 }$ &-                              & $ 1.85 ^{+ 0.22 } _{- 0.22 }$ & 0.43\\\hline
		B&NGC5005 & $ 0.43 ^{+ 0.06 }_{- 0.11 }$ & $ 0.50^{+ 0.07 }_{- 0.08 }$   & $ 52.94 ^{+ 32.80 } _{- 49.57 }$ & 0.08\\\hline
		B&NGC6503 & $ 0.45 ^{+ 0.02 }_{- 0.03 }$ &-                              & $ 1.99 ^{+ 0.20 } _{- 0.22 }$ & 1.91\\\hline
		C&UGC06983 & $ 0.51 ^{+ 0.11 } _{- 0.16 }$ &-                            & $ 1.66 ^{+ 0.29 } _{- 0.36 }$ & 0.69\\\hline
		C&UGC07261 & $ 0.53 ^{+ 0.12 } _{- 0.21 }$ &-                            & $ 0.41 ^{+ 0.11 } _{- 0.14 }$ & 0.17\\\hline
		C&UGC07690 & $ 0.68 ^{+ 0.11 } _{- 0.06 }$ &-                            & $ 0.13 ^{+ 0.05 } _{- 0.05 }$ & 0.89\\\hline
		C&UGC07866 & $ 0.38 ^{+ 0.06 } _{- 0.08 }$ &-                            & $ 0.02 ^{+ 0.07 } _{- 0.02 }$ & 2.52\\\hline
		C&UGC08490 & $ 0.78 ^{+ 0.02 } _{- 0.01 }$ &-                            & $ 0.60 ^{+ 0.08 } _{- 0.10 }$ & 0.78\\\hline
		C&UGC08550 & $ 0.49 ^{+ 0.08 } _{- 0.17 }$ &-                            & $ 0.18 ^{+ 0.04 } _{- 0.04 }$ & 1.02\\\hline
		C&UGC08699 & $ 0.71 ^{+ 0.05 } _{- 0.05 }$ & $ 0.67 ^{+ 0.03 }_{- 0.05 }$& $ 6.86 ^{+ 1.16 } _{- 1.34 }$ & 0.86\\\hline
		C&UGC09992 & $ 0.43 ^{+ 0.10 } _{- 0.13 }$ &-                            & $ 0.03 ^{+ 0.03 } _{- 0.03 }$ & 1.98\\\hline
		C&UGC10310 & $ 0.53 ^{+ 0.10 } _{- 0.22 }$ &-                            & $ 0.28 ^{+ 0.06 } _{- 0.08 }$ & 1.25\\\hline
		C&UGC12506 & $ 0.78 ^{+ 0.02 } _{- 0.01 }$ &-                            & $ 17.77 ^{+ 2.53 } _{- 2.22 }$ & 1.22\\\hline
		D&NGC7331 & $ 0.32 ^{+ 0.01 }_{- 0.01 }$ & $ 0.49 ^{+ 0.08 } _{- 0.18 }$ & $ 20.56 ^{+ 0.86 } _{- 0.79 }$ & 0.87\\\hline
		D&NGC7793 & $ 0.41 ^{+ 0.05 }_{- 0.05 }$ &-                              & $ 1.01 ^{+ 0.21 } _{- 0.24 }$ & 0.95\\\hline
		D&NGC7814 & $ 0.76 ^{+ 0.04 } _{- 0.03 }$ & $ 0.60 ^{+ 0.03 } _{- 0.03 }$& $ 21.39 ^{+ 2.07 } _{- 2.09 }$ & 0.82\\\hline
		D&UGC02259 & $ 0.72 ^{+ 0.08 } _{- 0.05 }$ &-                            & $ 0.75 ^{+ 0.09 } _{- 0.11 }$ & 2.84\\\hline
		D&UGC03546 & $ 0.55 ^{+ 0.04 } _{- 0.04 }$ & $ 0.38 ^{+ 0.04 } _{- 0.04 }$& $ 9.33 ^{+ 0.66 } _{- 0.64 }$ & 1.05\\\hline
		D&UGC06446 & $ 0.50 ^{+ 0.09 } _{- 0.19 }$ &-                            & $ 0.56 ^{+ 0.08 } _{- 0.09 }$ & 0.25\\\hline
		D&UGC06930 & $ 0.40 ^{+ 0.06 } _{- 0.10 }$ &-                            & $ 1.19 ^{+ 0.21 } _{- 0.20 }$ & 0.62\\\hline
		D&UGC06983 & $ 0.40 ^{+ 0.05 } _{- 0.09 }$ &-                            & $ 1.65 ^{+ 0.21 } _{- 0.23 }$ & 0.67\\\hline
		D&UGC07261 & $ 0.49 ^{+ 0.09 } _{- 0.18 }$ &-                            & $ 0.34 ^{+ 0.08 } _{- 0.10 }$ & 0.11\\\hline
		D&UGC07690 & $ 0.66 ^{+ 0.13 } _{- 0.07 }$ &-                            & $ 0.10 ^{+ 0.03 } _{- 0.04 }$ & 0.72\\			
		\hline		
	\end{tabular}
	\caption{\label{tabelaresultado1}The maximum likelihood estimation for the $\Upsilon_{\text{*D}}$,$\Upsilon_{\text{*B}}$ and $M_\text{200}$ parameters, and the goodness of fit $\chi_\text{red}^2$ for each galaxy, for the Yukawa model.}
\end{table}
\clearpage
\begin{table}[tbp]
	    \centering
		\small
		\begin{tabular}{|l|c|c|c|c|c|c|}
			\hline
			Set &Galaxy &\multicolumn{3}{|c|}{Best-fit values}&$\chi^2_{\text{red}}|_{\beta=0}$\\\hline
			
			&&$\Upsilon_{*\text{D}}$&$\Upsilon_{*\text{B}}$&$M_{200}(10^{11}M_{\odot})$&\\\hline
			A&F568V1 & $ 0.63 ^{+ 0.16 }_{- 0.10 }$ &-                              & $ 5.43 ^{+2.53 } _{- 3.57 }$ & 0.63\\\hline
			A&NGC0024 & $ 0.79 ^{+ 0.01 }_{- 0.01 }$ &-                             & $ 2.75 ^{+ 0.23 } _{- 0.29 }$ & 2.31\\\hline
			A&NGC2683 & $ 0.68 ^{+ 0.05 }_{- 0.04 }$ & $ 0.52^{+ 0.11 }_{- 0.21 }$  & $ 4.38 ^{+ 0.74 } _{- 0.96 }$ & 1.20\\\hline
			A&NGC2915 & $ 0.32 ^{+ 0.02 }_{- 0.02 }$ &-                             & $ 1.31 ^{+ 0.12 } _{- 0.14 }$ & 1.17\\\hline
			A&NGC3198 & $ 0.52 ^{+ 0.01 }_{- 0.01 }$ &-                             & $4.65 ^{+ 0.09 } _{- 0.11 } $ & 1.44\\\hline
			A&NGC3521 & $ 0.51 ^{+ 0.01 }_{- 0.01 }$ &-                             & $ 17.65 ^{+ 3.59 } _{- 3.36 }$ & 0.29\\\hline
			A&NGC3769 & $ 0.36 ^{+ 0.03 }_{- 0.06 }$ &-                             & $ 2.66 ^{+ 0.31 } _{- 0.39 }$ & 0.68\\\hline
			A&NGC3893 & $ 0.49 ^{+ 0.04 }_{- 0.03 }$ &-                             & $ 12.05 ^{+ 2.34 } _{- 2.52 }$ & 1.27\\\hline
			A&NGC3949 & $ 0.37 ^{+ 0.03 }_{- 0.06 }$ &-                             & $ 18.98 ^{+ 10.56 } _{- 14.10 }$ & 0.29\\\hline
			A&NGC3953 & $ 0.65 ^{+ 0.07 }_{- 0.07 }$ &-                             & $ 3.69 ^{+ 2.24 } _{- 2.99 }$ & 0.54\\\hline
			B&NGC3992 & $ 0.77 ^{+ 0.03 }_{- 0.02 }$ &-                             & $ 15.28 ^{+ 1.36 } _{- 1.60 }$ & 0.82\\\hline
			B&NGC4051 & $ 0.43 ^{+ 0.07 }_{- 0.10 }$ &-                             & $ 2.72 ^{+ 1.40 } _{- 2.20 }$ & 0.92\\\hline
			B&NGC4088 & $ 0.31 ^{+ 0.01 }_{- 0.01 }$ &-                             & $ 4.62 ^{+ 0.66 } _{- 0.71 }$ & 0.60\\\hline
			B&NGC4100 & $ 0.72 ^{+ 0.03 }_{- 0.03 }$ &-                             & $ 5.18 ^{+ 0.57 } _{- 0.64 }$ & 1.28\\\hline
			B&NGC4138 & $ 0.71 ^{+ 0.08 }_{- 0.04 }$ & $ 0.53 ^{+ 0.17 }_{- 0.22 }$& $ 4.04 ^{+ 1.24 } _{- 1.60 }$ & 1.50\\\hline
			B&NGC4157 & $ 0.38 ^{+ 0.03 }_{- 0.03 }$ & $ 0.46 ^{+ 0.09 }_{- 0.16 }$& $ 8.21 ^{+ 1.09 } _{- 1.00 }$ & 0.55\\\hline
			B&NGC4183 & $ 0.67 ^{+ 0.12 }_{- 0.06 }$ &-                             & $ 1.43 ^{+ 0.15 } _{- 0.19 }$ & 0.18\\\hline
			B&NGC4559 & $ 0.33 ^{+ 0.02 }_{- 0.03 }$ &-                             & $ 2.33 ^{+ 0.15 } _{- 0.14 }$ & 0.24\\\hline
			B&NGC5005 & $ 0.44 ^{+ 0.07 }_{- 0.09 }$ & $ 0.51 ^{+ 0.08 }_{- 0.08 }$& $84.90 ^{+ 53.76 } _{- 81.03 }$& 0.09\\\hline
			B&NGC6503 & $ 0.53 ^{+ 0.01 }_{- 0.01 }$ &-                             & $ 2.36 ^{+ 0.04 } _{- 0.05 }$ & 2.80\\\hline
			C&UGC06983 & $ 0.65 ^{+ 0.14 }_{- 0.07 }$ &-                             & $2.16 ^{+ 0.24 } _{- 0.31 } $ & 0.71\\\hline
			C&UGC07261 & $ 0.57 ^{+ 0.17 }_{- 0.15 }$ &-                             & $ 0.64 ^{+ 0.17 } _{- 0.22 }$ & 0.21\\\hline
			C&UGC07690 & $ 0.70 ^{+ 0.10 }_{- 0.06 }$ &-                             & $ 0.21 ^{+ 0.06 } _{- 0.08 }$ & 0.67\\\hline
			C&UGC07866 & $ 0.43 ^{+ 0.09 }_{- 0.13 }$ &-                             & $ 0.03 ^{+ 0.04 } _{- 0.02 }$ & 0.61\\\hline
			C&UGC08490 & $ 0.79 ^{+ 0.01 }_{- 0.01 }$ &-                             & $ 0.87 ^{+ 0.05 } _{- 0.06 }$ & 1.52\\\hline
			C&UGC08550 & $ 0.63 ^{+ 0.16 }_{- 0.08 }$ &-                             & $ 0.27 ^{+ 0.03 } _{- 0.03 }$ & 0.69\\\hline
			C&UGC08699 & $ 0.77 ^{+ 0.03 }_{- 0.02 }$ & $ 0.67 ^{+ 0.02 }_{- 0.02 }$ & $ 8.22 ^{+ 0.61 } _{- 0.68 }$ & 0.69\\\hline
			C&UGC09992 & $ 0.47 ^{+ 0.10 }_{- 0.17 }$ &-                             & $ 0.03 ^{+ 0.03 } _{- 0.02 }$ & 0.32\\\hline
			C&UGC10310 & $ 0.54 ^{+ 0.15 }_{- 0.23 }$ &-                             & $ 0.43 ^{+ 0.09 } _{- 0.12 }$ & 0.58\\\hline
			C&UGC12506 & $ 0.79 ^{+ 0.01 }_{- 0.01 }$ &-                             & $ 19.76 ^{+ 1.70 } _{- 1.50 }$ & 1.73\\\hline
			D&NGC7331 & $ 0.35 ^{+ 0.01 }_{- 0.01 }$ & $ 0.48 ^{+ 0.09 } _{- 0.17 }$  & $ 20.83 ^{+ 0.74 } _{- 0.72 }$ & 0.83\\\hline
			D&NGC7793 & $ 0.54 ^{+ 0.04 }_{- 0.04 }$ &-                              & $ 1.33 ^{+ 0.27 } _{- 0.32 }$ & 0.90\\\hline
			D&NGC7814 & $ 0.77 ^{+ 0.04 }_{- 0.02 }$ & $ 0.66 ^{+ 0.03 } _{- 0.03 }$  & $ 25.32 ^{+ 1.76 } _{- 1.67 }$ & 1.42\\\hline
			D&UGC02259 & $ 0.77 ^{+ 0.05 }_{- 0.02 }$ &-                             & $ 1.39 ^{+ 0.08 } _{- 0.09 }$ & 6.35\\\hline
			D&UGC03546 & $ 0.65 ^{+ 0.03 }_{- 0.03 }$ & $ 0.37 ^{+ 0.03 } _{- 0.04}$ & $ 9.15 ^{+ 0.61 } _{- 0.60 }$ & 0.98\\\hline
			D&UGC06446 & $ 0.69 ^{+ 0.15 }_{- 0.08 }$ &-                             & $ 0.98 ^{+ 0.09 } _{- 0.11 }$ & 0.44\\\hline
			D&UGC06930 & $ 0.51 ^{+ 0.14 }_{- 0.16 }$ &-                             & $ 1.53 ^{+ 0.27 } _{- 0.30 }$ & 0.28\\\hline
			D&UGC06983 & $ 0.66 ^{+ 0.12 }_{- 0.09 }$ &-                             & $ 2.13 ^{+ 0.22 } _{- 0.27 }$ & 0.71\\\hline
			D&UGC07261 & $ 0.57 ^{+ 0.16 }_{- 0.15 }$ &-                             & $ 0.62 ^{+ 0.15 } _{- 0.19 }$ & 0.19\\\hline
			D&UGC07690 & $ 0.71 ^{+ 0.11 }_{- 0.07 }$ &-                             & $ 0.21 ^{+ 0.06 } _{- 0.07 }$ & 0.64\\	
			\hline			
		\end{tabular}
	\caption{\label{tabelaresultado2}The maximum likelihood estimation for the $\Upsilon_{\text{*D}}$,$\Upsilon_{\text{*B}}$ and $M_\text{200}$ parameters,	and the goodness of fit $\chi_\text{red}^2|_{\beta=0}$ for each galaxy in the case of $\beta =0$.}
\end{table}
\clearpage
\clearpage
\clearpage
\clearpage

\section{Discussion and conclusions}

\label{section6}

In this work we have used observational data from the SPARC catalogue to constrain the properties of modified gravity models in the presence 
of  dark matter, and assuming that the fifth force couples weakly to baryons but with unrestricted strength to dark matter.
Contrary to some previous work, our aim is not to replace dark matter with modified gravity but to see how much modified gravity can improve the rotation curve fit. Since baryons are assumed to be weakly coupled, we do not need to invoke a screening mechanism, and the Yukawa term is left free and constant for every galaxy.
We considered four different sets of 10 galaxies each and we found the region in the parameter space for $\lambda$ and $\beta$ that are 
allowed by the data. To the best of our knowledge, this is the largest set ever analysed in the context of modified gravity.  We found that in all the data sets the standard  $\beta = 0$ model gives a much worse fit than a value different from zero, with preference for a positive value, corresponding to an attractive Yukawa force.  
We have also calculated for each galaxies the values of the parameters $\Upsilon_{*\text{D}}$, $\Upsilon_{*\text{B}}$ and 
$M_{200}$. We find that the presence of the attractive fifth force reduces the need for dark matter by 20\% in mass, on average. 

We have then combined all the data sets together to find the allowed region in the parameter space.    The values for the parameters $\beta$ and $\lambda$ 
are: $\beta = 0.34\pm0.04$ and $\lambda = 5.61\pm0.91 $ kpc. 
The Bayesian evidence ratio strongly favors the Yukawa model, to more than 8$\sigma$ for the combined dataset, with respect to the $\beta=0$ case.

We notice that the $\beta$ value is remarkably close to $\beta=1/3$, the value predicted by one of the simplest modified gravity model, the $f(R)$ theory. However, as mentioned in Sec. 2, we should interpret $\beta$ as the product of a small baryon coupling times a large dark matter coupling, neither of which would be close to the $f(R)$ prediction. So the underlying model can be identified with a scalar-tensor theory with non-universal coupling, rather than the specific form $f(R)$.

It is clear that we cannot conclude that standard gravity is ruled out. Rather, we found that a model of the baryon components (gas, disk and bulge), plus a NFW profile for the dark matter, plus an attractive Yukawa term, fits much better the rotation curves of our sample than a similar model but without the Yukawa correction. The SPARC catalog contains normal galaxies as well as LSB and dwarfs. The latter two types are known to be poorly fitted by a NFW profile\cite{moore1994evidence,mcgaugh1998testing,cote2000various}, so it will be interesting in a future work to try to fit them separately with different profiles. In our set of 40 galaxies, however, only 10 galaxies are of this kind so we believe our choice of NFW for all galaxies was justified. Whether this results holds assuming different modelling for the baryon or the dark matter component, remains to be seen.

\section*{Acknowledgments}

This work was supported by the SFB-Transregio TR33 "The Dark Universe". 
The work of AOFA was supported by CAPES, grant number 88881.135537/2016-01. 
AOFA wants to acknowledge discussions with Davi Rodrigues about mass-to-light ratio in galaxies. 

\bibliographystyle{JHEP}
\bibliography{report}

\end{document}